# HALLEY'S COMET OVER NEW ZEALAND IN 1910: AN OVERVIEW OF A SPECTACULAR APPARITION

**John Drummond**
*University of Southern Queensland, West Street, Toowoomba, Queensland 4350, Australia.*
E-mail: kiwiastronomer@gmail.com

**Wayne Orchiston**
*University of Science and Technology of China, Hefei, Anhui, China; and University of Southern Queensland, West Street, Toowoomba, Queensland 4350, Australia.*
E-mail: wayne.orchiston@gmail.com

**and**

**Carolyn Brown and Jonathan Horner**
*University of Southern Queensland, West Street, Toowoomba, Queensland 4350, Australia.*
E-mails: carolyn.brown@unisq.edu.au
jonti.horner@unisq.edu.au

**Abstract: Halley's Comet (1P/Halley) returns to the inner Solar System approximately every 76 years. In 1910, the comet put on a particularly wonderful display, whereby the comet attained an apparent magnitude of 1 or brighter, and had a tail of up to 120° in length. Many New Zealanders saw 1Comet P/Halley and submitted articles to newspapers across the land. These articles were mostly from lay-people who reported their observations of this, the most famous of all comets. There were, however, numerous serious amateur and professional astronomers who made detailed observations and took world-class photographs of the comet. In this paper, we present the results of a thorough search through historical records and media publications in Aotearoa/New Zealand, to investigate how extensively Comet 1P/Halley was observed from this island nation. Where appropriate, we compare the observations described with international accounts, and find that those made from Aotearoa/New Zealand often echo what observers were witnessing around the world. We find that thousands of New Zealanders observed Comet 1P/Halley in 1910, and that the comet raised the interest in astronomy across the land resulting in the establishment of a number of astronomical societies that remain active to the current day.**

## 1 INTRODUCTION AND METHODOLOGY

1910 (CE) was an exceptional year for naked-eye comets since two great comets appeared within months of each other. Drummond et al. (2025) have written extensively on the appearance of the Great January Comet (C/1910 A1) in January 1910.1 They revealed that the second recorded sighting of this comet (after it was first spotted by diamond mine workers in South Africa) was possibly made from Gisborne, Aotearoa/New Zealand (referred to simply as New Zealand hereafter). They found that few New Zealanders observed this comet, and concluded that the apparition of C/1910 A1 was something of a missed opportunity for New Zealand astronomers. Perhaps this was due to the comet appearing in the summer dawn sky and the Sun rising around 5 am, before New Zea-land adopted Daylight Saving. In addition, some of those who did witness the Great January Comet incorrectly misidentified it as 1P/Halley. A few months later, in May 1910, Halley's Comet reached its brightest magnitude for that apparition. Comet Halley was much more extensively observed by lay observers and amateur and professional astronomers than C/1910 A1. New Zealand localities mentioned in the text are shown in Figure 1.

Drummond et al. (2025: 689–690) highlight the excitement that was building in New Zealand (and the world) in late 1909 and early 1910 as people waited for the expected return of Halley's Comet (henceforth 1P/Halley—unless in a quote). In Section 2, we show that this burgeoning excitement was bolstered through books, newspaper articles, and public lectures. The elderly recalled the 1835 apparition of 1P/Halley with fond memories, stating it was a "… magnificent object …" (Smith, 1986: 2) and that it had a 20° tail (A Famous Comet, 1894). Many New Zealanders wanted to view the 1910 apparition of this, the most famous of all comets, for themselves.

In this paper, we present a detailed review of the 1910 apparition of 1P/Halley, as seen from New Zealand. Section 3 sets the scene,

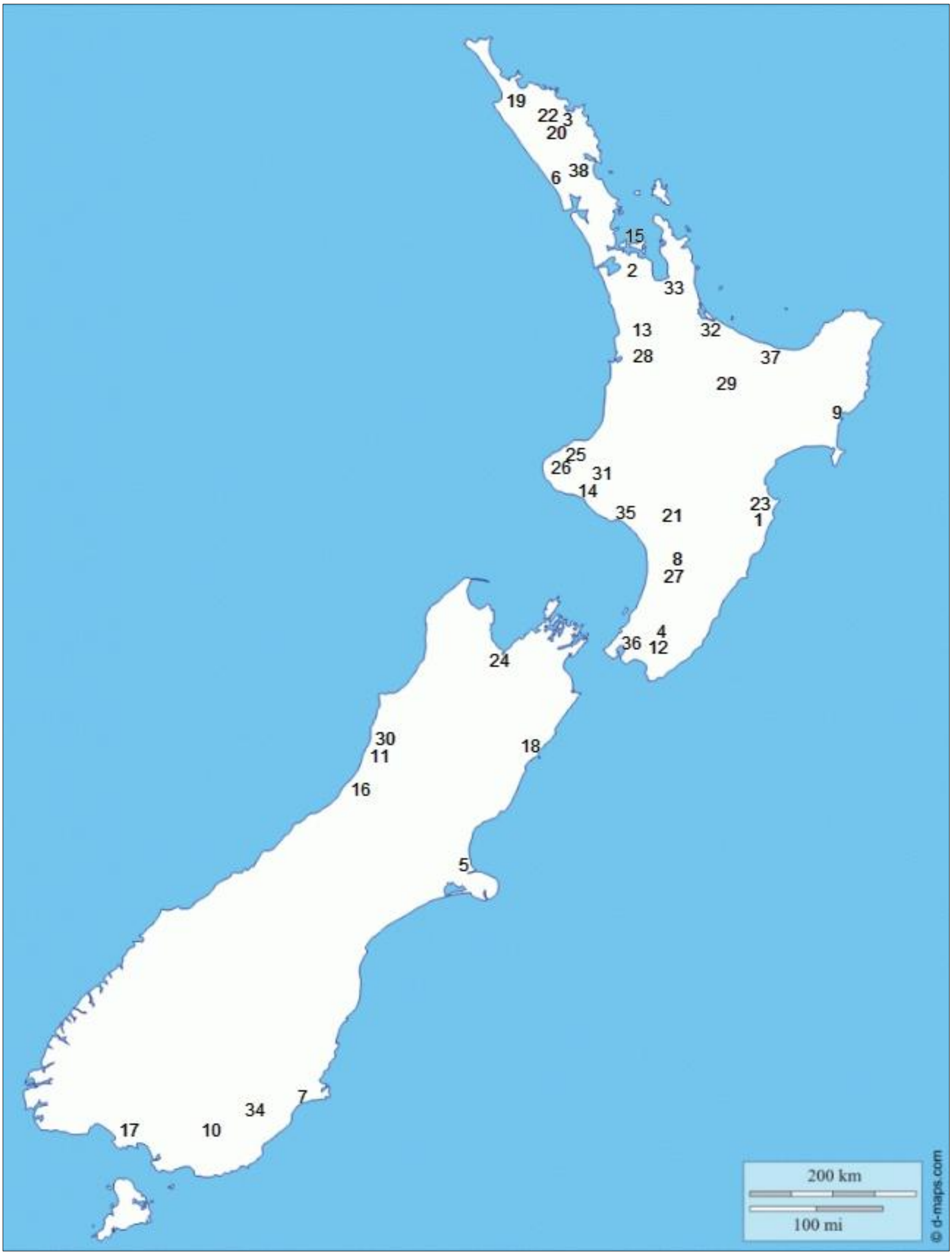

Figure 1: New Zealand localities mentioned in the text. Many New Zealanders, from the top of the North Island to the bottom of the South Island, had awe-inspiring views on Comet 1P/Halley in 1910. Key: 1 = Amuri; 2 = Auckland; 3 = Bay of Islands; 4 = Carterton; 5 = Christchurch; 6 = Dargaville; 7 = Dunedin; 8 = Fielding; 9 = Gisborne; 10 = Gore; 11 = Greymouth; 12 = Greytown; 13 = Hamilton; 14 = Hawera; 15 = Hibiscus Coast; 16 = Hokitika; 17 = Invercargill; 18 = Kaikoura; 19 = Kaitaia; 20 = Kerikeri; 21 = Manawatu; 22 = Mangonui; 23 = Napier; 24 = Nelson; 25 = New Plymouth; 26 = Okato; 27 = Palmerston North; 28 = Rakaunui; 29 = Rotorua; 30 = Runanga; 31 = Stratford; 32 = Tauranga; 33 = Thames; 34 = Waitahuna; 35 = Wanganui; 36 = Wellington; 37 = Whakatane; 38 = Whangarei (source map: http://toursmaps.com/wp-content/uploads/2017/07/blank-map-of-new-zealand_1-776x1024.gif; map modifications: John Drummond).

through the introduction of the first international observations of 1P/Halley by photography and visual use of large telescopes. Max Wolf photographically detected the comet in September 1909, and then, a few days later, Professors Burnham and Barnard visually observed it with the 40-inch refractor at Yerkes Observatory, USA. Section 4 describes the first observations of 1P/Halley from New Zealand in its 1910 apparition. It was reported as being photographed by astronomers at Meeanee Observatory, before being detected visually by Joseph Ward of Wanganui. Section 5 takes an in-depth look at observations of the comet that were made by professional and amateur New Zealand astronomers, as well as the public. Where appropriate, we compare domestic observations with those made internationally, and note similarities and differences. We also discuss the astrophysical traits of 1P/Halley as it reached first perihelion, and then perigee, where the tail is believed to have swept across the Earth.

In order to find observations of 1P/Halley from New Zealand, we used the online search engine, Papers Past, to search New Zealand newspaper articles using the keywords 'Halley' and 'Comet'. The search dates used were confined to the period from January 1908 to December 1912 (inclusive). In all, we found 1,157 articles relating to 'Halley's Comet', although a number of these were duplicate reports, where one newspaper often repeated another newspaper's content (or their common source). We also investigate the photography of 1P/Halley from New Zealand, and discuss the 'key players' in that endeavour.

In Section 6, we look at the influence that the appearance of 1P/Halley had on the general public and some of the fears and non-scientific concerns that it generated. We also present details of humorous events surrounding the observation of the comet. In Section 7, we discuss whether the appearance of 1P/Halley was influential in propagating an interest in astronomy across New Zealand, and whether local astronomical societies were formed as a result. Finally, in Section 8, we present our concluding comments based on the key findings within this paper.

## 2 ANTICIPATION BUILDS FOR THE APPEARANCE OF COMET 1P/HALLEY

1P/Halley is undoubtedly the most well-known comet in our Solar System—ever since Sir Edmund Halley (1656–1742, Figure 2) recognised the repetitive return of a naked-eye comet every 75–76 years and then correctly predicted its next return in 1758. Halley (1705: 21) stated in his book *Astronomy of Comets* that

> ... there are many Things which make me believe that the Comet which Apian ob[s]erv'd in the Year 1531 was the [s]ame with that which Kepler and Longommtanm took Notice of and de[s]crib'd in the Year 1607. and which I my [s]elf have [s]een return, and ob[s]erv'd in the Year 1682. All the Elements agree, and nothing [s]eems to contradid this my Opinion …

On the next page Halley (1705: 22) predicted, "Hence I dare venture to foretell, That it will return again in the Year 1758." It did. On Christmas night in 1758, a comet was observed by Johann Georg Palitzsch (1723–1788) (Vsekhsvyatskii, 1964: 128), a German amateur astronomer. Halley, using Newtonian laws of gravity, was justifiably cemented into the pages of history, and the comet quickly became known as 'Halley's Comet', in fitting recognition of his achievement. Unfortunately, Halley died in 1742, sixteen years before his predicted comet returned.

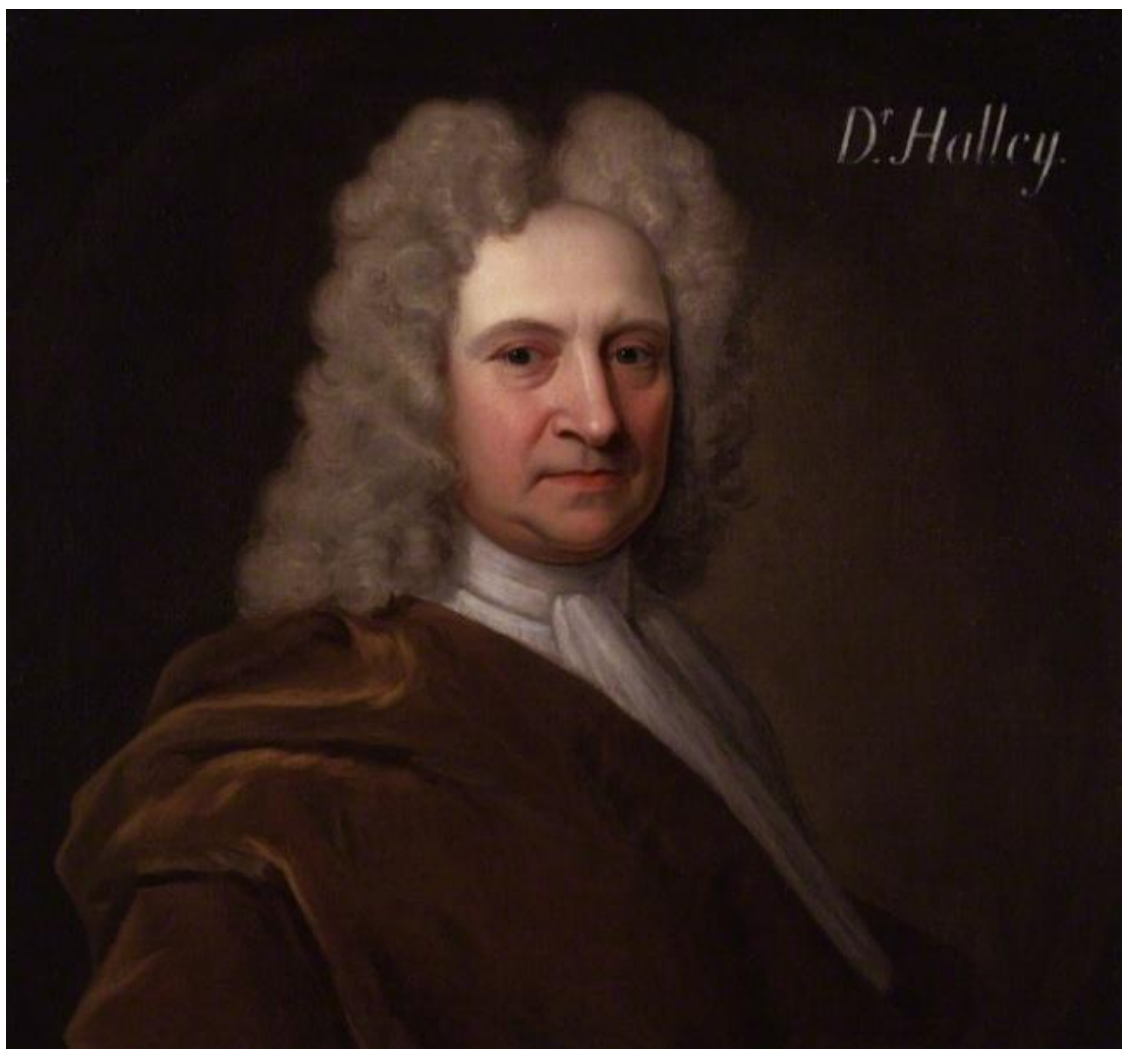

Figure 2: Edmund Halley correctly identified the periodicity of a comet that returned to the inner Solar System every 75–76 years and predicted its appearance in 1758. Portrait of Halley by Richard Phillips (source: Wikipedia, 2016).

After 77 years, 1P/Halley returned in 1835. The 1835 appearance is described in Kronk (2003) and Vsekhsvyatskii (1964), as well as in other literature. Vsekhsvyatskii (1964: 161–164) states that by September 1835 it was "... seen everywhere with the naked eye …" until February 1836. Kronk (2003: 108–116) provides a more thorough description, describing not only the magnitude but also the tail length. 1P/Halley was first seen without optical aid on 19 September 1835 (UT) and brightened to magnitude 1 in October with a tail of up to 45° long in the middle of that month. The comet re-mained a naked-eye target until the middle of March 1836.

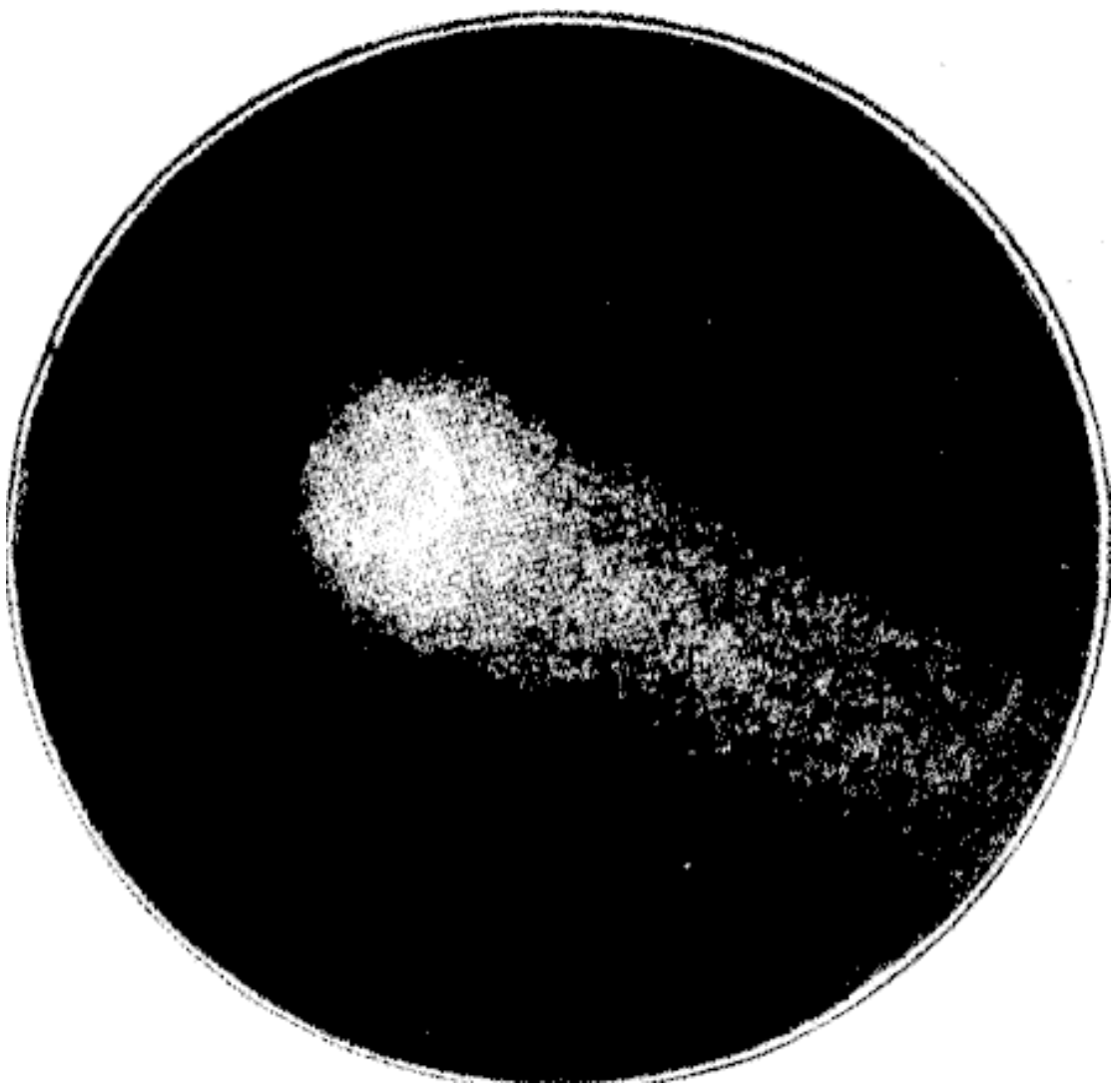

Figure 3: Comet 1P/Halley as sketched in 1835 by an unknown observer (after Halley's Comet as Seen in 1835, 1910).

With such a splendid appearance in 1835, the next generation was eagerly anticipating its return in 1910. An example is that four years before its expected arrival, a lecture on 1P/Halley was given by Victoria University College of Wellington's Professor Maclaurin to a large audience in the Manawatu, New Zealand (Professor Maclaurin's Lecture, 1906).

## 2.1 1835 Apparition of Comet 1P/Halley

Many witnessed the spectacular appearance of 1P/Halley in 1835. Both Vsekhsvyatskii's *Physical Characteristics of Comets* (Vsekhsvyatskii, 1964) and Gary Kronk's (2003) *Cometography Volume 2* record the visit in detail, stating that at its brightest, the comet reached an apparent magnitude of +1 (Kronk, 2003: 112–113; Vsekhsvyatskii, 1964: 163), and had a tail up to 45° long (Kronk, 2003: 113). The coverage of the Papers Past system does not extend back to 1835, so no direct New Zealand observations can be gleaned from the newspapers of that time; Papers Past coverage starts from 1 January 1839, four years after the 1835 visit. There were, however, recollections of the fortunate few who managed to see both the 1835 and 1910 apparitions. Several of these New Zealand witnesses were interviewed in 1910 and their descriptive comments published. 1P/Halley was stated to be a naked eye object for the whole of October 1835 with a 20° tail (Comets and Their Ways, 1909). Some, such as Mr Woolford, who saw it in 1835, "... did not think so much of it." (Local and General, 1910i). One elderly Māori lady was interviewed who saw the comet as a child in 1835 and then again in 1910 (Halley's Comet, 1910ag), and stated

> ... she had seen it [1P/Halley] that morning, and … a similar comet which the Natives termed "Hapainatia" when she was a girl about ten years of age. She also stated that there was a long spell of exceedingly dry weather at the time and the Natives attributed this to the influence of the comet. Her forefathers had handed it down, that the presence of "Hapainatia" in the heavens meant the absence of cold, wet weather, and a long spell of sunshine. I have since been reckoning it up and I find that this was somewhere about the time when Halley's comet last appeared …

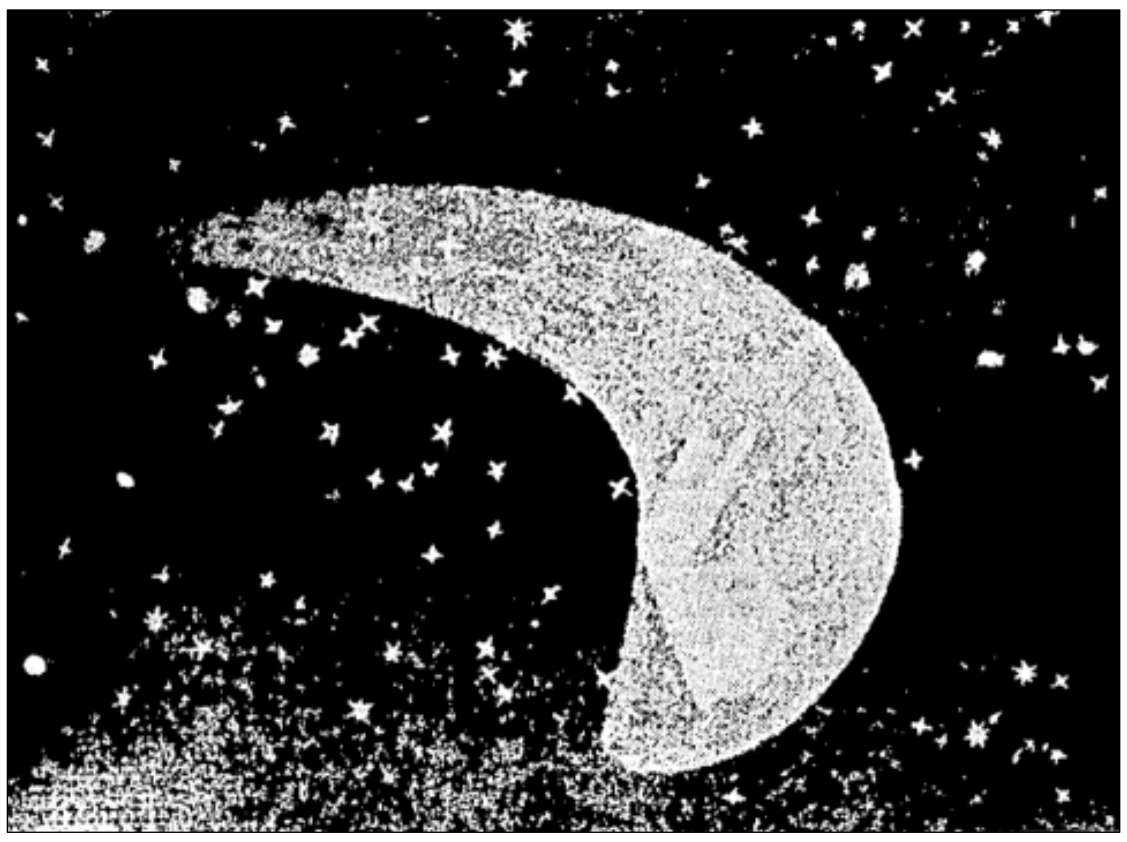

Figure 4: A sketch of 1P/Halley from 1835 published in a newspaper on 19 February 1910 (after Halley's Comet: Seventy Four Years Ago).

The *Otago Witness* published a drawing made of 1P/Halley's 1835 appearance, shown here in Figure 3. No details were provided, apart from the caption, "Halley's Comet as Seen in 1835" (Halley's Comet as Seen in 1835, 1910). The article title for the drawing in Figure 4 stated, "This drawing, by N. Rosenthal, is from a telescopic view, and affords an idea of the vivid impression the comet makes upon a beholder." Mackrell (1985: 112) quotes the views that Douglas MacIntyre had of 1P/Halley in 1835 when he was 22, stating,

> He [1P/Halley] was an awe-inspiring sight … The head looked like a huge star, the tail like a huge shaft of light. He was visible over Wellington in various lengths for about 3 or 4 weeks.

## 2.2 Awaiting the Arrival of the Comet

As stated, years before the return of 1P/Halley in 1910, people were anticipating its return. The first New Zealand newspaper reference to the forthcoming reappearance was in an article on comets published in the *Daily Southern Cross* newspaper on 1 February 1865. The author, discussing the appearance of historically bright

comets—including the previous appearances of the 1P/Halley and Halley's prediction of its 1758 return, stated, "We saw its last visit in 1835, and it will not again be seen until 1910." (Comets, 1865). Interestingly, the article omitted any reference to the Great Southern Comet (C/1865 B1), which was visible to the naked-eye in strong evening twilight around late January 1865 (Kronk: 2003: 335–336), weeks before the date when the newspaper was published.

Guillemin (1877: 106–108) discussed the question of 1P/Halley's periodicity, stating that it varied due to planetary perturbations, as did Halley himself (Ball, 1910: 7), and Clairaut in 1758 (Ball, 1910: 8–9). The period between the 13 March 1759 UT perihelion to the next perihelion on 16 November 1835 UT was found to be 76 years, 8 months and 3 days (Guillemin 1877: 106–107). Guillemin (1877: 107), using Pontecoulant's calculations (Anonymous, 1875: 175–176), suggested that the next perihelion would occur on 24 May 1910. In actual fact, perihelion occurred one month earlier than predicted, on 20 April 1910 UT (Kronk, 2007: 149), something that J.A. Hardcastle of the British Astronomical Association's Comet Section predicted in 1909 after Max Wolf's photographic recovery (and astrometric positions) of the comet in September 1909 (Hardcastle, 1909: 456)—see below. This variation reveals the difficulty that astronomers had at that time in making long-range perihelion predictions, because of the limited arcs over which most comets were observed. This was further exacerbated by the limited knowledge of cometary motion, preventing astronomers from allowing for the non-gravitational effects on comets caused out-gassing, for example (Hartmann, et al., 1997: 517–518).

In his book, *A Comet Called Halley*, Ian Ridpath (1985) wrote that the German Astronomical Association (the Astronomische Gesellschaft), offered a 1,000 Mark prize (then worth £50) for the best prediction of the comet's 1910 perihelion passage. Predictions were submitted anonymously from international astronomers. Philip H. Cowell and Andrew C.D. Crommelin of the Royal Observatory in Greenwich took up the challenge and suggested that 16 April 1910 would be the perihelion date. Ridpath (*ibid.*) states that "Observatories around the world raced each other to sight the returning comet." As stated below, Max Wolf of Heidelberg spotted it first on photographs he took early on the morning of 12 September 1909 UT. The first visual observations were made from Yerkes Observatory (USA) with the 40-inch refractor three nights later. With these observed astrometric positions, the competition predictions could be judged against the actual path of 1P/Halley. When the submitted envelopes were opened, the winners were Cowell and Crommelin.

Ridpath (1985) wrote that the winners decided to further refine their calculations by taking the gravitational effects of all the planets into account, and published their results in a 1910 book titled *Investigation of the Motion of Halley's Comet from 1759 to 1910*. Their revised conclusion was that perihelion would occur on 17 April 1910 UT. However, as Ridpath highlights, despite Cowell and Crommelin's best efforts, the comet arrived at perihelion three days later than predicted on 20 April UT. They concluded, "There are forces of an unrecognized kind influencing the comet's motion." These "unrecognised forces" remained a mystery until Fred Whipple (1906–2004) mooted his 'icy conglomerate' hypothesis, later to be called the 'dirty snowball' hypothesis, in 1949. He suggested that comets could have their orbits slightly altered as sublimating gas escaping from the nucleus produced a rocket-like effect with thrust, as Newton's Third Law states, "Every action has an equal and opposite reaction". This hypothesis became theory as the Giotto Space Probe photographed jets coming off 1P/Halley during it's 1986 apparition. Astronomers now take these non-gravitational forces into account when calculating the path of a comet.

### 2.3 The Great January Comet of 1910 (C/1910 A1): Was it Comet 1P/Halley?

Several months before 1P/Halley became visible to the naked eye in 1910, a bright comet (C/1910 A1) was discovered in the January dawn sky by diamond mine workers in South Africa. At discovery, the 'Great January Comet' (Figure 6) had an apparent magnitude of –1 and sported an impressive tail (Bortle, 1998). Drummond et al. (2025) describe observations of the comet from New Zealand and conclude that few people observed it (see Figure 7) which shows the locations where the comet was observed from in New Zealand)—and if they did, they did not submit observations to New Zealand newspapers. In addition, the majority of those who did observe this impressive comet incorrectly assumed that it was 1P/Halley. Prior to perihelion, C/1910 A1 was solely visible from the Southern Hemisphere, although it became an easy target for observers in the Northern Hemisphere in the days after its perihelion passage.

### 2.4 Public Lectures about Comet 1P/Halley

As 1P/Halley drew nearer to its perihelion passage in April 1910, and then perigee in May

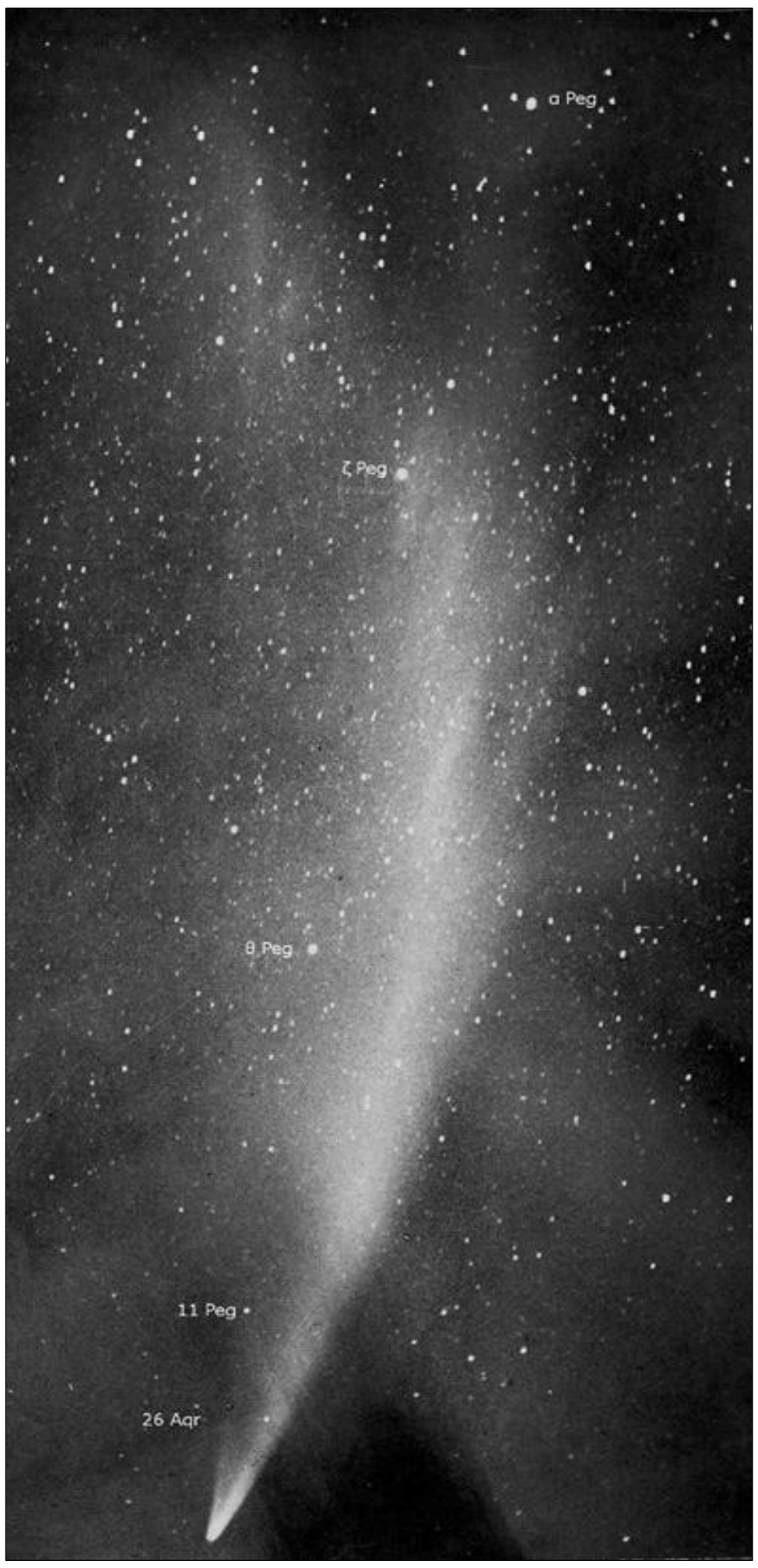

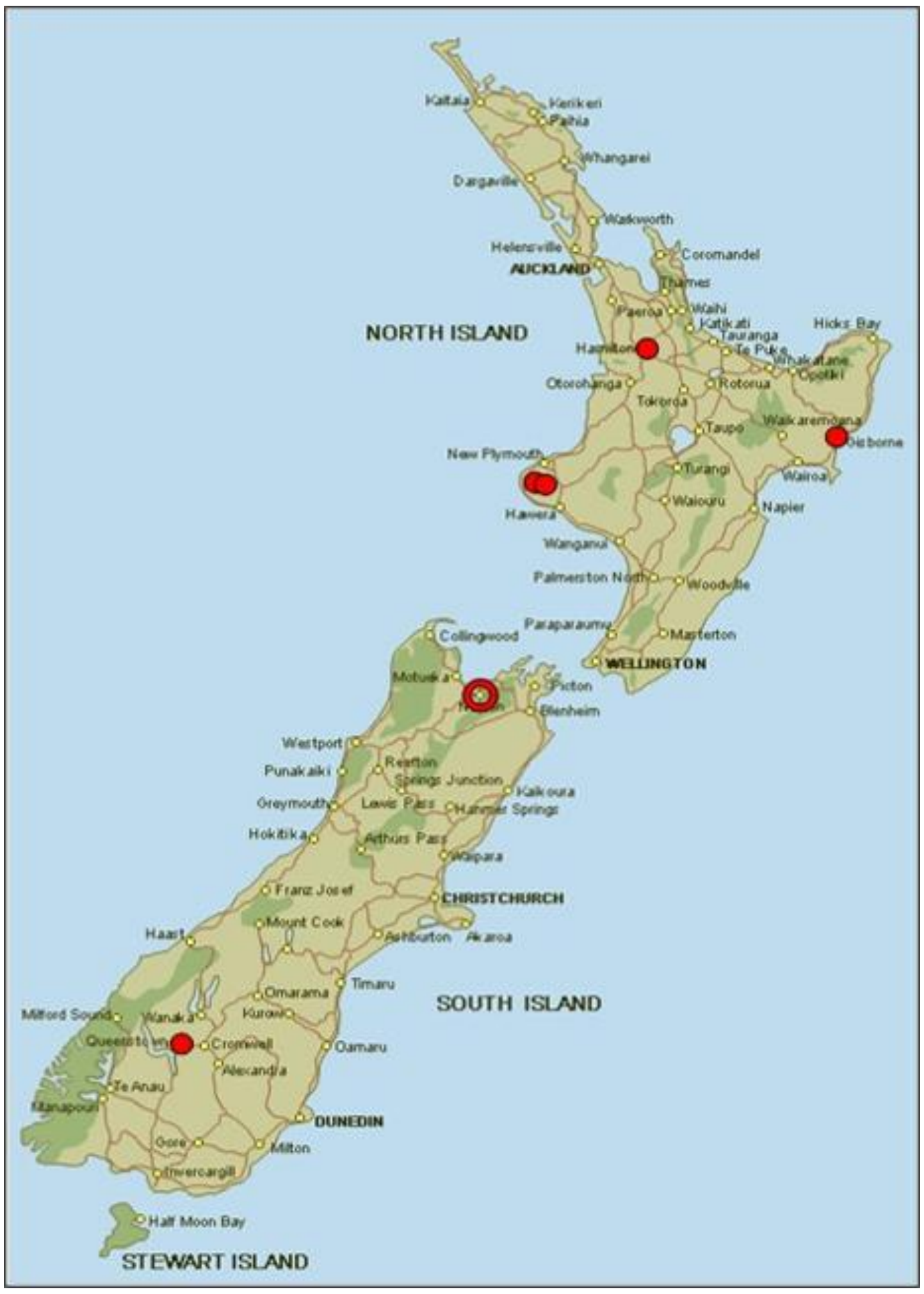

Figure 5 (left): C/1910 A1, The Great January Comet (Earthrise, n.d.). This remarkable comet was photographed from Lowell Observatory (Flagstaff, Arizona, USA) by Carl Lampland and Vesto Slipher on 28 January 1910 (UT), 11-days after the comet's perihelion (Bortle, 2010). Alpha Pegasi, at the top, is magnitude 2.48 in V-band, as detailed in the SIMBAD database (https://simbad.cds.unistra.fr/simbad/). Zeta (upper middle) is magnitude 3.1 in the V-band. In this photograph, the tail is approximately 27° long.

Figure 6 (right): New Zealand map showing the locations from which Comet C/1910 A1 was observed, based on newspaper records in the Papers Past archive. The open circle marks a non-newspaper report by Dennis Goodman's grandmother, Dennis being a Past-President of the Royal Astronomical Society of New Zealand. Several who saw C/1910 A1 assumed that it was 1P/Halley (base image: Back-pack New Zealand; map modifications: the authors).

1910, so too anticipation for views of this once in a lifetime (usually) comet increased. This was greatly encouraged through newspaper articles (e.g. Halley's Comet Approaching, 1908; Amusements, 1910; *Auckland Star*, 1910), books (e.g. Chambers, 1909: 124–125; Elson, 1910: 63–75; Guillemin, 1877: 107), and public lectures (e.g. *New Zealand Herald*, 1909; *Northern Advocate*, 1910; *Town and Country*, 1910). Based on a Papers Past search using the words, 'Comet, Halley, Lecture' between the dates of 1 January 1900 and 31 December 1911, public lectures on Halley's Comet were popular across New Zealand. In the 12 years investigated, 51 lectures were advertised in New Zealand newspapers. Of course, other talks were likely given that were not advertised in newspapers, and are therefore not included in this analysis. Table 1 displays the dates and locations where talks were given, as well as the speaker; whilst Table 2 reveals the most popular cities for talks. A brief discussion of the more prolific speakers is presented below.

The most prolific lecturers were Reverend Dr. David Kennedy (who gave 8 lectures), Joseph Taylor (7), Professor Hugh William Segar (6), Clement L. Wragge (5), Reverend Paul Wynyard Fairclough (4), Professor Richard Cockburn McLaurin (3), and Mr Frank H. Hitchings (3)—who owned a 6-inch refractor, which for many years was among the largest privately owned refracting telescopes in New Zealand. Other lecturers, such as Bickerton, Hudson, Scott-Bennett, Aldridge, Walker, Jellie, Scott,

Table 1: Public lectures on Halley's Comet in New Zealand between 1901 and 1911 as recorded in Papers Past. At least 46 lectures were given in 1909 and 1910. An asterisk (*) indicates that the lecture was repeated due to overcrowding and people being turned away the first time the lecture was given. HC = Halley's Comet.

| Date | Lecturer | Topic | Location |
|---|---|---|---|
| 21 June 1901 | Reverend P.W. Fairclough | 'The Comet, the Meteor, and the Moon' | Dunedin |
| 12 July 1901 | Professor McLaurin | Comets (including HC) | Wanganui |
| 15 August 1905 | Reverend Dutton | 'Comets and Meteors' | Gore |
| 04 October 1905 | Professor McLaurin | 'Comets' | Wellington |
| 22 March 1906 | Professor McLaurin | 'Comets' | Manawatu? |
| 08 March 1909 | Professor Wragge | Wonders of the Universe (including HC) | Wanganui |
| 16 August 1909 | Professor Segar | HC | Auckland |
| 24 August 1909 | Professor Segar | HC | Auckland |
| 01 September 1909 | Reverend J.S. Ponder | Comets | Waitahuna |
| 12 September 1909 | Reverend P.W. Fairclough | HC | Dunedin |
| 15 September 1909 | Professor T.H. Laby | HC | Victoria Uni. Wellington |
| 07 October 1909 | Professor Wragge | 'The Majesty of Creation' | Nelson |
| 16 October 1909 | Professor Wragge | Wragge's 'Farewell Lecture' | Wellington |
| 18 October 1909 | Captain Hewitt | HC | Manawatu |
| 25 February 1910 | Professor Bickerton | HC | Christchurch |
| 21 March 1910 | Professor Wragge | HC | Whangarei |
| 28 March 1910 | Professor Bickerton | HC | Auckland |
| 25 April 1910 | Mr Joseph Taylor | HC | Nelson? |
| 25 April 1910 | Professor Segar | HC | Auckland |
| 01 May 1910 | Mr W.W. Collins | HC | Wellington? |
| 01 May 1910 | Mr A.B. Thompson | HC | Wellington |
| 04 May 1910 | King/Adams/Kennedy | HC | Wellington |
| 10 May 1910 | Mr G. Hudson | HC | Wellington |
| 12 May 1910 | Reverend P.W. Fairclough | HC | Auckland |
| 14 May 1910 | Mr G. Hudson | HC | Wellington |
| 15 May 1910 | H. Scott-Bennett | HC | Auckland |
| 15 May 1910 | H. Scott-Bennett | 'HC and the end of the world'. Nothing to fear. | Royal Albert Hall (Auckland) |
| 16 May 1910 | Reverend P.W. Fairclough | HC | Dunedin |
| 16 May 1910 | Mr Joseph Taylor | HC | Nelson? Inangahua? |
| 17 May 1910 | Professor Segar | HC | St Andrews Hall, Symonds St (Auckland) |
| 18 May 1910 | Mr Joseph Taylor | 'Halley's Comet and the Significance of its Present Return' | Trinity Hall, Greymouth |
| 18 May 1910 | Professor Segar | HC | St Andrews Hall, Symonds St (Auckland)* |
| 18 May 1910 | Reverend Kennedy | HC: lantern slides | Town Hall (Wellington) |
| 19 May 1910 | Reverend Kennedy | HC: lantern slides | Town Hall (Wellington)* |
| 19 May 1910 | Mr Joseph Taylor | 'Halley's Comet and the Significance of its Present Return' | Trinity Hall, Greymouth* |
| 22 May 1910 | Reverend D.D. Scott | HC | Onehunga, Auckland |
| 22 May 1910 | Reverend W. Jellie | HC | Auckland |
| 22 May 1910 | Reverend M. Walker | HC | Auckland |
| 24 May 1910 | Professor Segar | HC | Auckland* |
| 26 May 1910 | Mr Joseph Taylor | HC | Runanga |
| 27 May 1910 | Mr Joseph Taylor | HC and the eclipse | Hokitika |
| 28 May 1910 | Mr Joseph Taylor | HC | Runanga* |
| 28 May 1910 | Mr Hitchings (BAA) | HC and the Tasmania eclipse | Christchurch |
| 30 May 1910 | Mr Hitchings | HC and Tasmanian eclipse | Christchurch |
| 03 June 1910 | Mr George Aldridge | History of comets – HC | Auckland |
| 07 June 1910 | Mr Hitchings (BAA) | HC and Tasmanian eclipse | Christchurch |
| 08 June 1910 | Mr George Aldridge | History of comets – HC | Auckland |
| 11 July 1910 | Reverend Kennedy | HC | Wellington* |
| 19 July 1910 | Reverend Kennedy | HC | Christchurch |
| 17 August 1910 | Reverend Kennedy | HC | Wellington |
| 29 September 1910 | Reverend Kennedy | HC | Wellington |
| 05 June 1911 | --- | Movie – Planetary System (including HC) | Dunedin |
| 18 June 1911 | Clement Wragge | 'The Majesty of Creation' (including HC) | Wellington |

Table 2: Key New Zealand cities where public lectures about Halley's Comet were presented between 1901 and 1911, as revealed in Papers Past.

| Location | No. of Lectures | Key Speakers (sometimes spoke on several occasions) |
|---|---|---|
| Auckland | 14 | Segar, Bickerton, Scott-Bennett, Scott, Jellie, Walker, Aldridge |
| Wellington | 12 | McLaurin, Wragge, Laby, Collins, King, Adams, Kennedy, Hudson, Wragge |
| Christchurch | 5 | Bickerton, Kennedy, Hitchings |
| Dunedin | 3 | Fairclough |
| Nelson | 3 | Taylor, Wragge |
| Greymouth | 2 | Taylor |
| Manawatu | 2 | McLaurin, Hewitt |
| Runanga | 2 | Taylor |
| Wanganui | 2 | McLaurin, Wragge |
| Gore | 1 | Dutton |
| Hokitika | 1 | Taylor |
| Whangarei | 1 | Wragge |
| Waitahuna | 1 | Ponder |
| Unsure | 2 | Fairclough, Taylor |
| TOTAL | 51 | |

Dutton, Ponder, Laby, Hewitt, Collins, Thompson, King, and Adams, gave either one or two talks based on Papers Past reports. Of note is that John Grigg of Thames, who discovered three comets (see Drummond et al., 2026b; Orchiston, 1993; 2001b; 2016: 481–508), and Joseph Ward, the Director of Wanganui Observatory, insofar as newspaper publicity and accounts are concerned, never gave talks about 1P/Halley that we know of. By this stage, Grigg was advanced in years, dying at the age of 82 in 1920, ten years after 1P/Halley's appearance. However, he did photograph the comet in 1910 (Orchiston, 1995).

Following is a brief overview of the lives of the most prominent New Zealand lecturers on 1P/Halley, namely Kennedy, Taylor, Segar, Wragge, Fairclough, McLaurin, and Hitchings.

### *2.4.1 Reverend Dr. David Kennedy (8 lectures)*

David Kennedy (1864–1936, Figure 7; Orchiston, 2022) was a zealous amateur astronomer who funded the purchase of the 9-inch (22.9-cm) Cooke photovisual refractor depicted in Figure 8 for Meeanee Observatory, near Napier (23 in Figure 1), when he was stationed at the Marist Seminary there (see Orchiston, 2016: 346–363; 2022). At the time of the official opening of Meeanee Observatory in July 1907, in New Zealand the refractor was surpassed in aperture only by the 9.5-inch (24.1-cm) refractor at Wanganui Observatory (Orchiston, 2001a) that Joseph Ward had purchased second-hand from a UK amateur astronomer for £450 in 1902 (Orchiston, 2016: 301–305).[1] Meeanee Observatory and the 9-inch Cooke refractor are described in more detail below. Unfortunately, one year before the perihelion of 1P/Halley, Kennedy was transferred to Wellington to become the Rector of Saint Patrick's College in May 1909 (Orchiston, 2016: 348).

Kennedy, now living in Wellington, gave his first lecture about 1P/Halley in Wellington on Wednesday 4 May 1910. It was a joint presentation with Mr A. Hamilton, who spoke about

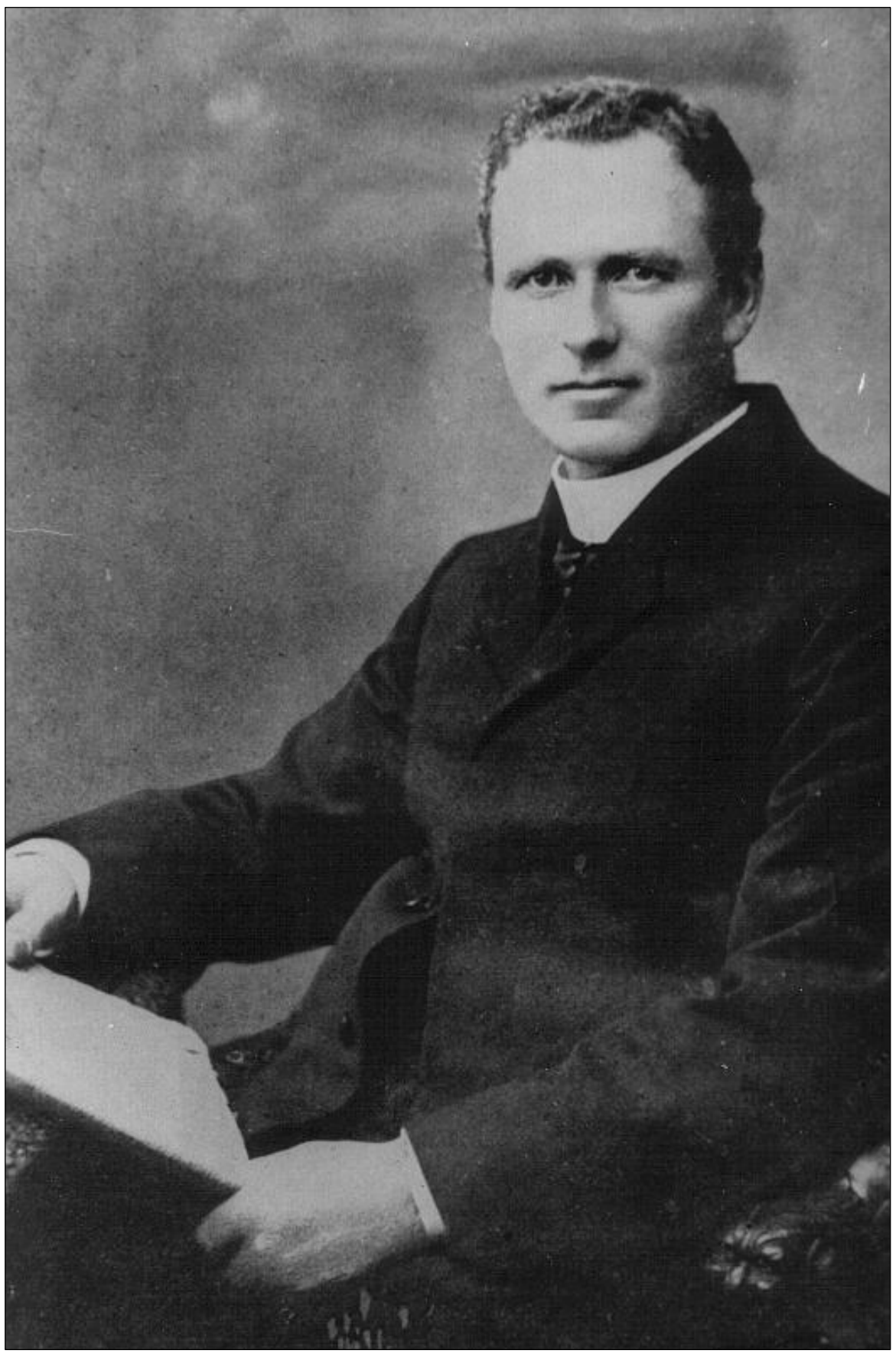

Figure 7 (right): The Reverend Dr David Kennedy, who established Meeanee Observatory, near Napier (courtesy: Marist Archives, Wellington).

the historical visits by 1P/Halley, and Mr Thomas King, who spoke about how Sir Edmund Halley determined the orbit of the comet using Sir Isaac Newton's celestial mechanics. Kennedy spoke on the astrophysics of comets and illustrated his talk with photographs of Halley's Comet taken from Meeanee Observatory only months before. Of note is that the newspaper article stated that

> ... the speaker [Kennedy] presented and explained an excellent series of photographic slides of the comet which had been taken by himself at the Meanee Observatory quite recently. (Lecture on Halley's Comet, 1910; our underlining).

Mackrell (1985: 141) points out that Kennedy travelled to Napier from Wellington (270 km to the southwest—see Figure 1) to help with positional calculations of 1P/Halley and probably to use the Cooke telescope to photograph the comet. At this stage, railways were the dominant mode of travel in New Zealand, but Wellington and Napier were both port cities so sea travel also was popular.

Kennedy spoke about 1P/Halley two weeks later at the Wellington Town Hall, on Wednesday 18 May, and again on Thursday 19 May 1910 because "... thousands of persons …" were turned away the previous night due to overcrowding (Halley's Comet Lecture, 1910). During these lectures Kennedy highlighted "The absence of facilities for astronomical observations in Wellington …" (News of the Day, 1910a), stating "... I find that in this big city there is no telescope to observe the phenomena of the heavens." The author of the article (based on a *New Zealand Times* article) raised the question as to why the funds donated by the Carter bequest "... of some thousands is still lying idle." (see also Mackrell, 1985: 141–142). Based on the purchase date of the 9-inch Meeanee refractor thirteen years later in 1923 by the Wellington City Corporation (below) and the official opening of Carter Observatory in December 1941 (see Orchiston, 2002) these funds remained dormant for some time!

Kennedy spoke for over two hours about "Astronomy and Halley's Comet" in Wellington on Monday 11 July (Halley's Comet. Popular Lecture, 1910) and then on Wednesday 17 August 1910. In an address "... under the au-

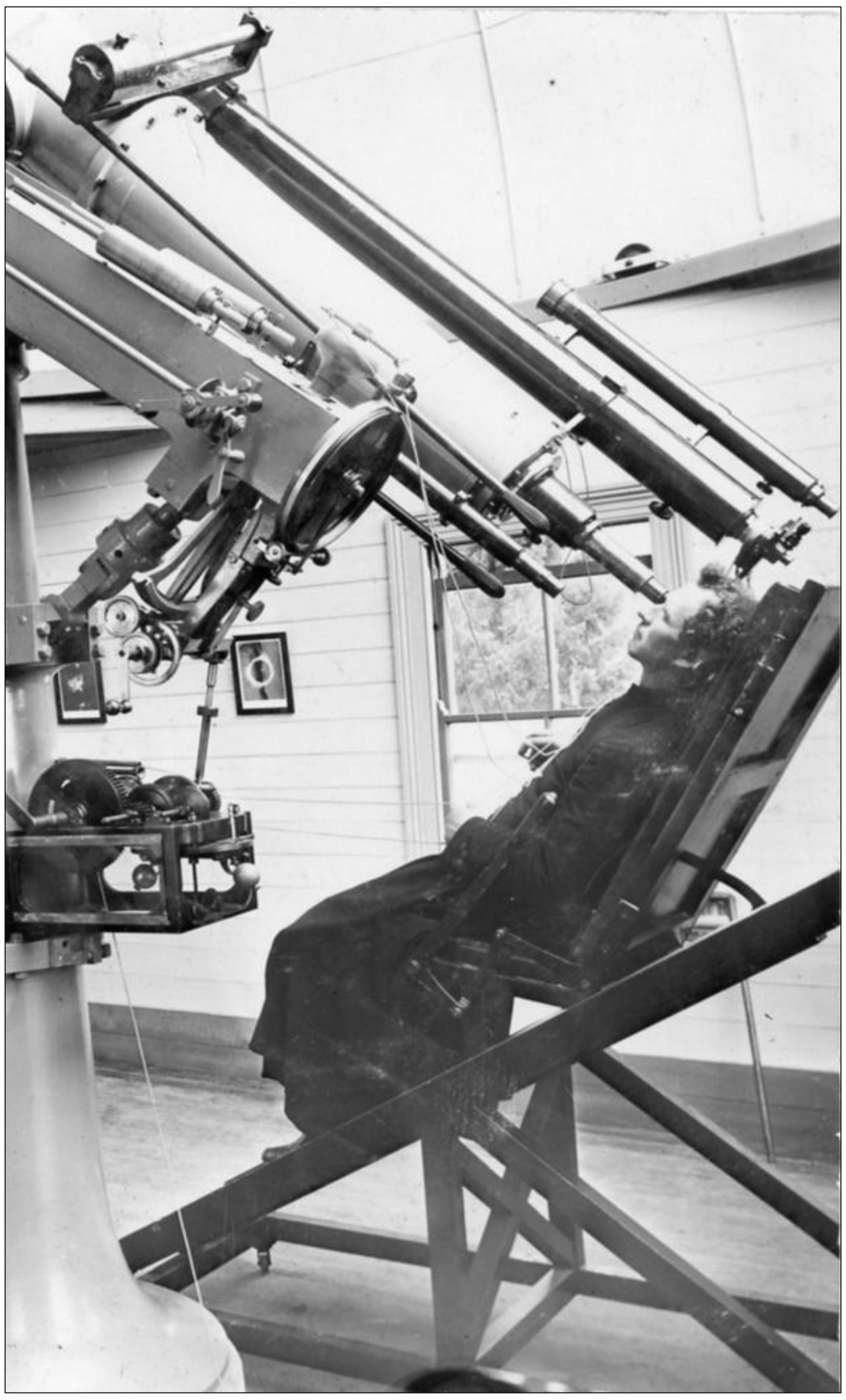

Figure 8 (right): Dr. Kennedy seated at the 9-inch (23 cm) Cooke refractor at Meeanee Observatory in 1907 (courtesy: Marist Archives, Wellington).

spices of the astronomical section of the Wellington Philosophical Society …" at the Wellington Town Hall on 29 September 1910 he again raised the issue of Wellington not having an observatory. He provided some historical insight into astronomy in New Zealand when he said:

> And wherever you go, whether to Europe, America, or Australia, you will find an up-to-date observatory almost in every city. If you go to Sydney, Melbourne, Adelaide or Perth, there you will find observatories well equipped. When you come to New Zealand - well, you go down to Christchurch and find a small observatory there to which the public are admitted from time to time. If you go to Palmerston North you see a small one, and also a small one at Wanganui. Up the East Coast at Meanee (sic), about five miles from Napier, you will find an observatory. It is not a very large one; still it possesses a very fine telescope. But when you come to Wellington and look around and ask, 'Where is the observatory?' you are told there is none. (For the Study of Astronomy, 1910).

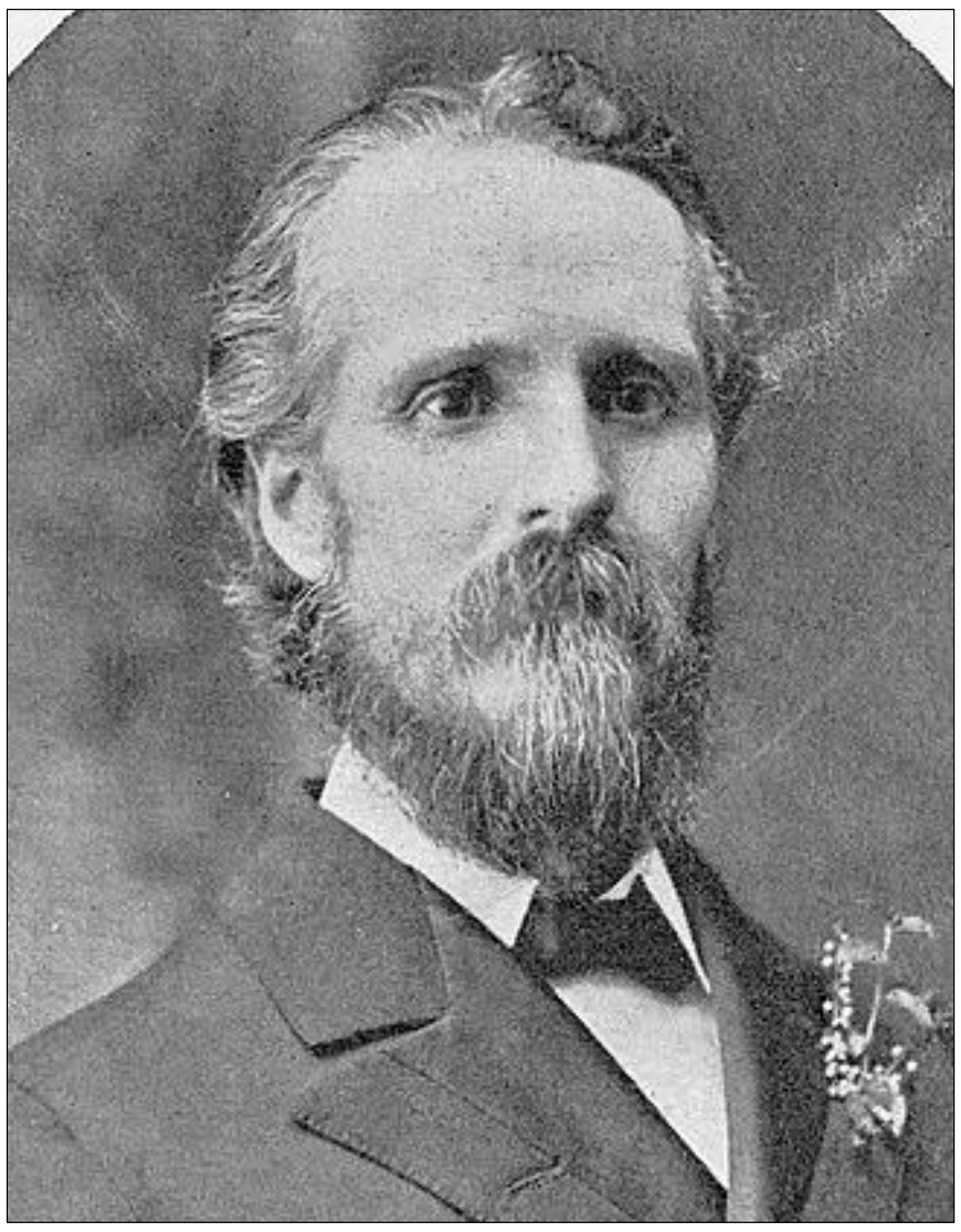

Figure 9: Joseph Taylor in about 1909. He and his family emigrated to New Zealand in 1894, settling in Collingwood, Golden Bay (https://en.wikipedia.org/wiki/Joseph_Taylor_(mining_engineer)#/media/File:Joseph_Taylor_(1858-1942).jpg).

## 2.4.2 Joseph Taylor (7 lectures)

In 1910, the second most prolific speaker about Halley's Comet, based on Papers Past, was Joseph Taylor (1858–1942. Taylor gave seven lectures on the West Coast of New Zealand's South Island (in Nelson, Greymouth, Runanga, and Hokitika) between 25 April 1910 and 28 May 1910. In his first lecture (25 April), when Halley's Comet was magnitude 2.5 (Kronk, 2007: 150), Taylor

> ... predicted that the comet would cause magnetic disturbances, heavy wind and rain storms, and in July or August next a great heat wave causing famine and pestilence in tropical countries. (A Lecture, 1910).

The science behind these claims was not explained, however, numerous New Zealand newspaper articles were quoting international speculations that recent floods were caused by 1P/Halley (Seeing the Comet, 1910); perhaps Taylor was basing his meteorological prophecies on these.

English-born Taylor was described as a "... well-known mining engineer, of Nelson …" where he lived for 40 years (*Greymouth Evening Star*, 1910). His town of residence and occupation would explain why his lectures were all on the West Coast of the South Island, a region well known for gold and coal mining. Taylor was responsible for discovering and developing the Puponga coalfield at the very top of the South Island, adjacent to Farewell Spit (Dyer, 2003). At the time of his last lecture, on 29 May 1910, Halley's Comet was past both perihelion and perigee, and fading at magnitude 2 with a ~30° tail (Kronk: 2007: 157).

## 2.4.3 Professor Hugh William Segar (6 lectures)

Professor Hugh William Segar (1868–1954; Figure 10; Nield, 1996) gave six lectures on 1P/Halley, all in Auckland. At the time, Auckland had the largest population in New Zealand (as it does today), with the city containing approximately 82,000 residents. With such a large population, it stands to reason that Segar would remain in the same city to lecture. His first lecture on 1P/Halley was on 16 August 1909 and his last on 24 May 1910. Segar, a Professor of Mathematics at Auckland University, "... spoke in a monotone which was reputed to leave his classes drowsy …" (*ibid.*). He regularly wrote newspaper articles on astronomy and economics.

As stated, Segar's first lecture was on 16 August 1909, at the Auckland Museum. He used a number of 'lime-light' slides and diagrams to explain the theory about comets. He spoke of 1P/Halley's historic encounters with the Earth and pointed out that Josephus' 'sword of fire' over Jerusalem in 66 CE was Halley's Comet and that it was woven into the Bayeux Tapestry when depicting the Norman invasion of England in 1066 CE (Meetings and Entertainment, 1910). Rather than leaving the audience drowsy, the newspaper reporter stated that Segar "... spoke in a popular style …", that his listeners followed with "... great clearness [and that] The lecture was listened to with the keenest attention …" (*ibid.*). Indeed, Segar's topic was so interesting to the general public that he needed to repeat the lecture the following night at Saint Andrews Church which permitted a greater seating capacity than Auckland University (The Tail of Halley's Comet, 1910). Owing to overcapacity, he again repeated the lecture one week later, on 24 May, at Saint James' Hall, which again had to turn people away (*Auckland Star*, 1910d). By 24 May 1910, the date of Segar's last reported lecture, 1P/Halley was approximately magnitude +1 and sporting a tail of around 30–40° in length (Kronk, 2007: 155–156).

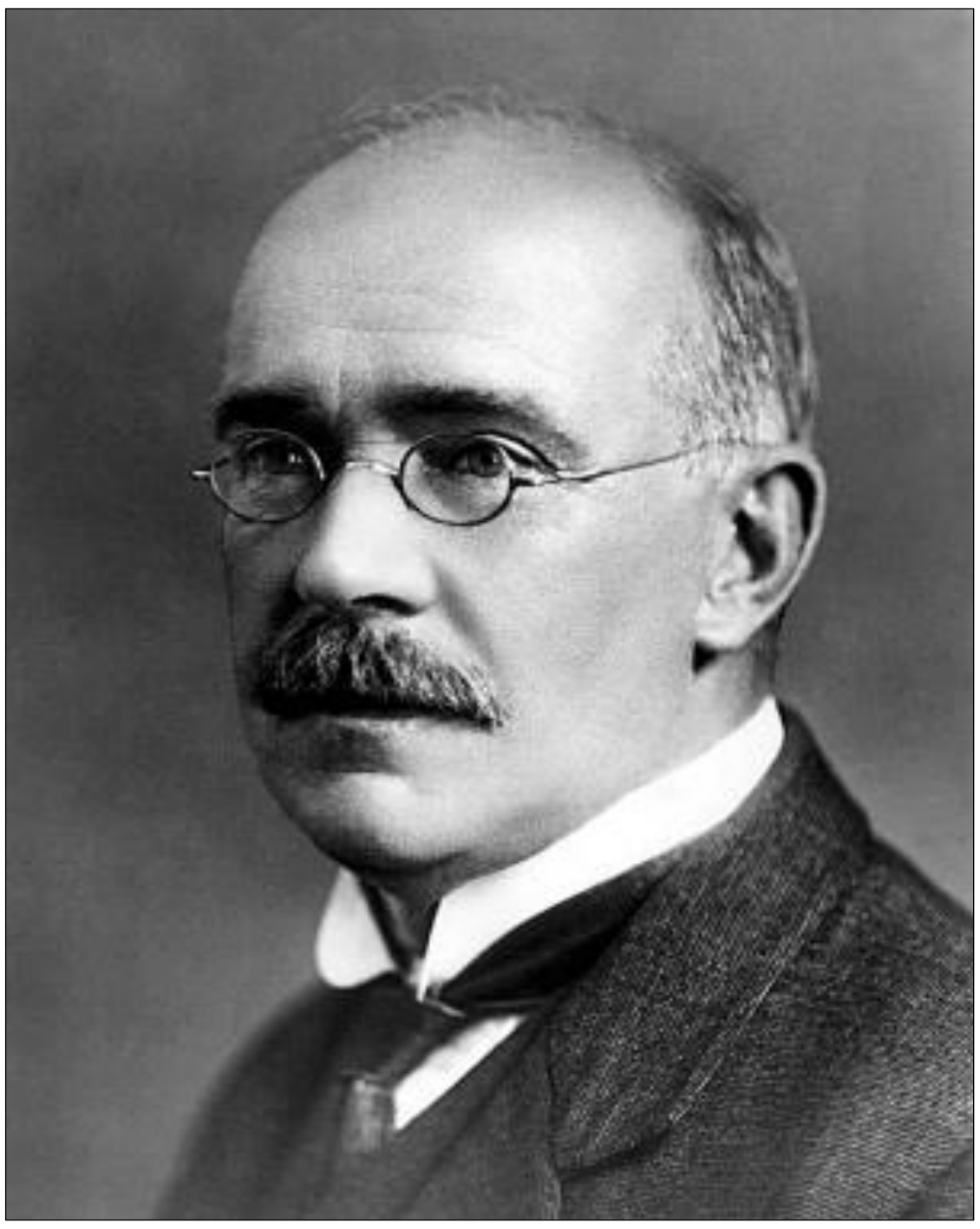

Figure 10: Hugh William Segar, was Professor of Mathematics at Auckland University College and gave six lectures on 1P/Halley in Auckland between August 1909 and May 1910. Being a popular speaker on a topic of high interest, Segar had to repeat his lectures due to overcrowding (https://teara.govt.nz/en/photograph/2236/hugh-william-segar).

#### *2.4.4 Professor Clement Wragge (5 lectures)*

English-born Clement Lindley Wragge (1852–1922, Figure 11; Wilson, 1990), a Fellow of the Royal Geographical and Royal Meteorological Societies was the Queensland Government Meteorologist from 1887 to 1903 (Mr Clement Wragge, 1910). In 1908 he moved to Auckland, New Zealand, when passed over as the founding Director of the new Australian Commonwealth Meteorological Bureau. As a meteorologist Wragge had a special research interest in tropical cyclones, and was the one who introduced the convention of naming them after people (*ibid.*). Apart from meteorology he also took an active interest in astronomy and maintained an observatory at his home in Auckland (Wilson, 1990).

Newspapers report that Wragge gave five lectures on 1P/Halley across New Zealand, namely Wanganui on 8 March 1909, Nelson on 7 October 1909, Wellington(?) on 16 October 1909, Whangarei on 21 March 1910, and finally Wellington on 18 June 1911 (over one year after 1P/Halley was a naked eye target).

Not a lot is divulged in the New Zealand newspapers about Wragge, however, what is highlighted is that he was also a religious figure, for on 15 April 1909 his topic to an audience in Te Kuiti was "Omnia ad Dei gloriam" ("All Things to the Glory of God") in which he often made reference to God and God's creation (Clement L. Wragge, 1909). At that stage, Halley's Comet was not reported to have been mentioned. He repeated the lecture in Gisborne on 11 May 1909 (Entertainments, 1909: 4) and Nelson on 8 October 1909 (The Majesty of Creation, 1909a). He also spoke in other centres such as Dunedin, Mangonui, and Auckland on astronomy and on some stones that he discovered, as stated in the *Auckland Star* (Mr Wragge's Discoveries, 1910):

> At St. James' Hall last evening Mr. Clement Wragge gave some interesting information regarding his recent discovery in the Bay of Islands of a group of stones so placed and shaped as to suggest that they were the work of a very ancient race.

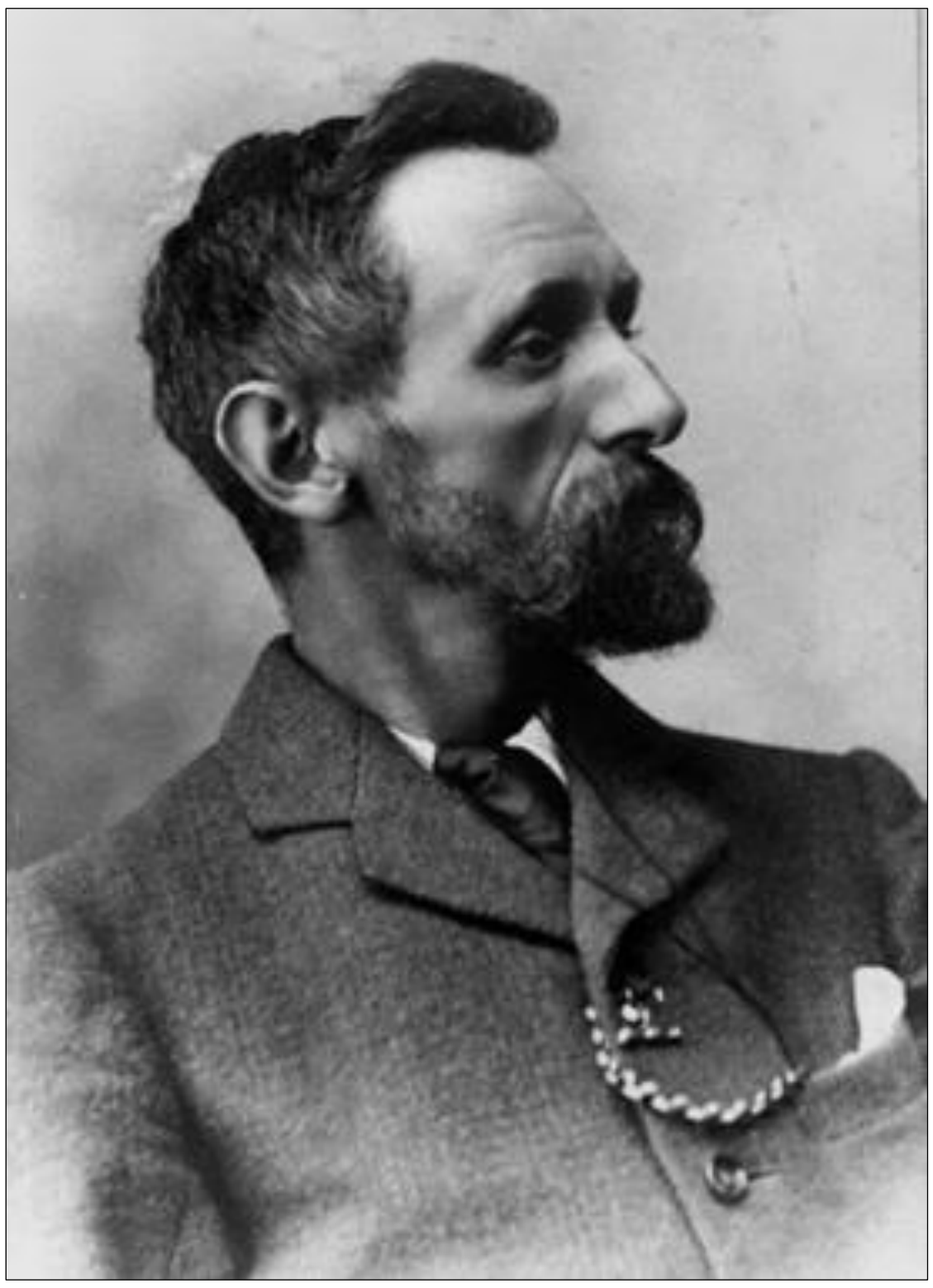

Figure 11: Clement L. Wragge moved to New Zealand in 1908, and gave lectures on 1P/Halley (https://en.wikipedia.org/wiki/Clement_Lindley_Wragge, 2026).

On 7 October 1909, Wragge began to introduce Halley's Comet in his lectures (The Majesty of Creation, 1909b). He spoke to the Nelson audience about comets in general and historically significant ones. In the talk he acknowledged Professor Alexander William Bickerton of Canterbury University College in Christchurch and stated that he agreed with Bickerton's partial impact theory (Gilmore, 2017) for crater formation on the Moon. Wragge encouraged the audience to frequent the local Atkinson Observatory and make use of its telescope. He also travelled to Tanzania to observe the 9 May 1910 solar eclipse when the New Moon covered the Sun and the transit of 1P/Halley across the Sun ten days later (the transit also being viewable from New Zealand). Using a solar telescope, he reported no observation of the comet as it did so (The Comet, 1910d).

Based on criticisms by other astronomers, such as Reverend Fairclough on 30 June 1909 (Fairclough, 1909), it would seem that Wragge was known as a spreader of unsubstantiated claims:

> Astronomical phenomena are rather scarce just now, unless we regard Mr Wragge as one. His explanations of the Antarctic coal beds by the simple process of altering the inclination of the earth's axis of rotation, his direct connection between a sun spot and a local storm, and his prophecies of earthquakes without specifying locality, are all phenomenal in their way. They remind us of a passage in a lecture descriptive of certain "Lick" photographs:—"O, ladies and gentlemen! Look! See! Behold! Beautiful! Splendid! Magnificent! Amazing! Great Scott!!" The prophecies, explanations, etc., were all that. They were, moreover, excellent as advertisements — but that's sublunary. As science, though they made the unskilful go early, they could not but make the judicious grieve.

An example is that Wragge is reported to have stated that

> Reverting to storms witnessed in the sun … these are favorable for further earthquake and volcanic action in regions of seismic activity. (Transit of the Sun, 1910).

In hindsight, we know that tectonic plate boundaries are regions of seismic activity that cause earthquakes and volcanic activity, but not sunspots.

### *2.4.5 Reverend Paul W. Fairclough, FRAS (4 lectures)*

Reverend Fairclough (1852--1917) was the first lecturer to give a talk on Halley's Comet in the Papers Past search, in Dunedin, on Friday 24 June 1901, nine years before the comet became a naked-eye object at its 1910 apparition (The Comet, the Meteor, and the Moon, 1901). The lecture mentioned 1P/Halley within the context of comets in general. He delivered three other lectures on 1P/Halley. After Fairclough died in 1917 aged 65, in his summary of Fairclough's life, Reverend C.E. Ward wrote of "His noble, outstanding, "Cedar"-like ability was seen in his sermons, his lectures, his talks, his conference debate, and his writing …" and that astronomy was his hobby (Late Rev P.W. Fairclough, 1917). Even though astronomy was one of Fairclough's hobbies, he was so dedicated to it that he was made a Fellow of the Royal Astronomical Society in England (The Comet. An Interesting Lecture. 1910) and that he was "One of the most instructive popular lecturers on astronomical subjects …" (*ibid.*). For a while, Fairclough served as a Methodist minister in the Christchurch region (Sarah J. Wilson, n.d.).

Fairclough also spoke about 1P/Halley on 12 September 1909, again in Dunedin. He elaborated in detail on the comet and the calculations as to when it would transit the Sun. Humorously, a comet-shaped prop fell off the wall and onto the floor during his talk (Strangers of Heaven, 1910)—an omen of things to come? When 1P/Halley was at its best, Fairclough spoke again in Dunedin about the comet, in early May 1910 (The Comet. An Interesting Lecture 1910), and this acted as a catalyst for the formation of the Dunedin Astronomical Society (which will be discussed in Section 7). The next lecture he gave was in 1911 when he spoke about solar eclipses. This occurred on 27 April 1911, undoubtedly because a partial solar eclipse was visible from New Zealand two days later in the morning hours of 29 April 1911 (NZST). No mention was made of 1P/Halley in this talk. Fairclough passed away six years later from a heart attack (findagrave). His final resting place in Christchurch is shown in Figure 12.

### 2.4.6 Professor Richard MacLaurin (3 lectures)

According to Victoria University, Richard MacLaurin (Figure 13) was one of that University's Founding Professors and a Mathematics scholar. In 1907 he became a Professor of Law, Dean of the Faculty of Law, and an Honorary Professor of Astronomy. He later left Wellington to take up a position at Columbia, in the USA, and then took up the role of President of the Massachusetts Institute of Technology (MIT).

MacLaurin was a catalyst for the formation of the Wanganui Astronomical Society by giving a lecture on comets to interested persons (Gilkison, 1910). This was on 12 July 1901, nine years before the public would see 1P/Halley (The Winter Lectures: Comets, 1901: 3). His talk covered the theory of comets in detail, how Halley correctly predicted the return of his comet, and how the next apparition would be in 1910. This meeting and talk were intended to "... to consider the formation of an Astronomical Society [but] was postponed, owing to the lateness of the hour …", nonetheless, 1P/Halley seems to have possibly been instrumental in the formation of the Wanganui Astronomical Society in 1901. In addition, the appearance of 1P/Halley in 1910 may have been the reason why several other New Zealand astronomical societies formed just after its appearance. This will be discussed in Section 7.

MacLaurin's second lecture was on 4 October 1905 at Newton, Wellington, where he gave an elaborate speech on comets in general, and highlighted the fact that 1P/Halley would return in 1910 (Comets, 1905). During the talk he alluded to the beauty of a comet by quoting Keats, saying "a thing of beauty is a joy forever" (*ibid.*). Finally, on 22 March 1906, MacLaurin spoke at the Municipal Hall in Manawatu (Professor MacLaurin's Lecture, 1906). He stated that about five new comets were discovered each year (then) and that a nice bright one appeared approximately every five years. He broke the lecture into three parts, namely, what a comet looks like; how it moves; and what a comet is made of. Regarding the last point, the newspaper reporter wrote

> The head consisted of a huge cloud of stone, of small particles of matter, while the tail was composed of various gases thrown off from the head. The chemical components of the comet were mentioned, they having been ascertained by means of that marvellous instrument the spectroscope …

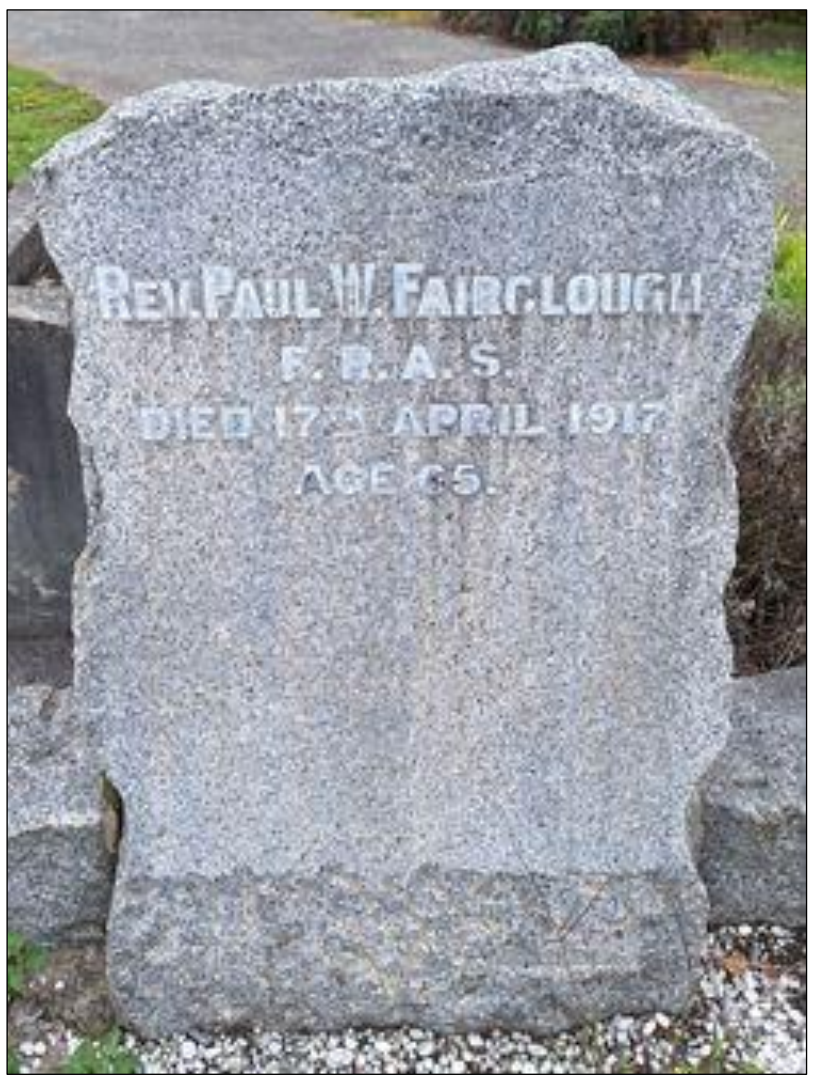

Figure 12: The final resting place of Methodist minister Reverend Paul W. Fairclough (FRAS), a popular lecturer on astronomy. According to Papers Past, Fairclough gave four lectures on 1P/Halley (source: findagrave).

### 2.4.7 Frank H. Hitchings (3 lectures)

Frank H. Hitchings (1843–1921) lived in Sydenham, Christchurch, after migrating from London in 1869. He gave three lectures on 1P/Halley (as well as solar eclipses and the transit of Venus). He was a bricklayer and chimney sweep by trade who built an observatory in his backyard and had a particular interest in sunspots (Archives). He was described as "... one of the unfortunate scientists who made the fruitless journey to Tasmania to view the recent solar eclipse …" (News of the Day, 1910c), however, observers at Queenstown in Tasmania witnessed the event (Rev. Leslie S. Macdougall Diaries, 2012). Hitchings may have been

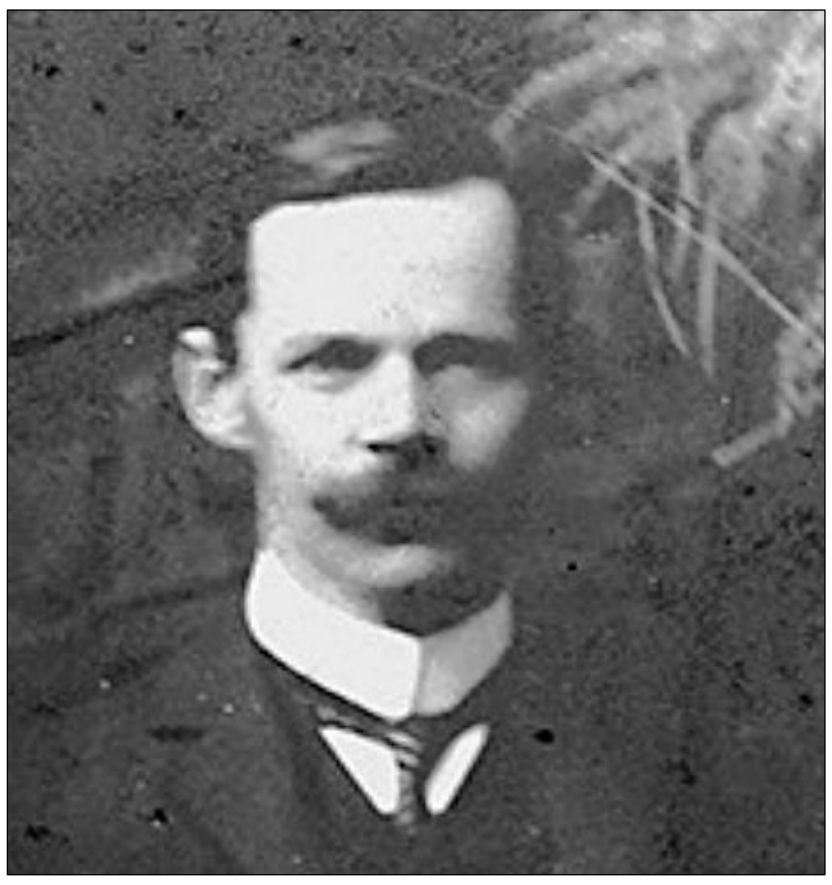

Figure 13: Professor McLaurin, who gave three lectures about 1P/Halley in New Zealand, all many years before the comet appeared in 1910 (source: Victoria University, New Zealand).

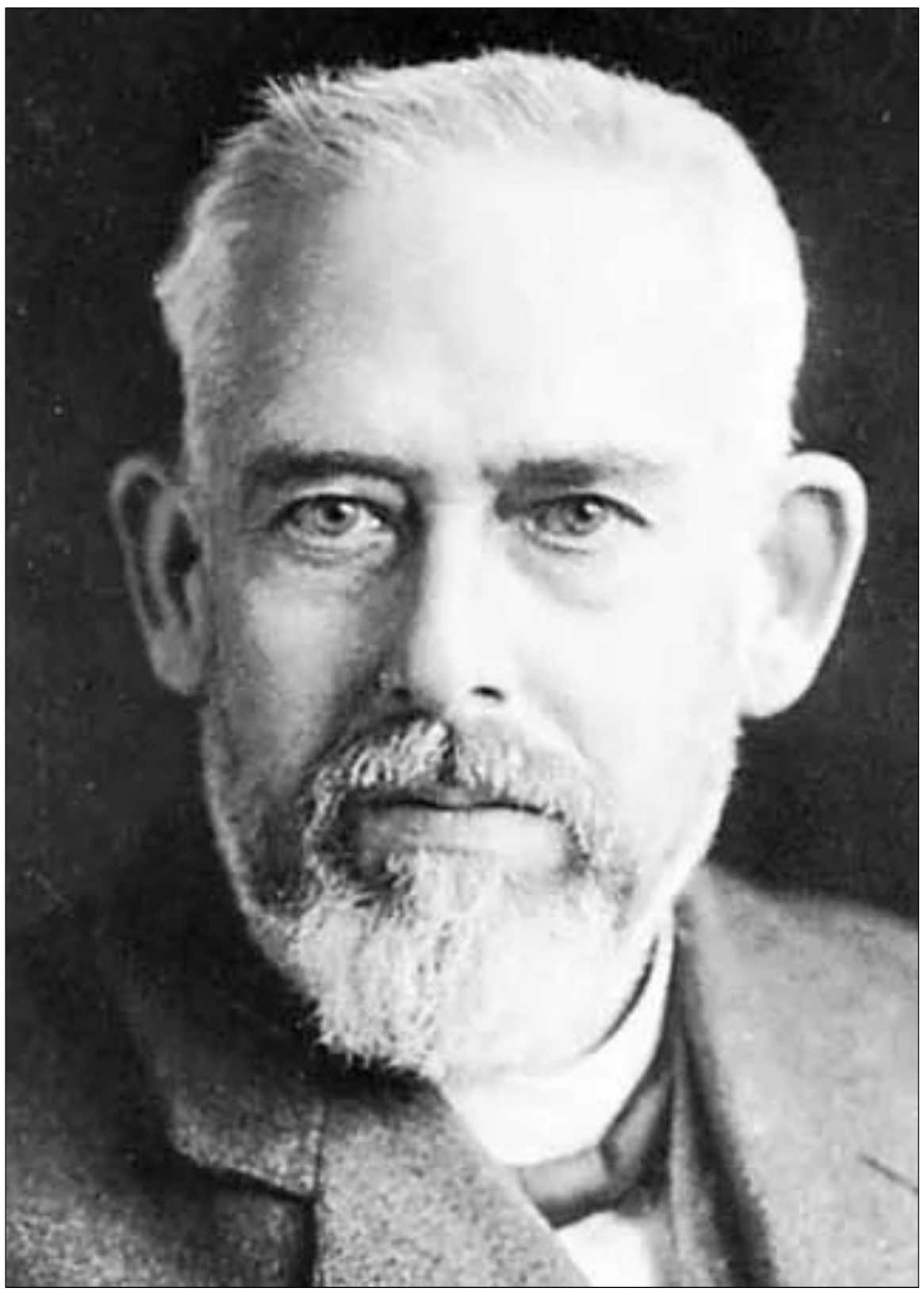

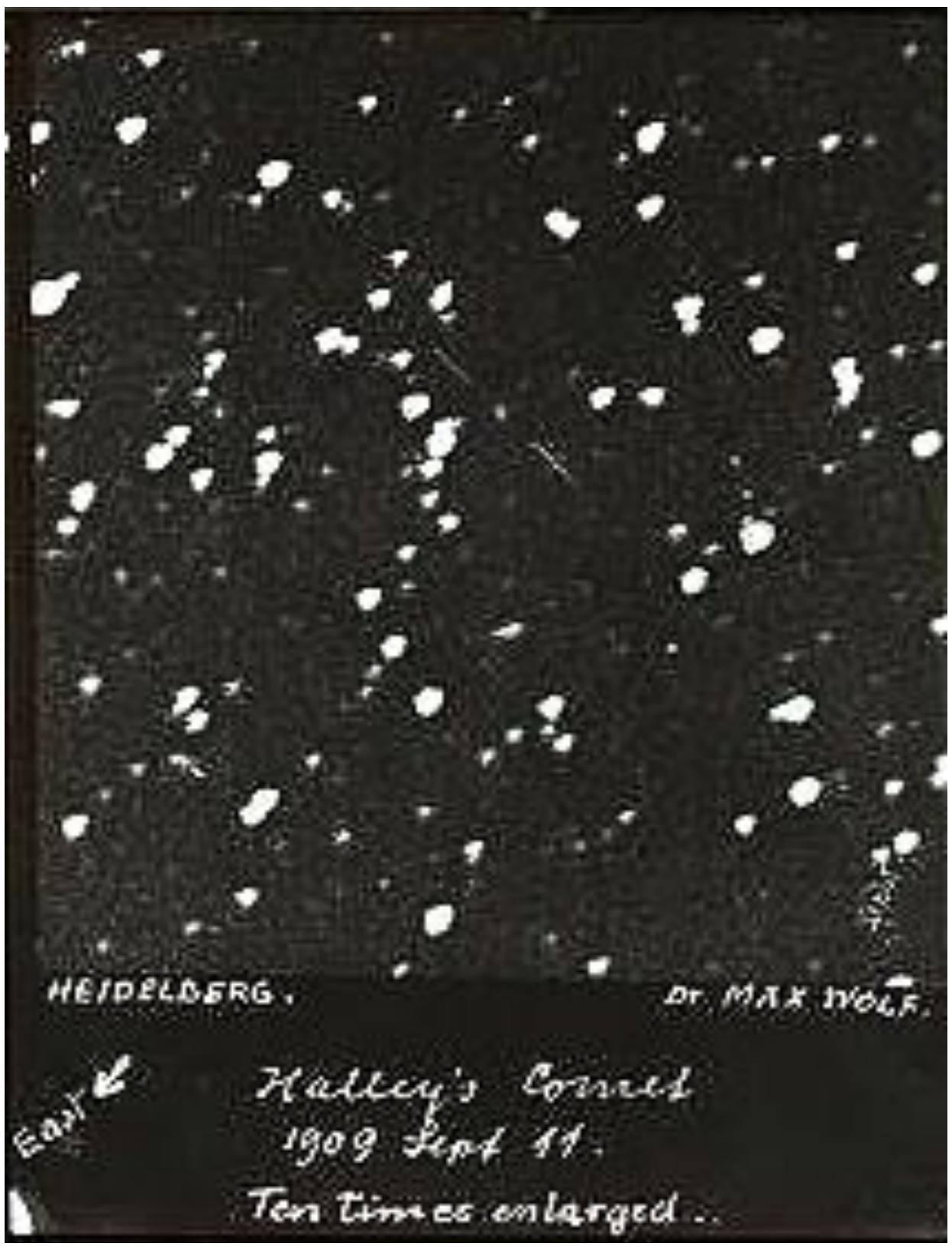

Figures 14a (left) and 14b (right): Max Wolf (left) was the first to identify Comet 1P/Halley in its 1909–1910 apparition. He photographically detected the comet from Germany in September 1909 (source: *Encyclopedia Britannica*, n.d.). Right: Wolf's photograph of Comet 1P/Halley (between the two diagonal lines just above centre), taken on 12 September 1909 (Drummond Collection).

been at another location on that island which succumbed to clouds.

Hitchings spoke at the Burnham Industrial school on the Saturday evening of 28 May 1911. The article states that, "Astronomy has a close interest for Burnham, inasmuch as the transit of Venus was observed by scientists from all over the world some thirty-six years ago, from a field in the grounds of the school." In reality, it was only astronomers from England who were based at Burnham in 1874 and 1882 (see Orchiston, 2016: 371–419).

Hitchings also spoke about 1P/Halley in Christchurch on Monday 30 May 1911. These talks were illustrated with 'limelight views' also known as 'lantern slides', many being taken by him (News of the Day, 1910b). Whether any of these were of Halley's Comet is uncertain. He spoke about the comet again on 7 June 1911, also in Christchurch. He was described as being a member of the British Astronomical Society, having become a member in 1902. The afore-mentioned Reverend Dr Kennedy thanked Mr Hitchings for his talk (*ibid.*).

## 3 FIRST INTERNATIONAL OBSERVATIONS OF 1P/HALLEY'S 1910 APPARITION

The noted British amateur astronomer, William F. Denning (1848–1931; Beech, 2023), predicted that 1P/Halley would be detected photographically before being seen visually (Denning, 1909: 62). He suggested that this would occur in either September or October 1909. He was correct. The comet was first identified in a 60-minute photograph taken by Maximilian Franz Joseph Cornelius (Max) Wolf (1863–1932; Figure 14) of Heidelberg Observatory, through a 28-inch (72-cm) reflector on 12.06 September 1909 UT (Kronk, 2007: 141; Wolf, 1909: 179), although Vsekhsvyatskii (1964: 382) states that the photograph was taken on 11 September UT. Wolf estimated the comet to be magnitude 16, with a coma 8–10" in diameter at a declination of +17°. Kronk (2007: 141) reveals that Wolf confirmed the comet with an additional photograph taken on 12.11 September UT. Although Wolf was the first to identify 1P/Halley in photographs, he was not the first to photograph it. Hardcastle (1909: 456) highlights that 25 and 30-minute exposure search plates taken with a 30-inch (76-cm) reflector on 10 September were initially scanned without success and then reexamined with success after Wolf's astrometry was announced (Kronk, 2007: 141–142). Additionally, a search plate taken on 25.05 August (18 days before Wolf) from Khedivial Observatory, Hel-Figure 15 (right): Yerkes Observatory, Wisconsin, where Professors Burnham and Barnard, using the 40-inch refractor, were the first astronomers to visually detect 1P/Halley during its 1909–1910 apparition (courtesy: Yerkes Observatory).

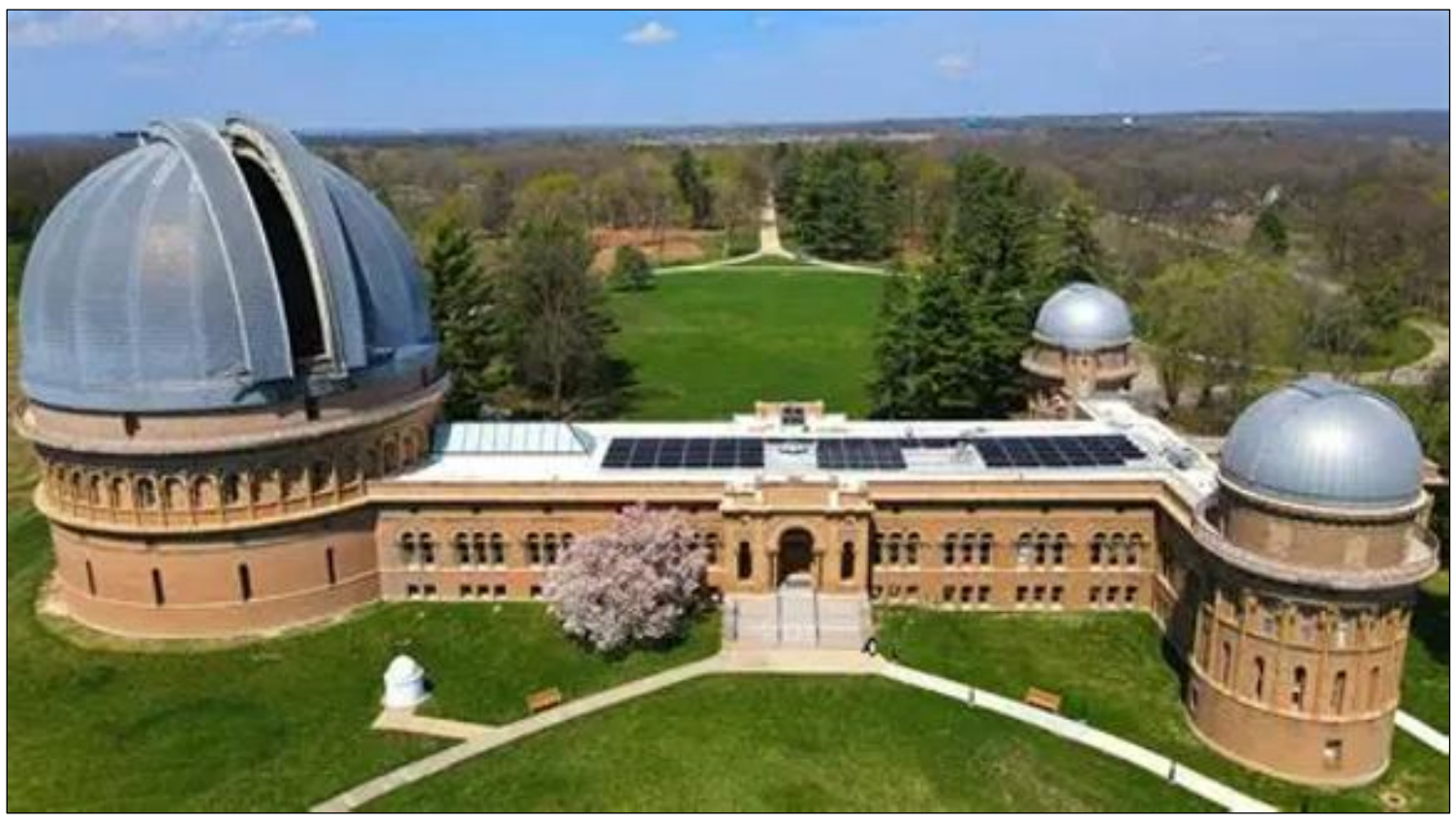

wan, Egypt, had recorded the comet, but it was not detected until the plate was later sent to the Royal Observatory, Greenwich, and scrutinised with an astrometric measuring machine (Vsekhsvyatskii, 1964: 382).

On 16 September 1909 (UT), four days after his recovery photographs, Wolf determined the photographic magnitude of 1P/Halley to be 15.5. The previous day (15 September 1909 UT), Professor Burnham used the Yerkes Observatory (Figure 15) 40-inch (102-cm) refractor to visually observe 1P/Halley (Comet Notes, 1910; Halley's Comet, 1910u). Vsekhsvyatskii (1964: 382) states that this was the first visual observation of the comet in the world for its 1910 apparition. Two days later, Professors Barnard and Burnham (Yerkes Observatory) used the same Yerkes 40-inch refractor (Figure 16) to visually observe the comet. Barnard (1909: 43–44) wrote:

> 1909 Sept. 17. The magnitude was estimated 15½, the diameter 5″+/-. Measured diameter 7″.2. There was possibly a faint nucleus of the 15th magnitude, or an indefinite brightening in the middle. It appeared as a "fleck" of light surrounded by a faint nebulosity for a few seconds. A little later it was quite easy to observe and was estimated to be 15th magnitude. Sky good, seeing good.

Barnard also made telescopic observations on 24, 26 September, 17, 19, 26 October, and 14, 28, 30 November 1909. Over the 2.5 months since his first observation, he noted that the magnitude of 1P/Halley brightened from magnitude 15 to magnitude 11 by 30 November (as seen with a 4-inch finder scope). He described the coma as becoming more stellar in the central region as time progressed and that the coma was 15″ in diameter when viewed without Moon interference. No tail was seen (Barnard, 1909: 43–44).

Kronk reveals the steady rise in the number of telescopic observations of 1P/Halley during October 1909. These submissions were all made from the Northern Hemisphere with refractors ranging from 12.6-inch (32-cm) to 30-inch (76-cm) apertures (Kronk, 2007: 142). By the end of October 1909 the comet was magnitude 14. Naturally, as time progressed, the comet's magnitude increased. By 30 November 1P/Halley was estimated to be magnitude 11. At the end of 1909 the magnitude was reported to be 10.

On 31 January 1910, the comet was determined to be magnitude 9 (see Figure 17). By the end of February it was magnitude 7, however, it should be noted that Wolf stated that he could see 1P/Halley with the naked-eye several weeks earlier on 9 February 1910 and that it was magnitude 7 (Kronk, 2007: 147; Moore and Mason, 1984: 67; Seargent, 2009: 55).

However, Vsekhsvyatskii (1964: 383) suggests that this first naked-eye observation occurred two days later on 11 February 1910 UT. Does this imply that Wolf had exceptional eyesight, that it was a false observation, or that other observers were reporting wrong magnitude estimates around the time of Wolf's observation? Kronk reveals that in the week either side of Wolf's 9 February observation, magnitude estimates for the comet ranged from 9.1 to

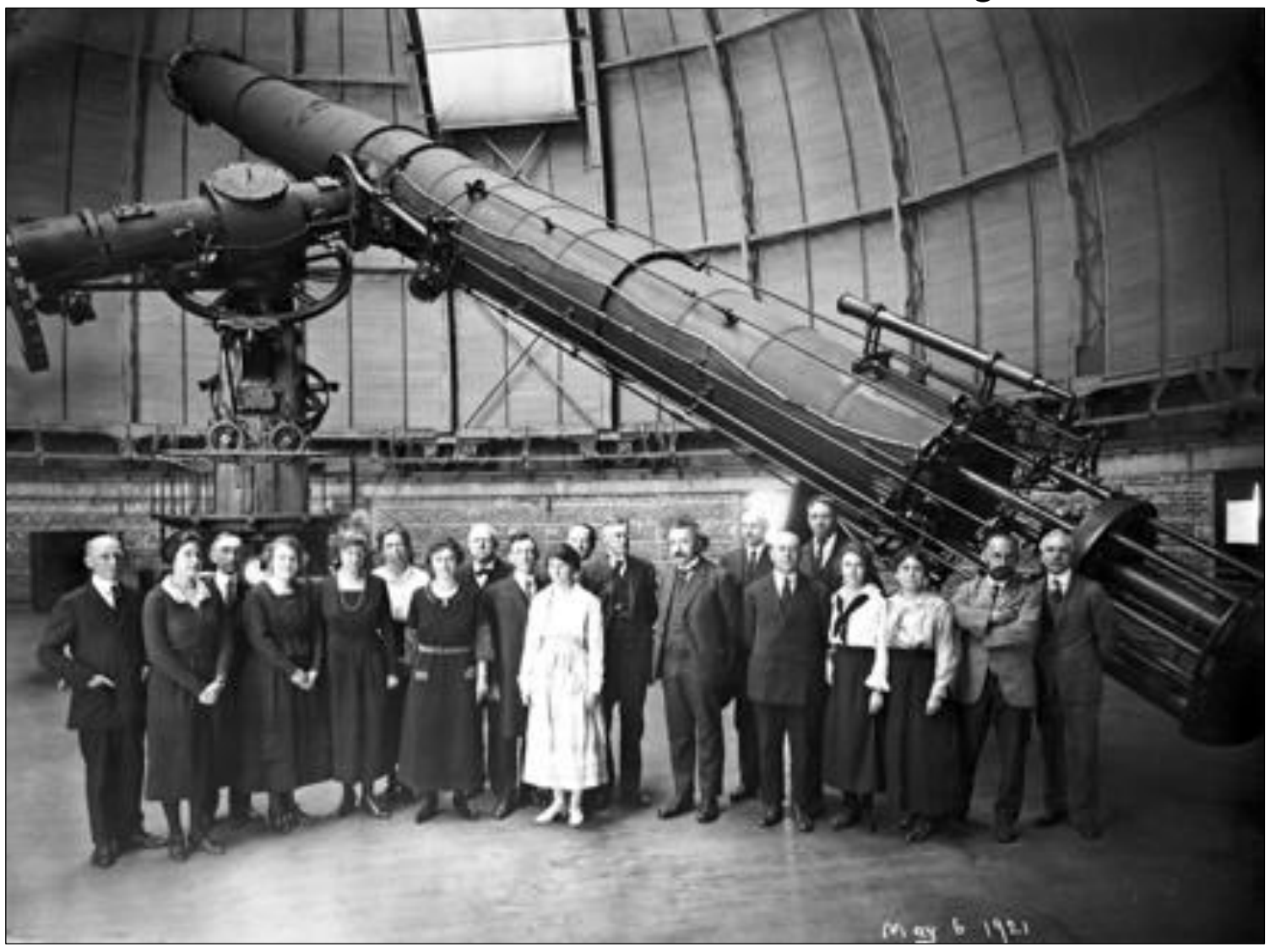

Figure 16 (right): Barnard (eighth from the left) and Einstein (eighth from the right) are pictured in this 6 May 1921 photograph posing with the 40-inch Yerkes refractor (source: Wikicommons, 2015).

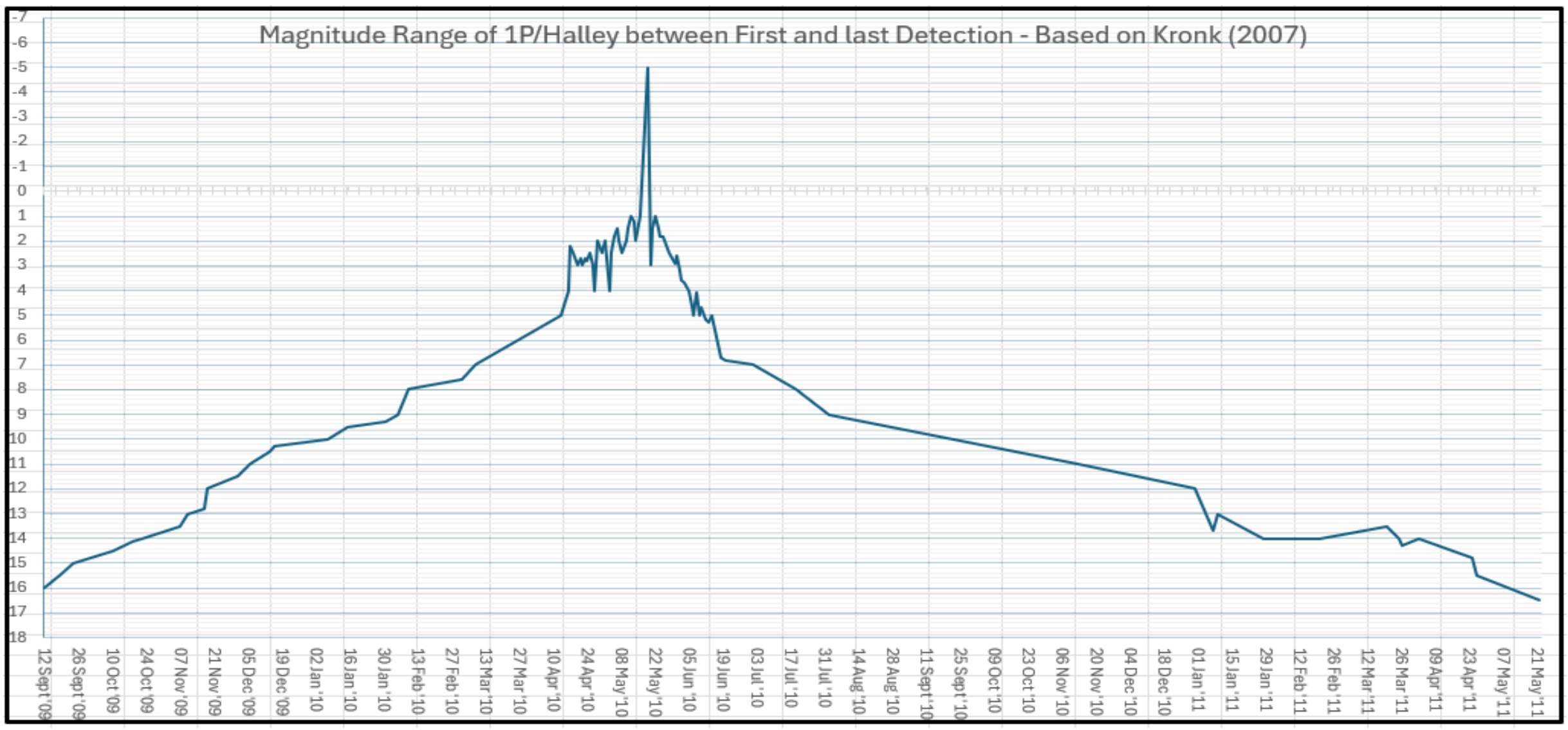

Figure 17: The change in the brightness of Comet 1P/Halley from mid-September 1909 until mid-September 1910. Based on Kronk (2007: 140–160). The dates are on the X-axis and the total magnitude on the Y-axis. Note that it may have reached a magnitude of around –5 on 19 May 1910 (UT), when seen during the daytime with the naked-eye (Kronk: 2007: 153–154). This most probably being due to the forward-scattering of light due to its apparent close proximity to the Sun (source: the authors).

Table 3: Key events in the 1910 apparition of Comet 1P/Halley (based mainly on Kronk, 2007).

| Date (UT) | Event | Observer | Location | Notes |
|---|---|---|---|---|
| 12 Sept 1909 | Photographed by Wolf (Germany) | Max Wolf | Germany | First identification of 1P/Halley |
| 18 Sept 1909 | First visual observation by Burnham | Burnham | USA | From the USA |
| 9 Feb 1910 | Naked eye (reported by Wolf) | Max Wolf | Germany | Wolf |
| Early March – 8 Apr 1910 | Too close to Sun for observations | — | — | Line-of-sight alignment |
| 8 Apr 1910 | Seen after solar conjunction | | | |
| 16 Apr 1910 | Naked eye after solar conjunction | Costa | | |
| 20 Apr 1910 | Perihelion | | | |
| 19 May 1910 | Comet transits the Sun; Earth passes through tail | | | Too close to Sun for observations |
| 19 May 1910 | Daylight comet—naked eye! | | | Seen during the day-time from Tasmania |
| 20 May 1910 | Perigee—closest to the Earth | | | |
| 22 May 1910 | Peak evening brightness: magnitude 1 | | | |
| 12 June 1910 | No longer a naked eye object | | | |
| 29 April 1911 | Last telescopic detection | Gonnessiat | Alger | |

7 in the ~week before his observation and 7 to 10 in the week after (Kronk, 2007: 146–147). This is a factor of 7–8 times, quite a range! A celestial object with a similar magnitude to Wolf's claim is NGC 5128 (Centaurus A) in the Southern Hemisphere. At magnitude 6.8 and subtended over approximately 12′ for the brighter section, it possibly resembles what Wolf saw when viewing 1P/Halley. The fourth author has seen reports of naked-eye observations of this enigmatic galaxy—as Google searches also reveal. On 9 February, 1P/Halley had a declination of +09°, so well suited to Wolf's Northern Hemisphere location, so it seems feasible that he may have detected it without optical aid.

There seems to be a consensus that the comet was sufficiently bright as to be an easy naked-eye object for the general public from April until mid-June 1910 (Kronk, 2007: 147–158). New Zealand observers reported similar brightness estimates and tail length descriptions. These will be discussed in Section 5.

To provide an overview of the 1910 visit by 1P/Halley, Table 3 shows the most important dates based on international observations. It states the dates of first photographic detection (12 September 1909 UT), first visual telescopic observation (18 September 1909), when it attained naked-eye visibility (9 February 1910), perihelion (20 April 1910), perigee (20 May 1910), the last visual telescopic observation (29 April 1911), and the last photographic detection (16 June 1911). The constellations it traversed from September 1909 to June 1911 were Orion,

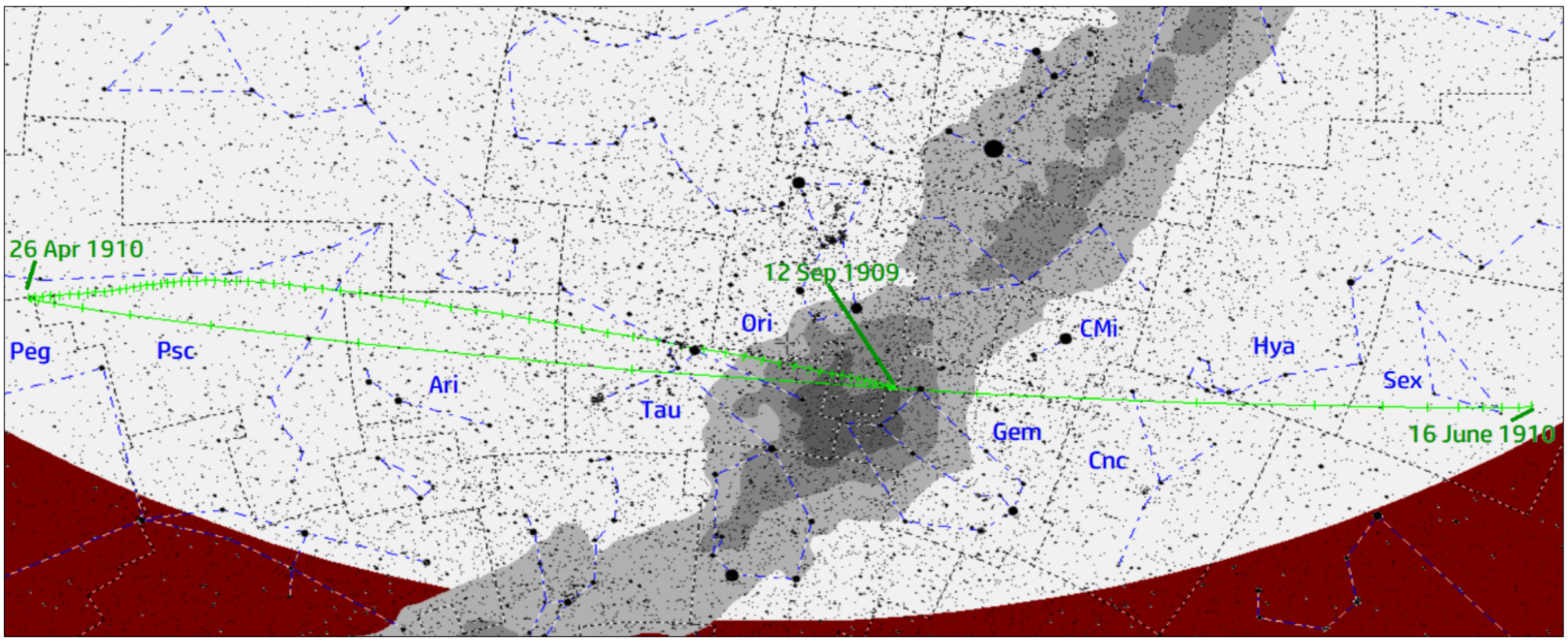

Figure 18: The path (green lines) of Comet 1P/Halley from Wolf's recovery date on 12 September 1909 UT until the comet was last photographed on 16 June 1911. North is down to match the view from New Zealand. The red fill shows the location of the northern horizon as seen from New Zealand when the comet was visible. Clearly, the position of the horizon relative to the background stars would change in the course of a single night (as the Earth rotated), and over the course of the nine months covered by this diagram (as the Earth moved around the Sun, but the red fill does give an indication of how far north an observer could see (sources: GUIDE, and Kronk, 2007: 140).

Taurus, Aries, Pisces, Pegasus, Pisces, Aries, Taurus, Orion, Gemini, Canis Minor, Cancer, Hydra, Sextans, Leo, Crater, Corvus, Crater, Hydra, and Sextans (Kronk, 2007: 140). Figure 18 reveals the path through the constellations using GUIDE. Figure 19 displays photographs of 1P/Halley taken by Ferdinand Ellerman from Diamond Head in Honolulu (Hawaii) from 26 April to 11 June 1910 UT. The brightening of the comet and increasing tail length, and then the reverse, are evident.

## 4 PRE-NAKED-EYE OBSERVATIONS OF COMET 1P/HALLEY FROM NEW ZEALAND IN 1909–1910

The first observation of 1P/Halley from New Zealand was photographic. The Marist Brothers at Meeanee Observatory, Napier (Figure 20), photographed the predicted positional field with a 3.5-inch (8.9-cm) Dalmeyer portrait lens and spotted the faint comet on 14 October 1909. The *Auckland Star* published the following on Wednesday 20 October 1909:

> Halley's comet was photographed last Thursday [14 October] at the Meeanee Observatory. The impression, however, was so extremely faint that its discovery was not made public. This morning [20 October] another attempt was made, increasing the exposure from one and a-half to two and a-half hours. This time the comet was clearly defined on the negative. Some time will elapse before the comet is visible through even the largest telescopes. The exposure this morning began at 1.15 and ended at 3.45, when the near approach of dawn rendered further work impossible. (Halley's Comet Photographed, 1909).

This successful observation of 1P/Halley from New Zealand with a 3.5-inch portrait lens was approximately four weeks after Wolf's first recovery photograph with the 28-inch refractor in Germany. On 14 October 1909, the date of Meeanee Observatory's first successful photograph, 1P/Halley was magnitude 15 (Kronk, 2007: 142). Unfortunately, the authors were unable to source a copy of this photograph for inclusion in this paper.

Father J.J. Cullen, of Meeanee Observatory, further described the background details of the photographs when he submitted the following to the *Dominion* newspaper which was published on 25 October 1909 (Halley's Comet. Taking its Photograph, 1909):

> I enclose an enlargement of a photograph of Halley's Comet, magnified from the original negative eleven diameters. Of course, the first photograph of an expected comet is necessarily very faint. Were it large, and very clearly defined, the comet could have been photographed before. It was not till within the last week or so that so long an exposure as two and a half-hours was possible. During the whole of this time, the sky in that direction must be perfectly cloudless. The plate used was extremely sensitive, and the lens was a three and a half inch Dalmeyer portrait lens at open aperture of 3.5 [inches]. Later on, when the comet becomes telescopically visible, photographs for reproduction can be taken.

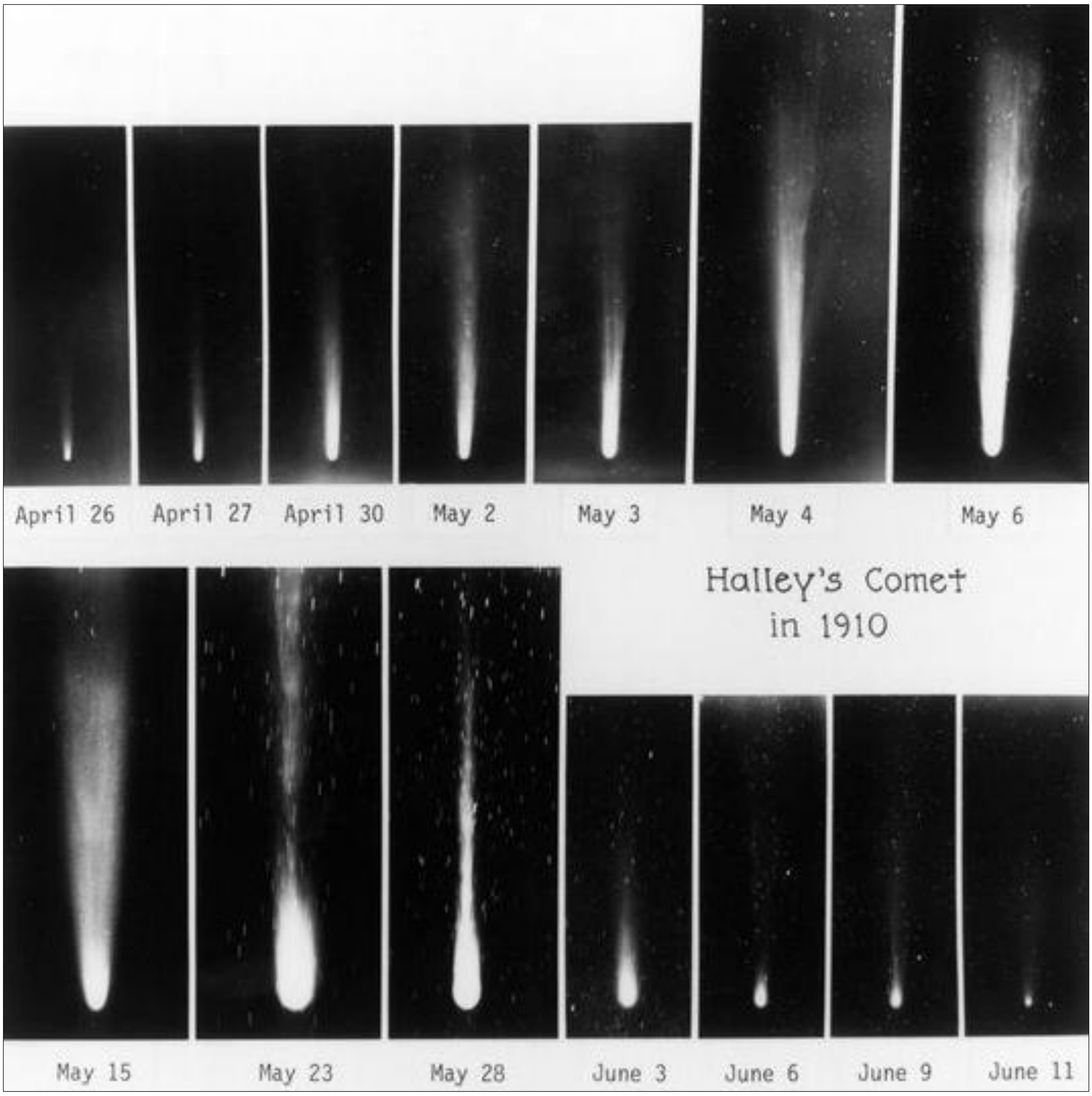

Figure 19: Photographs of Comet 1P/Halley taken from Diamond Head in Honolulu by Ferdinand Ellerman. Note the change in comet brightness and tail length in the six weeks between 26 April and 11 June 1910 (photograph: Orchiston Collection).

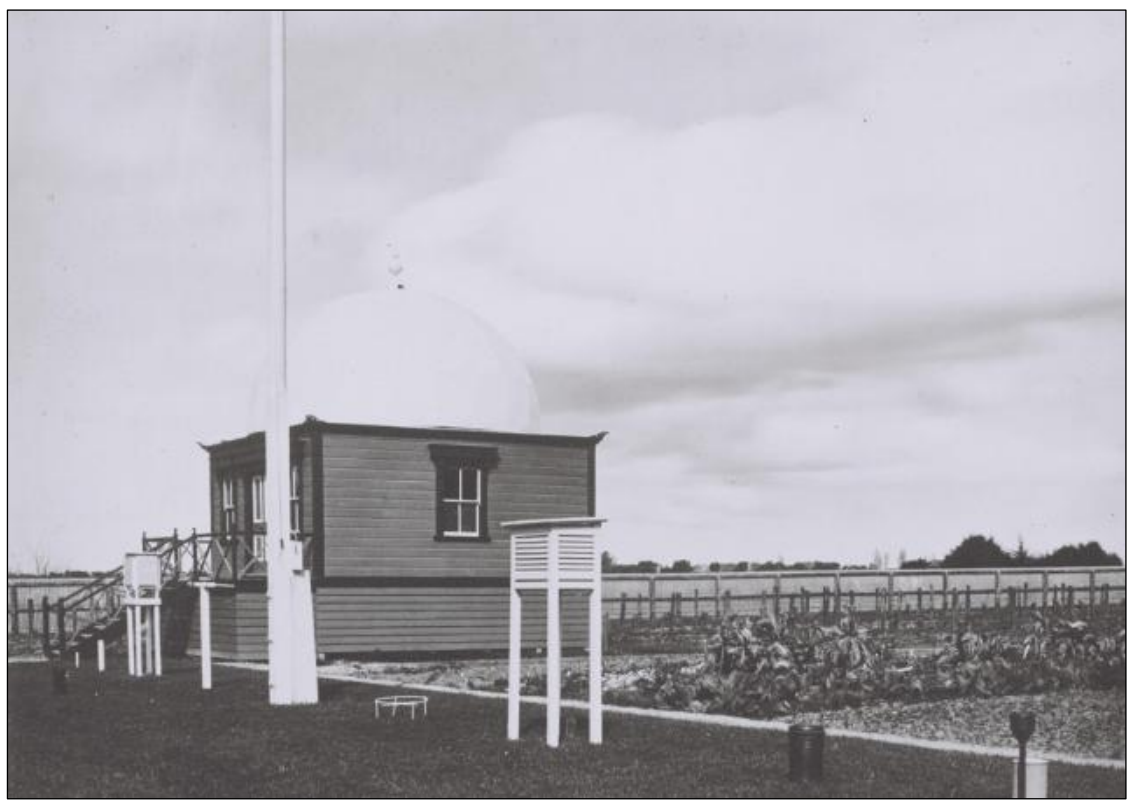

Figure 20: Meeanee Observatory, near Napier, was where Comet 1P/Halley was first photographed from, in New Zealand, during its 1909–1910 apparition (Orchiston Collection).

Then the camera will be able to follow the comet itself, and not the adjacent stars, as it must at present. So accurately was its position gauged at the time of exposure, that it appears almost in the centre of the plate which photographs clearly a portion of the heavens the size of the Southern Cross only. The comet rises at a quarter past twelve at present, just where the sun rises in winter [thus, in the North-East]. It is, however, invisible as yet, even through the largest telescopes. Its feeble light was allowed to add minute by minute to an extremely faint impression, till at dawn a distinct, though very small, image of the comet revealed itself on the developed plate.

This account reveals some of the techniques that the Brothers used to photograph 1P/Halley. The Dalmeyer lens, such as the one shown in Figure 21, had an aperture of 3.5-inches (8.9-cm) and with the camera's photographic plate dimensions had a field of view of approximately 5° × 5° ("... the size of the Southern Cross …"). When 1P/Halley is brighter, "... the camera will be able to follow the comet itself, and not the adjacent stars, as it must at present." This suggests that the 3.5-inch lens was piggybacked on a larger telescope (probably the 9-inch refractor in the Meeanee Observatory) and the main telescope was guided on a visible star close to the comet. Later on, when the comet was bright enough to see with the 9-inch refractor, the telescope could be used to guide on the comet as the lens and camera were taking the photographs. The exposures were 2.5 hours long, manually guided the whole time—a true test of patience and endurance as any astrophotographer before autoguiding can attest!

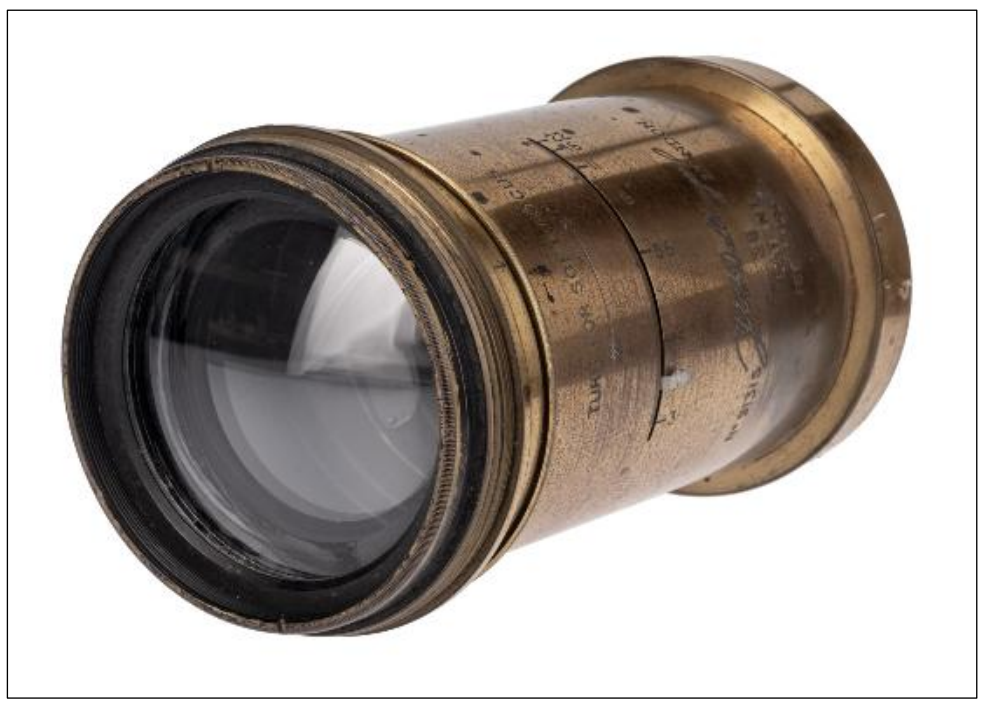

Figure 21: A 3.4-inch (8.5-cm) J.H. Dallmeyer portrait lens, similar to the one used by the Marist Brothers at Meeanee to photograph Comet 1P/Halley. With an overall height of 17.5 cm and diameter of approximately 8 cm, the pictured lens possibly had a focal ratio of f/2.5 to f/3 (after Flints, 2024).

The statement that "… it is, however, invisible as yet, even through the largest telescopes …" indicates that Comet 1P/Halley could not be seen with the largest telescope at Meeanee, i.e. the 9-inch refractor. The Marist Brothers had possibly not heard of Burnham and Barnard's visual telescopic observation of the comet five weeks previously. Thus, "… the largest telescopes …" was applicable to their observatory (and probably the Wanganui Observatory (Figure 22) and its 9.5-inch refractor (Figure 23) and not overseas observatories such as the 40-inch refractor at Yerkes Observatory. Theoretically, the visual limiting magnitude ($m$) of the 9-inch refractor (229-mm) was 14.5 ($m = 2.7 + 5 \log D$, where D is the aperture in millimetres, thus, $m = 2.7 + 5 \log 229 = 14.5$) (Ridpath, 2004: 28–29). Drummond et al. (2026) point out that non-stellar comets are more difficult to see than point-like stars. The telescope could possibly see comets down to magnitude 13, however, with Comet 1P/Halley being at a declination of +17° (NASA/JPL) and New Zealand having a latitude of 34° to 47° South, atmospheric extinction would have reduced the limiting magnitude fairly significantly. In the ensuing weeks, as 1P/Halley brightened, more New Zealand astronomers (both professional and amateur) began to photograph it (this will be discussed in Section 5).

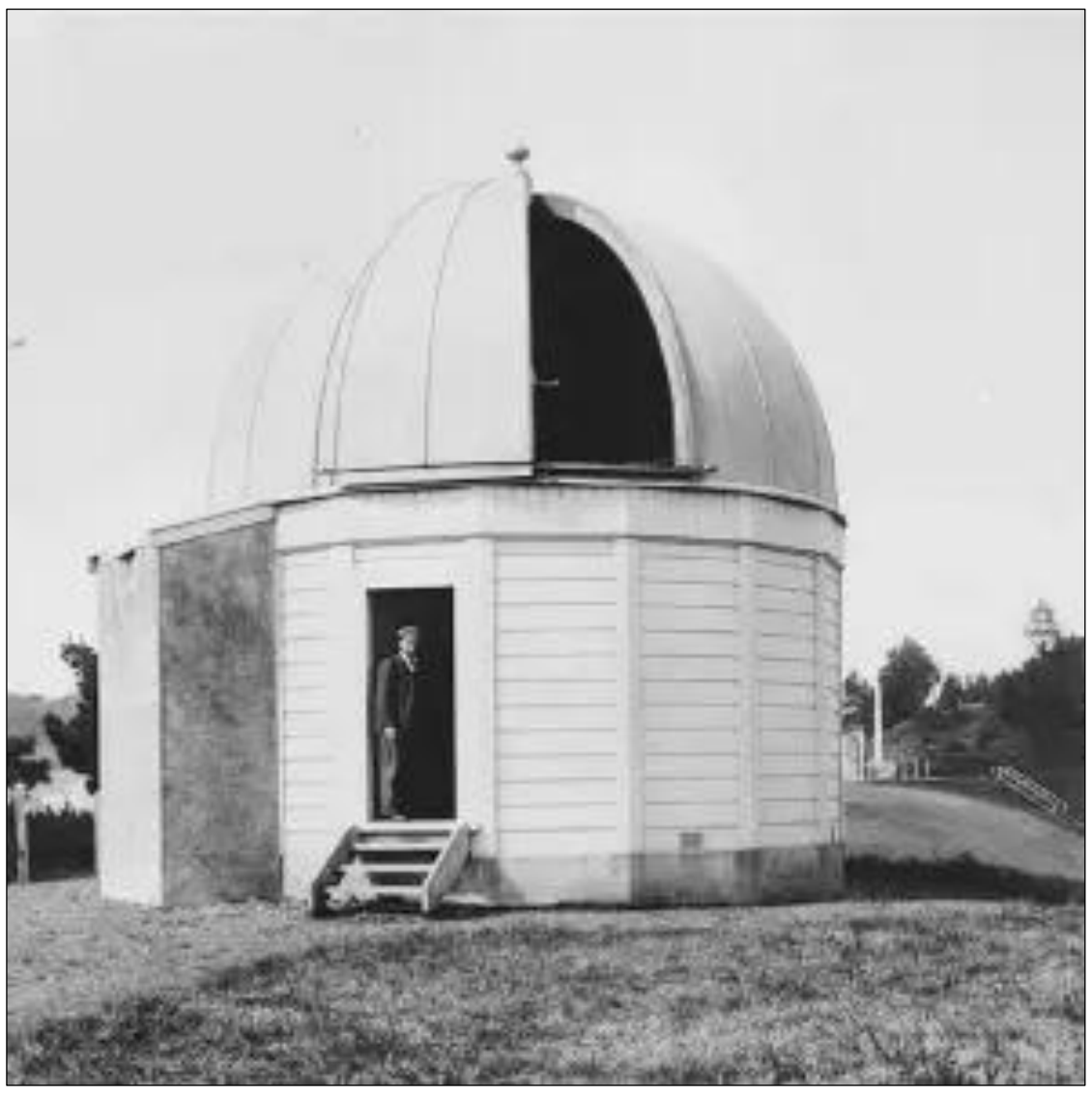

Figure 22: The Ward Observatory at Wanganui soon after its official opening in May 1903 (courtesy: Wanganui Observatory Archives).

The lunar eclipse on Saturday 27 November 1909 was taken as an opportune time to try to observe the comet visually because the sky was unimpeded from the bright light of the full Moon for 1.5 hours. Joseph Ward, the Director of Wanganui Observatory, unsuccessfully attempted to visually observe 1P/Halley with the Observatory's 9.5-inch (24-cm) refractor. By late November, the comet was estimated to be magnitude 11 (Kronk: 2007: 143–144), and should have been viewable, however, by the time totality began at 8:15 pm local time, 1P/Halley was only 9° above the NW horizon (3° east of Aldebaran). Such a low elevation, with the associated atmospheric extinction, would have rendered a positive observation difficult, if not impossible.

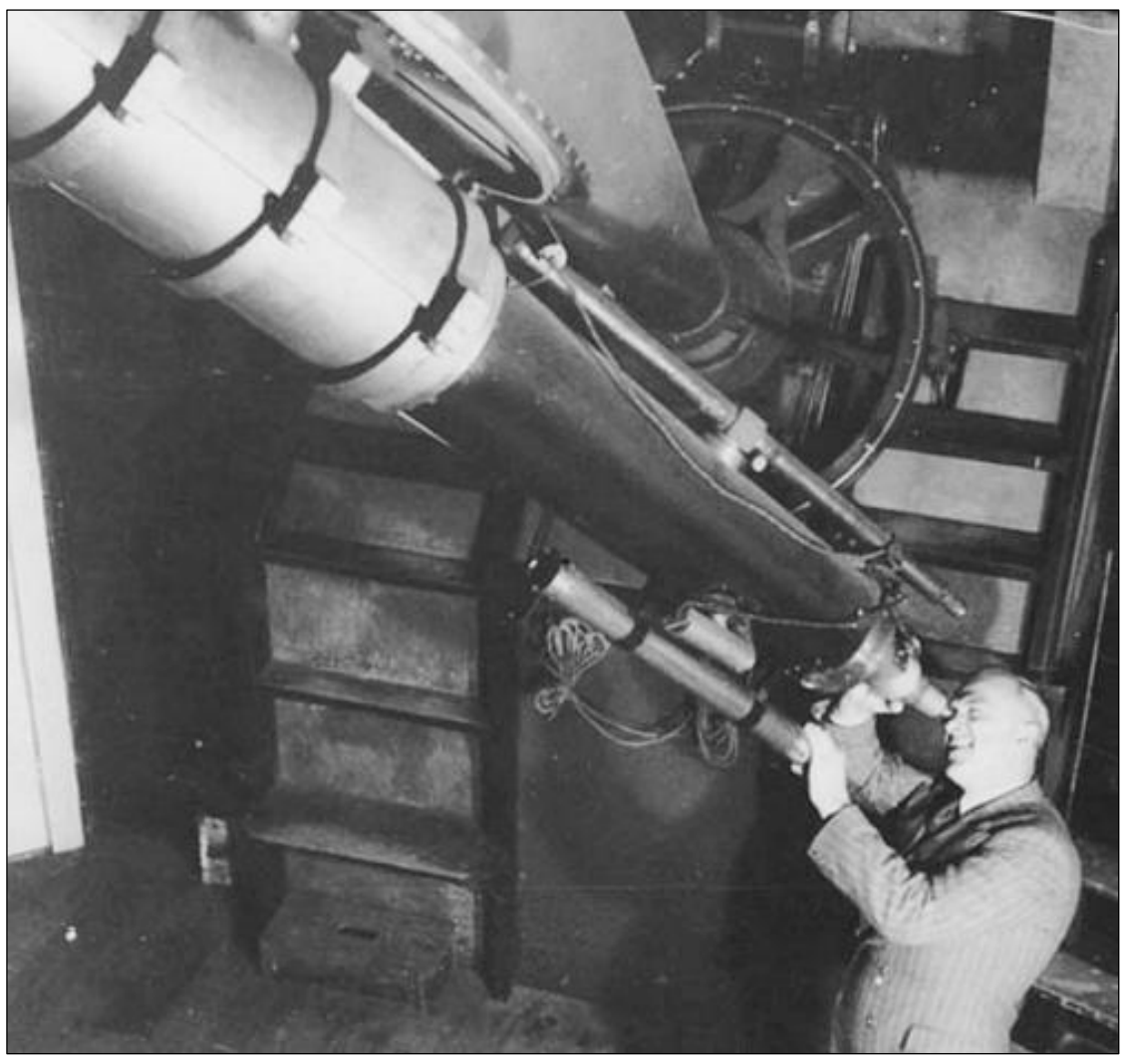

Figure 23: William Herschel Ward (Joseph Ward's son) observing with the 9.5-inch Cooke refractor at Wanganui Observatory. He was the Observatory's Honorary Director from 1927 to 1959. This telescope was most probably the first to be used to visually detect Comet 1P/Halley during its 1910 apparition (photograph: Orchiston Collection).

Mackrell (1985: 106) states that Joseph Ward of Wanganui Observatory may have been the first to visually observe 1P/Halley from New Zealand. Mackrell writes that on 3 January 1910, five weeks after Ward's first unsuccessful attempt, Ward noted in the observatory log, "Halley's Comet seen … as a very faint nebulosity between intervals of cloudiness …" The following evening (4 January) he wrote, "... the comet seen much better than on last night. The nucleus is quite star-like …"; an attribute confirmed by Graff in Kronk (2007: 145). Ward wrote to the *Otago Witness*, which published the following on 19 January 1910:

> Halley's Comet now lies near Mars or at a point between the stars in the constellations Aries and Pisces. It may be seen in a telescope of from 4 to 5 inches diameter. Fastening a 3-inch refractor to the large telescope of the observatory (9½ inches in diameter) and bringing it into collimation, it was directed exactly on to the comet's position, the large telescope serving meanwhile as finder, but without result. No trace of the comet could be seen in the 3-inch. This was at the beginning of the month [i.e. early January 1910]. As seen in the 9½ inch, the coma might easily be passed over if the sky were the least bit hazy, or the observer's eye fresh from using the observatory lamp. (Ward, 1910).

So, Ward could visually detect 1P/Halley with the 9.5-inch (24-cm) on 3 and 4 January 1910, but not with a 3-inch (7.5-cm) refractor. The limiting magnitude for a comet with a 3-inch (75-mm) telescope is approximately 9–10 ($m = 2.7 + 5 \log 75 = 12$) (Ridpath, 2004: 28–29), but fainter for a diffuse comet. Kronk (2007: 145–146) states that at the beginning of January 1910 1P/Halley was estimated to be around magnitude 10–11, so slightly too faint to be seen with the 3-inch refractor (note that magnitude estimates for 1P/Halley in Kronk during this period varied slightly).

Approximately one month after Ward's observation with the 9-inch refractor at Wanganui, newspaper articles stated that John Grigg of Thames observed 1P/Halley on Wednesday 2 February 1910 (Halley's Comet, 1910b). Grigg required a brightening of the comet to see it with his smaller aperture 3.5-inch (8.9-cm) refractor compared to Ward's earlier viewing with more light collecting power. Grigg was an enthusiastic amateur astronomer who had an observatory and 3.5-inch Wray refractor at Thames on the western side of the Coromandel Peninsula. His astronomical interests were varied, and he was the first person in New Zealand to photograph a comet (in 1901). For a greater treatise of his life and exploits, see Drummond et al. (2026), Orchiston (2016: 481–508, 597–607, 612–613, and 622–623) and Mackrell (1985: 75–79). On Friday 5 February, 21 newspapers across New Zealand reported Grigg's visual observation of the comet on 2 February. No details were supplied other than the fact that it was "... in the direction of the planet Saturn …" (Halley's Comet, 1910b). An article on 7 February stated that Ward observed 1P/Halley on the same evening as Grigg. They compared notes and found their recorded positions for the comet were "... almost identical …" (Halley's Comet, 1910a).

Approximately one week after Ward and Grigg's observations of 1P/Halley, Reverend F.W. Walker (Fellow of the Royal Astronomical Society) observed the comet on several occasions in mid-February 1910. (Local and General News, 1910a). He made telescopic observations of the comet as it was low in the west after sunset. He stated that it was about 4° west of Saturn. Walker described, "The comet, which at present is tailless, presents a beautiful nebulous appearance, is somewhat circular in shape, and in size is about twice the diameter of the planet Jupiter."

At this time, international observers were reporting a coma diameter of 1.5–2′ (Kronk, 2007: 147). Jupiter has an angular diameter of approximately 0.7′, so Walker's statement that 1P/Halley's coma was "… twice the diameter of the planet Jupiter …" was clearly in keeping with those international observations. Around the time of Walker's observations in mid-February, Joseph Ward suggested that the comet was still too faint to be seen with the naked eye from New Zealand and that "It should be seen by the naked eye at the end of April, or beginning of May, after it has passed the sun on the farther side, and [appear] in the early morning." (Ward, 1910b). By this stage the comet would be very much brighter due to its interaction with the Sun and its greatly reduced distance from Earth. Ward (1910c) further wrote in late February:

> Halley's Comet is now in Pisces, rather to the west, and beneath the planet Saturn. He is now setting rather soon after the Sun, and is a more difficult object in the telescope. His appearance has changed very little, as seen at this observatory, since the beginning of January, the coma being rather larger, but not brighter, owing to our increasing distance. The next favourable opportunity will occur about the middle of April, as it draws away from the sun in the early morning.

Ward's last observation of 1P/Halley prior to its 1910 perihelion was 17 February 1910 (Ward, 1910d) when it was 47° from the Sun. Each night thereafter the elongation was reduced until the comet–Sun separation was 6° in late March (*ibid.*). Ward predicted that the best time to see the comet would be in the few mornings until 19 May 1910 (after it reached an acceptable elongation of 41° on 7 May, *ibid.*). After 19 May and its transit of the Sun, it would move into the evening sky and fairly rapidly diminish in brightness. The passage of the comet in relation to the Sun and Earth can be seen in Figures 24 and 25.

The astronomers at Meeanee Observatory agreed with this prediction. Towards the end of March 1910, a journalist from the *Napier Daily Telegraph* paid a visit to Meeanee Observatory so as to better inform his readers about Comet 1P/Halley (Halley's Comet. Observations at Meeanee, 1910). The comet was stated to be "... not visible at present, as it is directly behind the sun …" Perihelion was expected to occur on 17 April (NZST), although Kronk (2007: 149) says 20 April 1910 UT (19 April NZST). The astronomers at Meeanee Observatory provided a nice summary of the predicted views of the comet by stating:

> ... it will rise forty-five minutes prior to our source of heat and light during April, and it will continue to travel further and further from the sun until about May 9th, when it rises about two hours and forty-five minutes earlier than the sun. This gives a good opportunity of observing the tail, which rises before the head. It will be possible for anyone to stand on the Marine Parade or hillsides during the second week in April, about three or four o'clock in the morning, looking directly eastward, which is practically at right angles with the parade, and observe the tail with the naked eye. Shortly afterwards, the early riser will see the complete comet. Its rise will become earlier and earlier every day until May 7th, when it will begin to descend towards the sun again. On May 19th the head transits the disc of the sun, and on that day or the next we will pass through its tail. The whole sky will then, it is said, be illuminated by a faint light soon after sunset, the tail shooting away from the sun and the head of the comet pointing downwards.

Around mid-March 1910, when 1P/Halley was too close to the Sun for viewing (NASA/JPL suggests it was 15° NE of the Sun on 15 March), some misinformed observers were reporting sightings of it, stating that it

> ... resembles very much in shape a silver bayonet suspended in the I sky so bright is it and so large in comparison to the rest of the stars in the firmament that they pale into insignificance altogether. (Late Locals, 1910).

This is a good description of Venus, which at that time was rising at approximately 3:15 am local time. One wonders if the observer was suffering from astigmatism with his describing Venus as a bayonet! 'WR' pointed out on 8 April 1910 (Halley's Comet and Venus, 1910) that the observer was describing Venus rising in the east before dawn. Other New Zealanders also reported "… a comet …" in the dawn sky from Gore, (*Oamaru Mail*, 1910a) however, 1P/Halley was not rising by the 3:30 am time they stated that they saw it on 9 April.

The first authentic observations of 1P/Halley after its close encounter with the Sun from mid-March 1910 were international. Observers in Melbourne and Cape Town reported seeing the comet in the dawn sky for a few minutes on 8 April 1910 UT (9 April NZST) until the view was obliterated by the glare of the Sun (Halley's Comet, 1910c). In the following days, other international observatories also reported seeing the dawn spectacle (see Kronk, 2007: 148–149).

As Comet 1P/Halley increased in elongation in early April 1910, observations from New Zealand began to populate newspapers. The first were made by Joseph Ward (Wanganui) and Mr A.W. Burrell (Stratford) (Observed at Wanganui, 1910). Ward observed the comet on 13 April 1910 at 5:10 am (NZST). The nucleus was described as very bright, and a 5′ tail was seen. The comet could be seen until 5:45 am, the Sun rising at 6:46 am (NZST) for Wanganui (Time and Date). Thus, it was visible until the start of nautical twilight. Burrell, who had an observatory near Stratford (20 km East of Mount Eg-mont, now Mount Taranaki, and ~93 km north-west of Ward), wrote a detailed account of his 13 April viewing, stating:

> Halley's comet was again located here with the 5in telescope this morning. After passing behind the sun, at 5.30 a.m., it was bearing about 14 degrees north of east, and at an altitude of about 5 degrees, rising about one-and-a-half hours before the sun. It is just discernible to the naked eye, fairly bright in a telescope of 1¾ in aperture, so that with a clear sky there should be no trouble in seeing it with ordinary field glasses now, in which it will appear as a fairly bright, fuzzy ball, with the 4th magnitude star Omega Piscium in the same field.

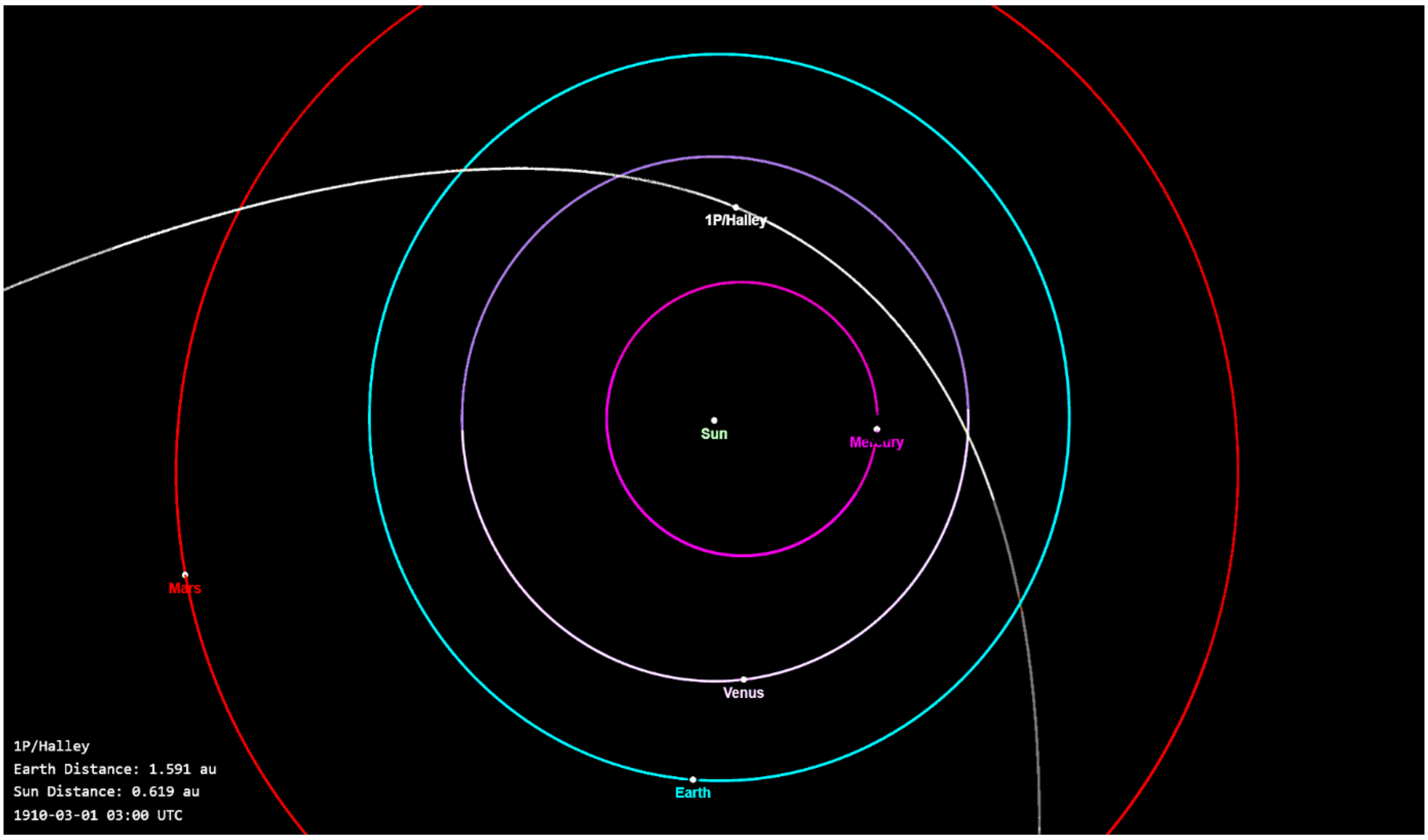

Figure 24: The positions of Comet 1P/Halley and Earth at the start of March 1910. Ward told readers that around this time, the comet would be too close to the Sun for viewing, as can be seen in this diagram. The planets move in an anticlockwise direction and Comet 1P/Halley in a clockwise direction in this diagram (source: NASA/JPL).

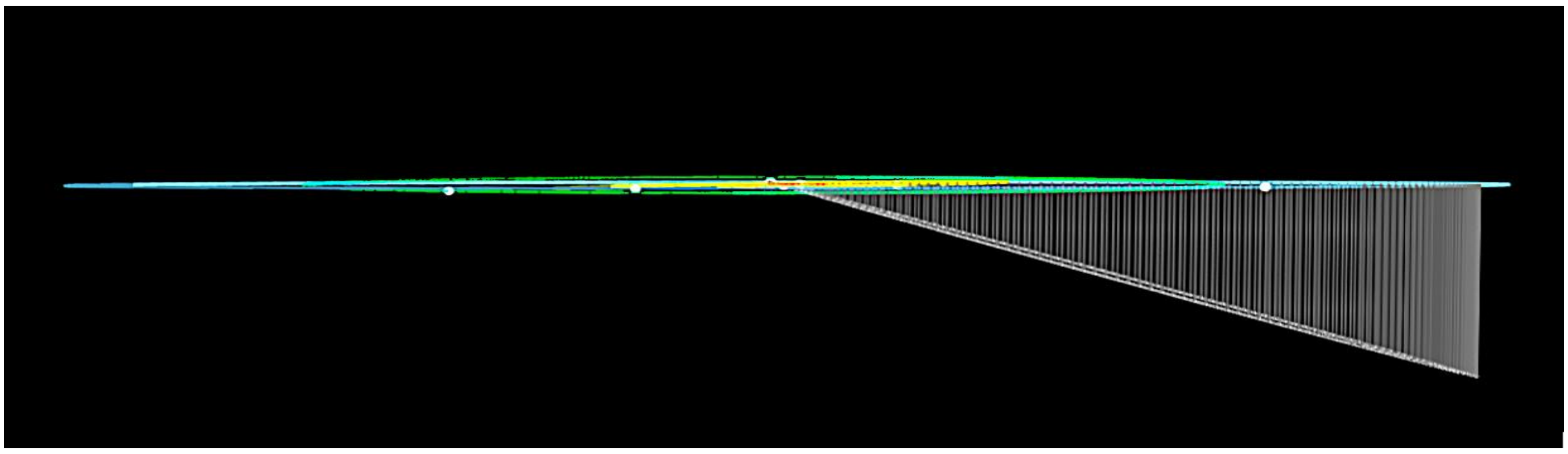

Figure 25: Comet 1P/Halley's orbit extends to that of Neptune's orbit and lies south of the ecliptic, as seen in this orbital diagram from JPL/NASA Small Bodies. The outer blue line is the orbit of Neptune (source: NASA/JPL).

Note that Burrell states, "Halley's comet was again located here with the 5in telescope this morning". Does this mean that he saw it on the previous morning(s) or that it was his first observation since it got too close in elongation to the Sun for observation in from early to mid-March 1910? On 13 April 1910, the comet was moving west, increasing its angular distance from the Sun, at about 14′ a day, so certainly could have been seen on 12 April, or even earlier. Ward confirms this when he wrote that he attempted to see the comet on 9 April but that a 'pinus insignia' (a tree) was in the way. Ward happily concluded, "The Borough Engineer having kindly seen to the decapitation of this interposing member, a free view was obtained this morning [13 April]." (*Wanganui Herald*, 1910). Kronk (2007) states that Howe observed it first on 8.49 April (UT), 8 April NZST. A Papers Past search from 5 to 14 April 1910

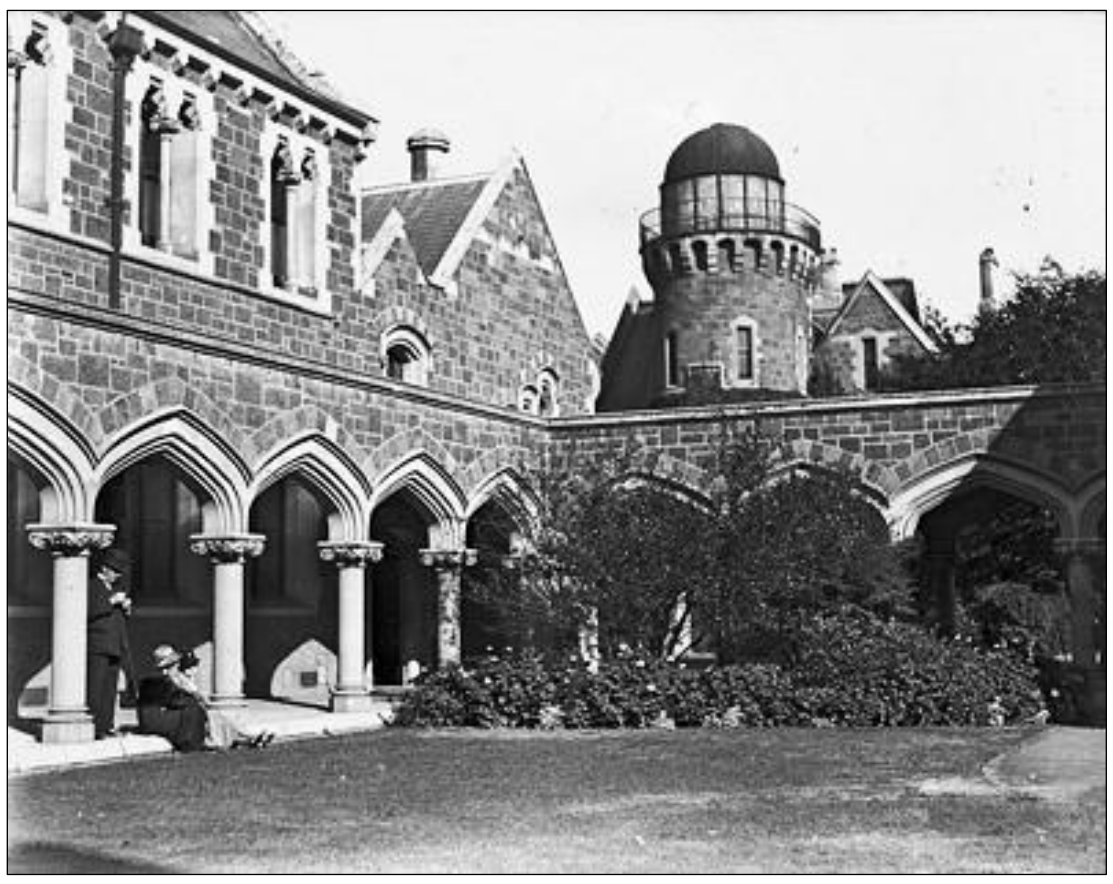
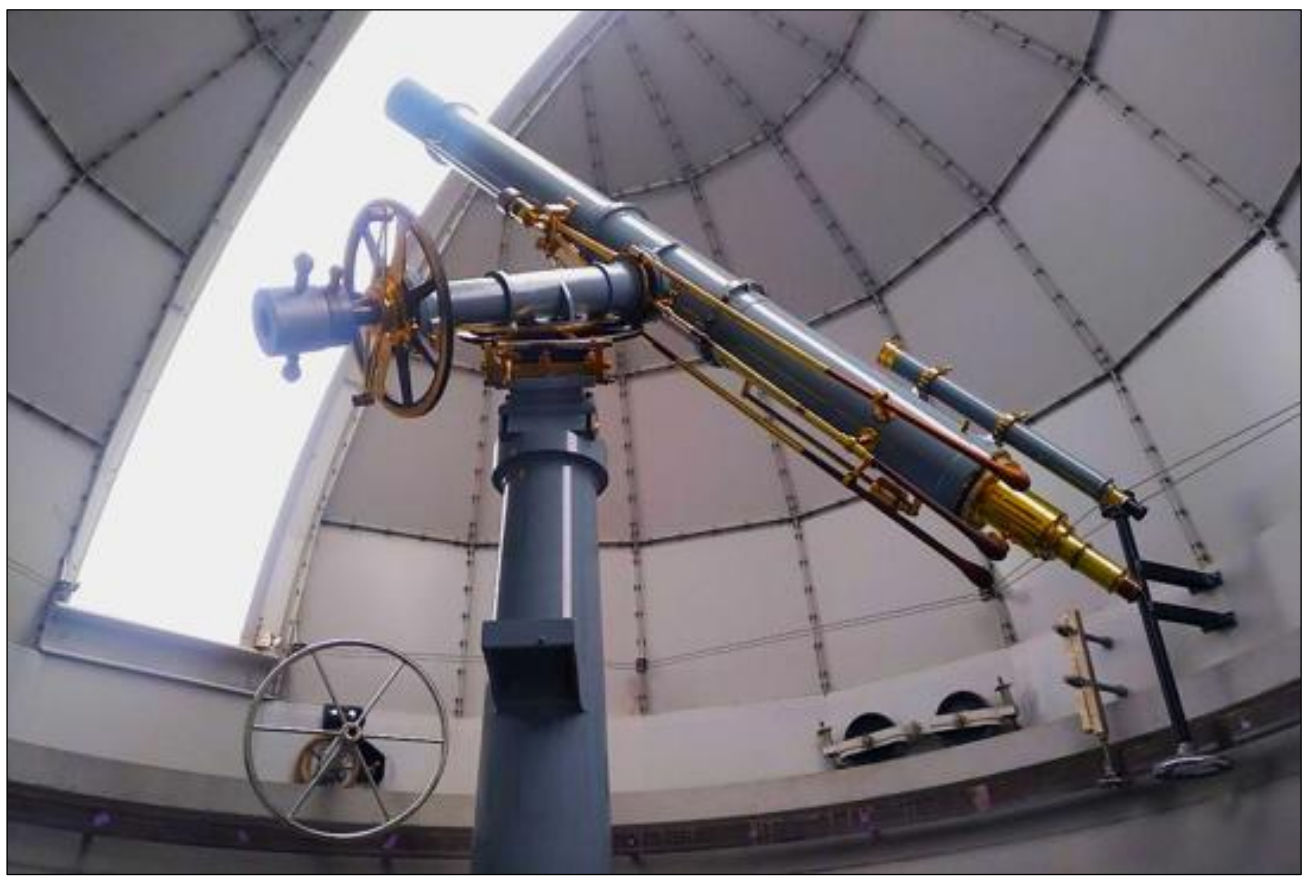

Figure 26a (left): Canterbury University College Observatory as photographed by an unknown photographer in the 1930s. 75 steep steps led up to the telescope. Figure 26b (right): The 6-inch (15-cm) Townsend Teece Telescope was made in 1864 and donated to Canterbury University College in 1891 by Christchurch amateur astronomer, James Townsend. The observatory and telescope were destroyed in the 2011 Christchurch earthquake. Amazingly, the objective survived the quake intact, and fall to the ground many metres below. It was restored and re-installed in 2023, as the photograph taken taken by Dr. Orlon Petterson shows (sources: https://eventshub.co.nz/events/heritage-stargazing-at-the-observatory-tower-2025-303-oleg-iudin and https://www.canterbury.ac.nz/research/about-uc-research/enabling-infrastructure/townsend-teece-telescope).

shows no evidence that Burrell sent in a positive report prior to his 13 April one. Perhaps he did see 1P/Halley but did not write to the newspapers. An alternative is that he saw it in February or March 1910, hence the word "again". If Burrell did view the comet on or before 8 April, he may have been the first in the world to see the appearance of the comet after it began emerging from the Sun's proximity. However, this is pure conjecture.

An interesting article surfaced that described A.W. Burrell's observatory in detail. An unnamed traveller in Taranaki wrote to the *Dominion* newspaper that he came across an "… up-to-date …" observatory next to a cottage in the countryside. He accepted an invitation to view the stars with his wife (being an amateur astronomer himself) (Astronomy Among the Cows, 1910). Burrell must have been quite a character with good improvisation skills. A description of his observatory by the unnamed traveller reveals that Burrell possessed a 5-inch (12.7-cm) Cooke refractor, imported from London. Since the equatorial head would have doubled the price of the telescope, Burrell, a dairy farmer by trade, built his own. For a telescope pier, he cut a red pine tree down and left the stump standing to a height of 8-feet (2.4-metres), upon which he mounted his refractor. Around this tree stump, Burrell built an observatory. In the corner of the observatory, a 24-hour sidereal clock revealed the right ascension.

Mr J.T. Stevenson (FRAS) saw 1P/Halley from Auckland on 15 April, though no details were given regarding instrumentation (Halley's Comet, 1910d). He anticipated that the comet's tail could brush Venus on about 2 May. On the same morning, Miss Marchant of Wellington described the telescopic view of the comet as "… a gently glowing but distinct nebulous mass …" (Evening Post, 1910a), and that its position agreed with that published by Ward two days prior. Mr H.F. Skey also saw the comet that same morning with the 6-inch (15-cm) telescope at the Canterbury University College Observatory in Christchurch (see Figure 26a and 26b). He said that the nucleus and tail appeared "… very bright …" (Halley's Comet. Observed from Christchurch, 1910). Skey nicely summarised the orbital mechanics of the comet, stating:

> Although the comet is approaching the sun it appears to be moving away from it, and on April 20 the distance of the comet west of the sun will be twenty-nine degrees. On April 30 the interval will have increased to nearly forty degrees, and by that time it will have passed perihelion and will probably have developed as much tail as it is likely to have, during the present apparition.

The following morning (16 April), Mr D.B. McLeod of the Physics Laboratory at Canterbury University College used the same 6-inch refractor as Skey the previous morning and described 1P/Halley as:

> ... a fairly distinct object, resembling a patch of nebulous or cloudy matter. The nucleus was bright, and was surrounded by a hazy luminosity, the whole covering an area of about the size of the full moon in the sky [30′]. The tail was not very distinct, but a decided luminosity extended from the nucleus in a direction away from the sun. The comet was bright enough to be seen in the small finder attached to the large telescope … the comet only appears as a dull white patch, showing no signs of twinkling. The position of the comet is in the constellation Pisces, and appears in the east, to the north of where the sun rises. The comet, Mr McLeod states, is at present approaching its perihelion, which it reached on April 19th. Until then, although it appears to us to be getting further from the sun, it is really approaching the sun. After the 19th it will turn and go back in about the same direction, only passing on this side of the sun instead of the other. On May 18th it will pass between us and the sun, and will then be nearest the earth, and as it happens that the sun, and the earth, and the comet will be probably in the same straight line, and as the tail points directly away from the sun, it is probable that the earth will pass through the comet's tail, and it may be that the tail will be visible at night beyond the earth. The comet will become brighter up to May 18th, when it will become an evening star, but after that it will quickly diminish in brightness, in four days dwindling to half this brightness, and in another four days it will only be one-tenth as bright as it was at its best. It will only be a good object for four or five days, and by the end of May will be very dim. (Halley's Comet, 1910e).

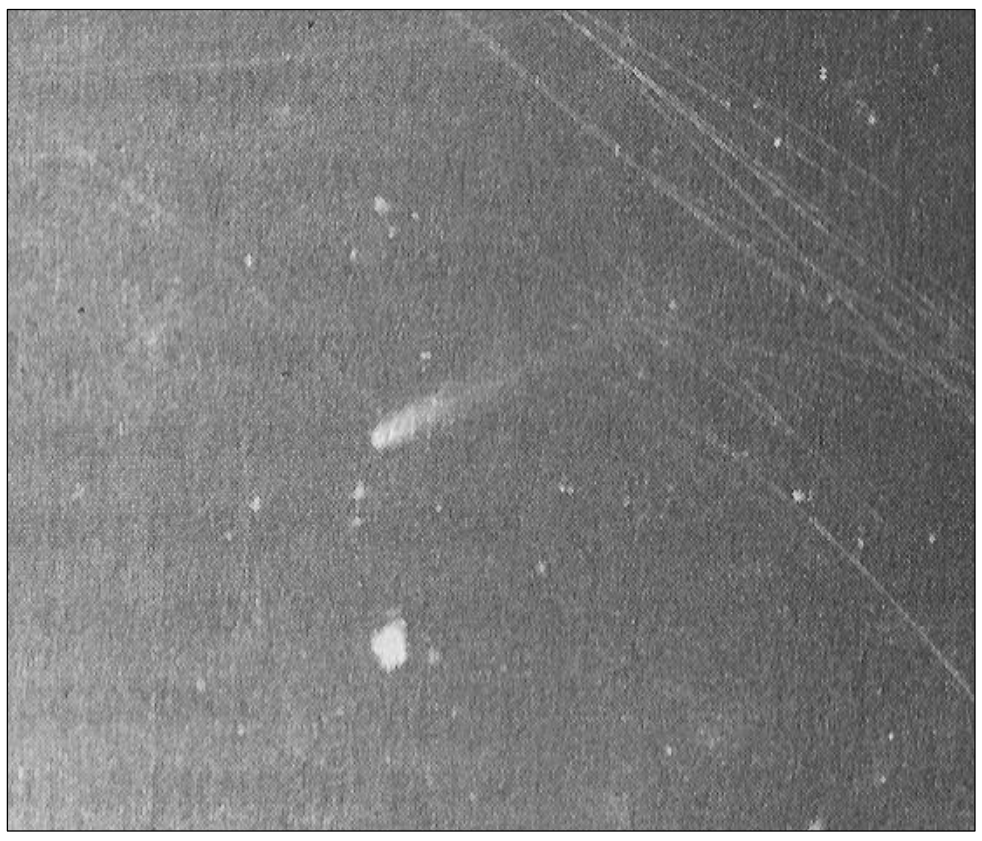

Figure 27: Comet 1P/Halley photographed from Meeanee Observatory on 18 April 1910, around the time when the comet was becoming a naked eye object after being too close to the Sun (line-of-sight) for observation. The criss-cross lines are scratches on the plate (after Mackrell, 1985: 110).

Joseph Ward also provided astronomical facts about 1P/Halley stating that the comet was travelling at over three million miles (five million kilometres) a day and, when at perigee, would be some 14 million miles (23 million kilometres) from Earth. The tail would possibly be 19–20 million miles (31–32 million km) long. When asked if the Earth passing through the comet's tail would have any effect on humanity, Ward replied, "... it will have about as much effect on us as a spoonful of whisky in a 400-gallon tank of water would have on a seasoned toper [*sic*]." (Late Locals, 1910b).

By mid-April 1910, numerous international observers stated that they were observing comet 1P/Halley telescopically (Kronk, 2007: 148). Kronk states that the first naked-eye observation after the comet passed 6° from the Sun on 27 March (UT) and then reappeared in the morning sky was made by D. Costa (National Observatory, Rio de Janeiro, Brazil) on 16 April (UT) (15 April NZST) (Kronk, 2007: 148). A newspaper article in the *Auckland Times* stated that 1P/Halley was also seen with the naked eye on 16 April 1910 UT from Cape Town by an unnamed observer (Halley's Comet, 1910f). However, remember Burrell's (Stratford, New Zealand) comment that "… it is just discernible to the naked eye …" on 13 April NZST (14 April UT), two days before Costa and the unnamed observer spotted it with the naked eye. Figure 27 shows a cropped version of a photograph of 1P/Halley taken by the Meeanee astronomers two days after Costa's 16 April (UT) observation was made.

## 5 1P/HALLEY BECOMES 'ANYONE'S COMET'

As stated, D. Costa from Brazil observed 1P/Halley with the naked-eye on 16 April 1910 (Kronk, 2007: 148). But who was the first to view the comet without optical aid from New Zealand?

### 5.1 Visual Observations of Comet 1P/Halley From New Zealand

Our study of the records in Papers Past suggests that the first New Zealand naked-eye observation of 1P/Halley at the 1910 apparition was made by Burrell on 13 April (NZST) (see above). The next was probably made on Sunday 17 April (NZST) by Mr G.V. Hudson of Wellington (Halley's Comet, 1910g; Halley's Comet Now Visible Here, 1910). Hudson was "... one of [the Dominion's] most enthusiastic observers …" (Halley's Comet Now Visible Here, 1910). Mackrell (1985: 109) quotes Hudson:

> I picked up Halley's comet at 4.54 this (Sunday) morning. It was then just clear of the hills on the north-east side of the harbour, and clearly visible to the naked eye. Its brilliancy was about equal to a star of 3½ magnitude, and the tail was Figure 28 (right): The rise times for 1P/Halley and the Sun for Wellington's location. The distance in millions of miles is also displayed. The table was prepared by Reverend Kennedy. The highlighted word 'comet' is a result of the Papers Past search (source: *Dominion*, Volume 3, Issue 794, 18 April 1910, Page 7).

| Date. | Comet rises. a.m. | Sun rises. a.m. | Distance of Comet from the Earth. Millions of miles |
|---|---|---|---|
| April 18 | 4.27 | 6.22 | 118 |
| April 19 | 4.23 | 6.23 | 115 |
| April 20 | 4.18 | 6.24 | 112 |
| April 21 | 4.14 | 6.25 | 108 |
| April 22 | 4.10 | 6.26 | 105 |
| April 23 | 4.6 | 6.27 | 102 |
| April 24 | 4.2 | 6.28 | 99 |
| April 25 | 3.58 | 6.29 | 95 |
| April 26 | 3.55 | 6.31 | 91 |
| April 27 | 3.51 | 6.32 | 88 |
| April 28 | 3.47 | 6.33 | 85 |
| April 29 | 3.46 | 6.34 | 81 |
| April 30 | 3.43 | 6.35 | 77 |
| May 1 | 3.41 | 6.36 | 73 |
| May 2 | 3.38 | 6.37 | 70 |
| May 3 | 3.38 | 6.38 | 66 |
| May 4 | 3.37 | 6.39 | 62 |
| May 5 | 3.36 | 6.40 | 59 |
| May 6 | 3.34 | 6.41 | 56 |
| May 7 | 3.35 | 6.42 | 52 |
| May 8 | 3.36 | 6.43 | 48 |
| May 9 | 3.37 | 6.44 | 45 |
| May 10 | 3.40 | 6.45 | 41 |
| May 11 | 3.44 | 6.46 | 37 |
| May 12 | 3.48 | 6.47 | 34 |
| May 13 | 3.57 | 6.48 | 30 |
| May 14 | 4.9 | 6.49 | 27 |
| May 15 | 4.28 | 6.50 | 24 |
| May 16 | 4.48 | 6.51 | 21 |
| May 17 | 5.17 | 6.52 | 18 |
| May 18 | 5.58 | 6.53 | 16 |
| May 19 | 6.49 | 6.54 | 15 |

On May 19 the comet will pass across the sun's disc between 1.50 and 2.50 p.m., and after that date will appear in the western sky after sunset.

| Date. | Comet sets. p.m. | Suns sets. p.m. | Distance of Comet from the Earth. Millions of miles |
|---|---|---|---|
| May 20 | 4.39 | 5.48 | 14 |
| May 21 | 4.38 | 6.32 | 15 |
| May 22 | 4.37 | 7.33 | 16 |
| May 23 | 4.36 | 7.55 | 18 |
| May 24 | 4.36 | 9.27 | 21 |
| May 25 | 4.35 | 10.0 | 24 |
| May 26 | 4.35 | 10.19 | 27 |
| May 27 | 4.33 | 10.22 | 30 |
| May 28 | 4.33 | 10.33 | 34 |
| May 29 | 4.32 | 10.38 | 37 |
| May 30 | 4.32 | 10.44 | 41 |

easily seen as a faint streak. It was well seen in a one-inch 'finder', and a good binocular would give a fine view. In a four-inch telescope the comet was a very brilliant object. The nucleus was well condensed, and the corona and tail were very distinct. The position was estimated at right ascension 23 hours 55 minutes, and north declination 7 degrees 55 minutes. The comet could not be seen with the naked eye after 5.26 a.m., but, with the telescope, it was clearly visible until 5.53 a.m., and was finally lost in the daylight at 5.56 a.m. This indicates a very great increase in brilliance since April 13, when the comet was first seen at the Wanganui Observatory.

On the same morning, Reverend Kennedy saw the comet without optical aid. He saw it from Saint Patrick's College (Wellington) at 5:30 am. The comet was visible until 5:40 am (when the Sun was 14° below the horizon—near the start of nautical twilight). The head of the comet was described as being magnitude 4 and the tail was faintly seen—but accentuated with field glasses (binoculars). Kennedy reminded readers that the comet would transit the Sun on 19 May 1910 between 1:50 and 2:50 pm (NZST). He also supplied a table of daily comet rise times, the Sun's rise times, and the distance between the Earth and 1P/Halley in millions of miles between 18 April and 30 May 1910 (Figure 28).

Auckland's J.T Stevenson saw 1P/Halley the same morning as Hudson and Kennedy (17 April). He also described the head as being a naked-eye object and the tail as displaying a greater length than his previous observation (which wasn't stated), but needed field glasses to discern it more readily. He predicted that "... it will become a striking object in the western sky, about May 21 [1910]." (Halley's Comet, 1910h). Professor Segar agreed that 1P/Halley would be best in May 1910, stating that the comet "... will then, owing to the sun's rays, rapidly disappear from view until well on in May, when it will appear as a great object in the western evening sky." (*ibid.*). That same morning, Mr S.E. Hughes of Auckland described the head as bright and of a reddish colour (possibly due to Rayleigh scattering) and that the tail was very luminous (Halley's Comet, 1910h).

An unnamed observer in Christchurch, who signed their name as '*The writer of astronomical telegrams*', also saw 1P/Halley with the naked eye on the morning of 17 April. They could trace the tail for 2° through a telescope and stated that the comet was distinctly brighter than Comet C/1908 R1 (Morehouse) at its best (Halley's Comet, 1910i). They photographed 1P/Halley the following morning and the negative image revealed a 5° tail. The photography of 1P/Halley is discussed in Section 5 below. Mr T. Scott also observed the comet with the naked eye that morning, this time from Nelson. He described it as a "... small dull star of fourth magnitude" (Halley's Comet Seen With the Naked Eye, 1910). According to GUIDE, 1P/Halley lay 26° west of the Sun on this date and rose nearly two hours before the Sun. Kronk (2007: 148–149) writes that international observers such as Costa and Ristenpart were also reporting the magnitude of 1P/Halley to be around 3–4 on 17 April (NZST).

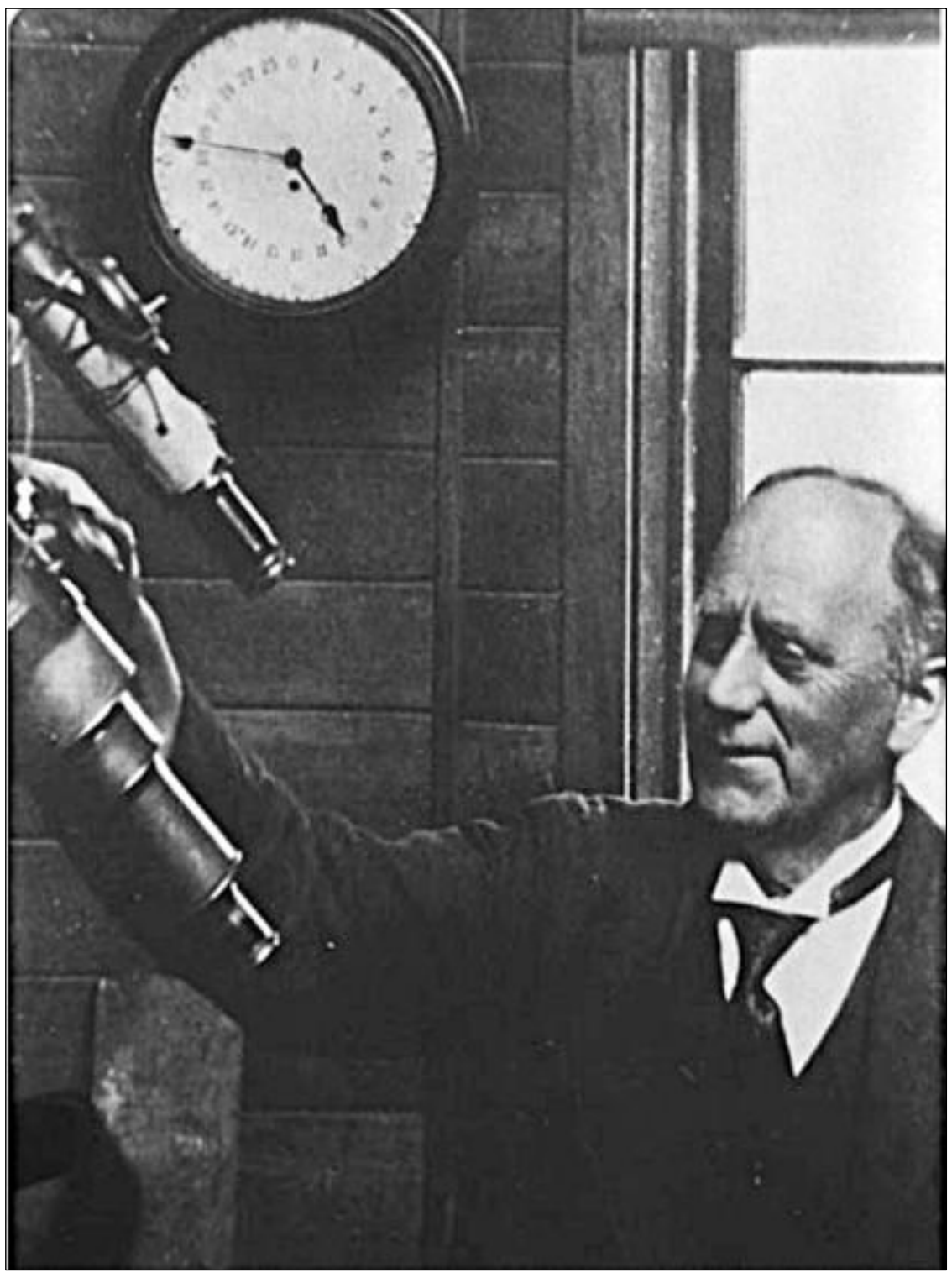

Figure 29: Joseph Ward at the eyepiece-end of his 9.5-inch refractor. Note the 24-hour sidereal clock on the wall which told Ward what right ascension was culminating at the meridian at any instant. After his 18 April 1910 observation of Comet 1P/Halley, Ward told the newspapers that the comet was now "any-one's comet" (photograph: Orchiston Collection).

On Monday 18 April 1910, two separate observation reports were published in the *Nelson Evening Mail* (1910). Mr T. Scott (Nelson) stated that 1P/Halley appeared as a "... small, dull star of the fourth magnitude …" with a tail pointing away from the Sun and towards Venus. Joseph Ward (Figure 29) of Wanganui Observatory offered a greater description stating that the comet was visible from 4:45 am that morning. He and some members of the public observed the comet with the 9.5-inch refractor. The Observatory was open to the public on Tuesday and Friday evenings from 8 to 9:30 pm, and possibly on other nights when 1P/Halley was visible (Gilkison, 1910). Ward stated:

> ... the telescope … revealed a well condensed nucleus of a somewhat irregular form, indicative of the influence of the sun upon this, the most solid portion of the body. The tail streamed away for quite 45 minutes of arc in a continuous band, no central division being detected so far. Its magnitude was about 3, and its rise in brilliancy during the last four days gives promise of making a fine appearance later on, when nearer the earth.

Based on this recent brightening, in the article, Ward described 1P/Halley as now being "… any-one's comet …", in that anyone could now see it, optical aid or not (*Wanganui Herald*, 1910b). Oscar Blundell, a student at Knox College (Dunedin), also determined the visual magnitude to be 3 on 18 April from Māori Hill (Dunedin) at 5:15am. He used a 3-inch (7.5-cm) telescope (*Evening Star*, 1910) to observe a tail of ¾-degree in length. Blundell would later spear-head the formation of the New Plymouth Astronomical Society in 1919 (Gladstone, 2018: 6). That same morning, the Reverend P.H. Pritchett saw the comet from Mount Somers (inland from Christchurch). From 1,250 feet (381 metres), he said that he could see a 1–2° tail with a pocket telescope (A Photograph Obtained, 1910). Pritchett's higher altitude (and less atmospheric extinction) at Mount Somers probably helped with a clearer sky to provide this longer tail description.

On Tuesday 19 April 1910 (NZST), two days before 1P/Halley's perihelion passage on 21 April (NZST, 20 April UT), several people, including Mr Coleman, Mr Harrison, Dr Reid Mackay and several unnamed ladies, observed the comet from Dannevirke with a 3-inch (7.5-cm) telescope. The nucleus was estimated to be magnitude 5. It also was described as being able to be seen by anyone (with the naked eye) but was still "… rather faint." (Halley's Comet, 1910j). J. Irvin (Massacre Hill, Nelson) also viewed the comet on 19 April, with a pair of binoculars. He stated that "... the comet could be seen until daylight." (Irvin, 1910; his underlining). At Paris Observatory in France, a Mr Giacobini also saw the comet with a 4-inch (10-cm) finder scope until after the Sun had risen (Kronk, 2007: 149). To see a comet during the daytime indicates that it must have been at least magnitude –4. An unnamed observer in Gisborne believed that they saw two tails (possibly a dust tail and a gas tail, or perhaps striae in the tail). On the same morning, others described the tail as "… bushy …" (Poverty Bay Herald, 1910). The view was reminding the less-young of their 1835 views of 1P/Halley 75 years earlier. A Kerikeri observer, Arthur Edmund, recalled that the head was then like a "… ball of fire, about a foot in diameter …", the tail stretching "… halfway across the heavens." (Rangitikei Advocate and Manawatu Argus, 1910).

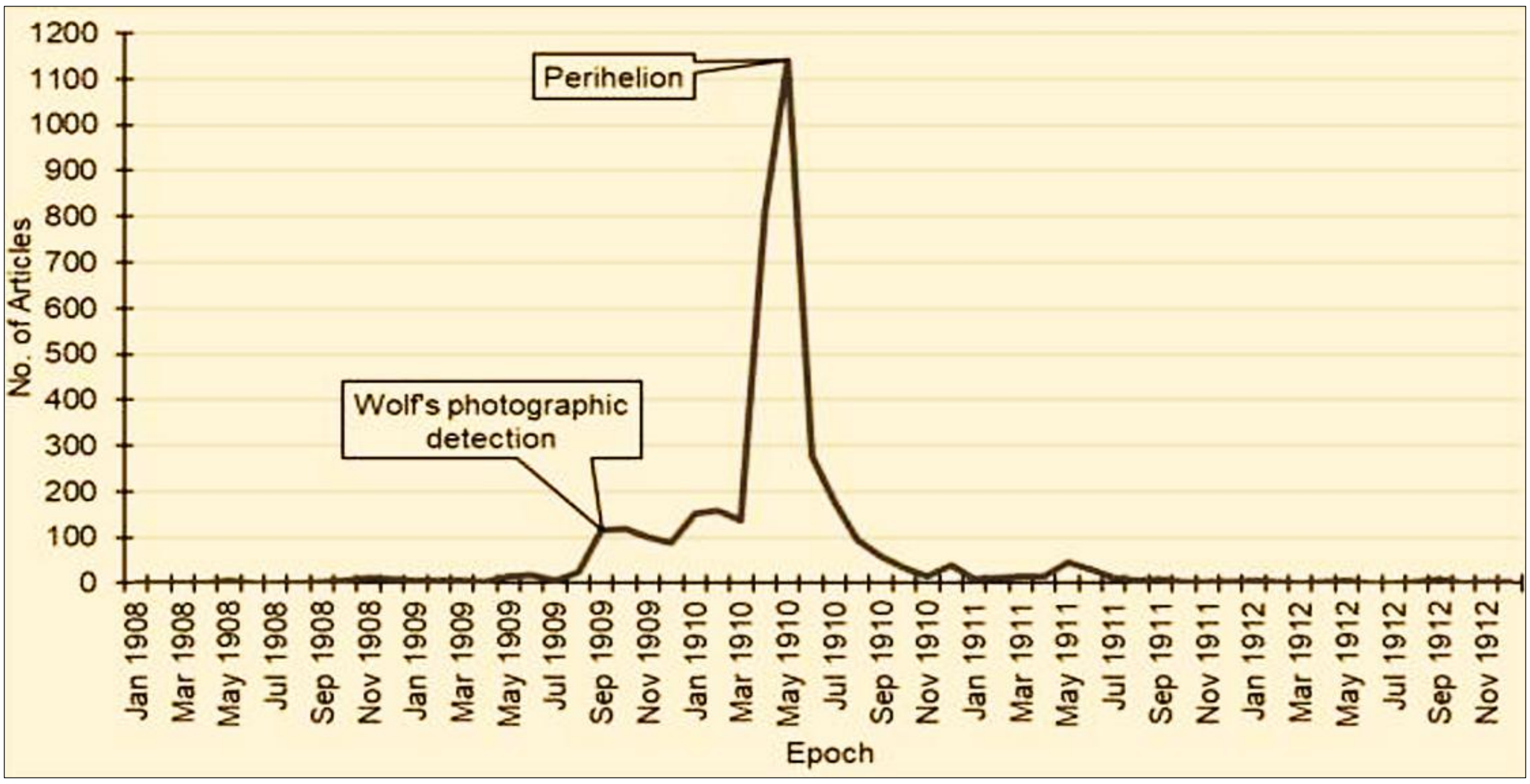

Figure 30: The number of newspaper articles relating to Comet 1P/Halley published across New Zealand between 1 January 1908 and 31 December 1912 using the Papers Past online search engine. Note how when Wolf photographically detected it in September 1909 (as well as Burnham and Barnard's visual detections), the number of articles rose sharply. After perigee on 20 May 1910 UT, as the comet faded, interest in the comet dropped dramatically (after Drummond et al., 2025).

On the same morning (19 April 1910), Joseph Taylor of Nelson said that the comet was in Pisces and that "... both nucleus and streamers were distinctly visible for some time after the usual stars had faded from sight in the growing glare of the rising sun." (Halley's Comet Observed at Nelson, 1910). Note the term "streamers". Was he observing the two tails (gas and ion, or perhaps striae) that the Gisborne observer witnessed? Kronk (2007: 148–149) writes that R.T.A. Innes (Transvaal Observatory, Johannesburg, South Africa) took photographs that revealed a 3° tail that was divided into five rays only days beforehand, this sounds more like striae than a dust and gas tail. Also on 19 April, Mr D.B. McLeod observed 1P/Halley from the Canterbury University College Observatory (Christchurch) with the 6-inch telescope. He stated that it had greatly increased in brightness from a few weeks before. A 1° tail was seen and the comet was observed until about 20 minutes before sunrise (Nearing the Earth, 1910).

By 20 April 1910, many residents were viewing the comet with the naked eye all around New Zealand, such as this amusing account,

> Between 4.30 and 5 o'clock this morning [20 April 1910], which was bright and clear, quite a large number of people were scattered along High-street leaning out of windows, etc., viewing Halley's Comet. (Halley's Comet, 1910k).

Interest in the comet ranged from the general public right up to the Prime Minister. The then Prime Minister of New Zealand, Sir Joseph Ward (1856–1930), and his wife, rose at 3 am to go to a good vantage point to see the comet (Woman in Print, 1910). On the same morning (20 April), Mr Henry B. France observed the comet from Levin. He described it as easily seen with the naked eye and that the tail was short and bushy and that it "… was brighter on the outside edges …" through a telescope. It had ".... greatly increased in brightness during the last two days …" and at 5:45 am NZST, the comet, Venus, Canopus and the Pointers were the only objects visible as dawn progressed (The Chronicle, 1910a). Professor Segar described it as a "... blurred luminous mass."

As 1P/Halley got brighter in late April and into May 1910, many New Zealanders reported seeing it to local newspapers; most were naked eye observations. As seen in Figure 30, the number of newspaper articles begins to rise after Wolf's photographic detection, with a dramatic increase around perihelion. During this time the number of articles exceeded 1100 articles within the month around perihelion.

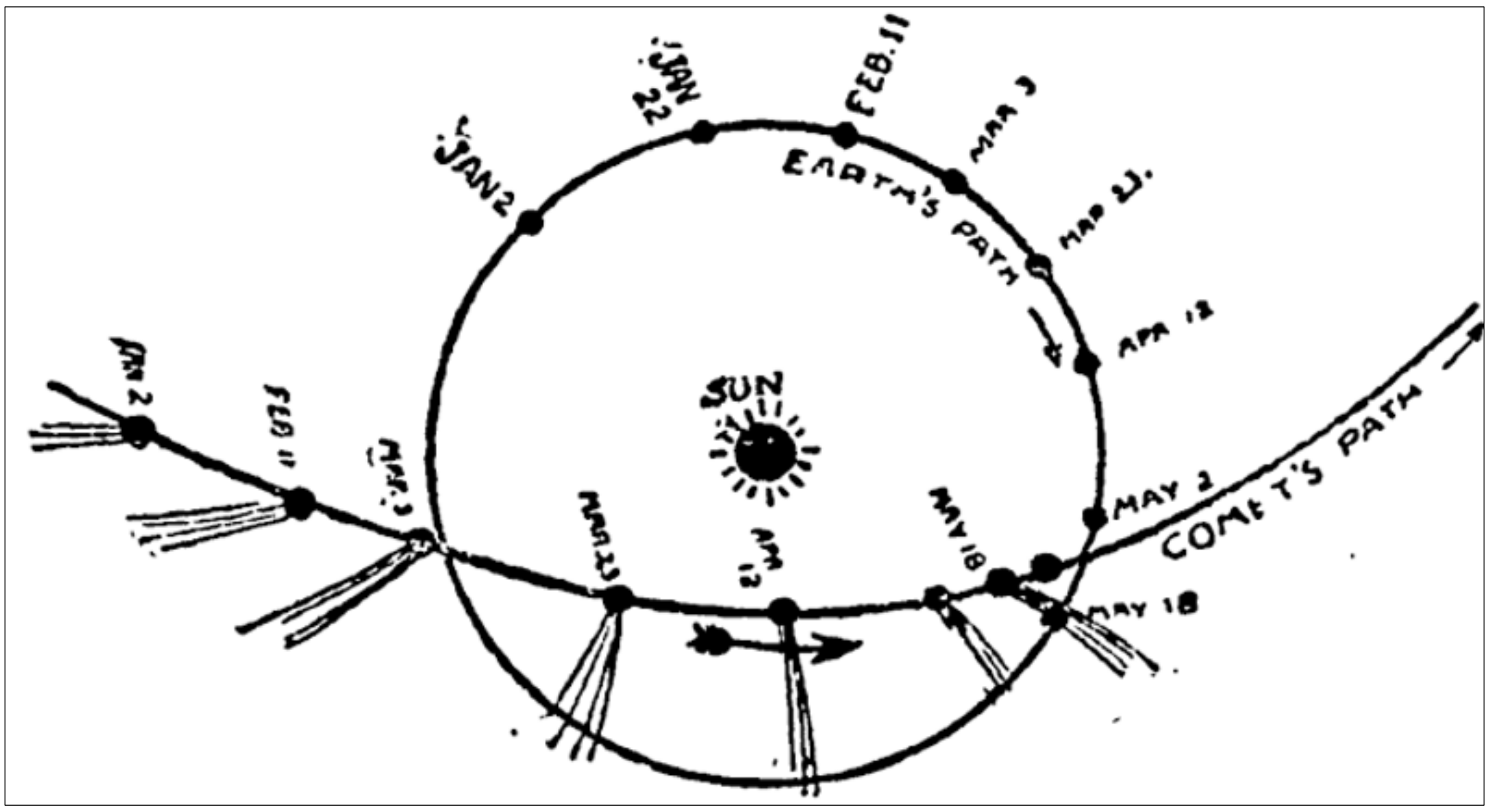

Figure 31: On 23 April 1910 the *Evening Post* published the above diagram of the orbits of Comet 1P/Halley and Earth from January to May 1910. Note the anticlockwise (from below the Southern Hemisphere) motion of the comet compared to the clockwise motion of Earth (as viewed from below Earth's South Pole). Astronomically speaking, Earth has a prograde and 1P/Halley a retrograde orbit (after Where the Earth and the Comet Collide, 1910).

## 5.2 Comet 1P/Halley Reaches Perihelion

On the day of perihelion, 21 April 1910 (NZST), Joseph Ward of Wanganui Observatory noted that the nucleus was well condensed and irregular in shape. The tail was described as "... stretching away in a continuous band, no central division being detected." He noted that it had increased in brightness over the last four days (Halley's Comet, 1910l). Mr F.G.A. Stucky observed it from Island Bay (Wellington) with a 4.5-inch (11.4-cm) refractor. He suggested that the head appeared as two entities at times. Joseph Ward in Wanganui wrote,

> The nucleus appears to be already under great strain as it appears of an irregular shape an oblong across the major axis & of very irregular outline. (Mackrell, 1985: 114–115).

R.T.A. Innes (Kronk, 2007: 149) also noted this double head, on 20 April (UT) he said that there was a "... secondary nucleus … situated 5″ towards PA 154°." This would suggest that the nucleus was fragmenting, however, other reports in Kronk (2007: 149–150) suggested that these were "... jets emanating from the nucleus, which might have been responsible for the reports Innes made of a "secondary nucleus"."

Another observation on 21 April by Mr Joseph Taylor from Nelson stated that both Comet 1P/Halley and Venus were "... brilliant …" He further provided a detailed description:

> The comet showed a definite nucleus nearly as large and bright as Venus herself; while her brilliant tail streamed upwards in a north-easterly direction about three degrees long and one degree wide at the termination (when it was bifurcated [divided into two branches or forks]). I secured two photographs, one at 5.30 and another at 5.40; but the comet was then rapidly fading in the light of the dawn, and by 6 o'clock it was lost even to an opera glass. (Halley's Comet, 1910m).

That same morning, a reporter from the '*Star*' newspaper was shown Halley's Comet (and Venus, which reached its greatest western elongation on Sunday 24 April) by Mr D.B. McLeod from the Canterbury University College Observatory. The tail was stated to be about 1.5° in length and extending away from the Sun. The comet was described as approaching the Sun at 3 million miles (5 million km) a day and growing in elongation until 6 May, after which the elongation would diminish until the transit on ~19 May 1910 (Halley's Comet: A Telescopic Observation, 1910).

On 23 April 1910 the *Evening Post* published the diagram reproduced here as Figure 31, which shows the orbits of Comet 1P/Halley and the Earth from January to May 1910 (Where the Earth and the Comet Collide, 1910). Note how the tail of 1P/Halley seems to brush the Earth on 18 May.

On 23 April 1910, Ward (Local and General, 1910a) wrote that the comet continued to "... increase rapidly in size and brilliancy." He described the nucleus as being star-like, and that the coma extended towards the Sun "... in greater volume than seen previously during this apparition." The tail was 3.5° long, and "... begins to show signs of its perihelion passage, in a distinct twist about half-way down its length, the tail from this to the end being directed at a different angle." Was Ward observing a solar magnetic effect on the tail? He noted that the Moon would start interfering with views of the comet as it was getting brighter for the next two weeks. According to GUIDE, the Full Moon would occur on 25 April, so would move into the morning sky and affect views of the tail. Reverend Kennedy also saw the comet from Wellington on 23 April and estimated the magnitude to be 3 (News Note, 1910).

An apparently skilled observer with the initials H.M.L, who was friends with Joseph Ward, made a detailed observation from Okato on 24 April. He said:

> The comet has increased during the last four days and has a short broad tail with a nucleus equal to the size of Jupiter (that is forty [arc] seconds*).* The breadth of tail shows the comet is approaching us. (Halley's Comet, 1910n).

He noted that both 1P/Halley (36° altitude) and α Centauri (magnitude –0.27) faded simultaneously as the dawn sky brightened. H.M.L also gave astrometric positions for the comet, which was 33° west of the Sun, and determined that on 24 April "The comet rises at 5hr. 23min. and passes the meridian at 9hr. 47min. and sets at 3hr. 23min. p.m."

The comet was seen until 5:55 am on Monday 25 April, this being well into morning twilight (Local and General, 1910c), sunrise being at ~6:25am (Halley's Comet, 1910o). On this date Reverend Kennedy published a comet/Sun rise table and stated that 1P/Halley would transit the Sun on 19 May 1910, between 1:50 and 2:50 pm (Halley's Comet: A Timetable, 1910), and that the Earth would pass through the comet's tail. The author pointed out that three types of comet tails were known, stating "... it is surmised that they are formed of different substances …" (*Oamaru Mail*, 1910b). Mr Chambers, at a meeting of the British Astronomical Association (England), highlighted that the last time Earth went through the tail of a comet (in 1861), a vicar reported seeing strange atmospheric effects such as the day being "gloomy" and "foggy" (News Notes, 1910). J.T. Stevenson suggested that due to the comet's tail being tens of millions of miles long, it would reach the Earth when 1P/Halley transited the Sun (Halley's Comet, 1910p). At this time, a somewhat scathing article was published in the *New Zealand Times* where a comparison between Canterbury University College opening up the Observatory to the public was compared to the lack of facilities in Wellington (News Notes, 1910). Orchiston (2016: 258), a former Director of Carter Observatory, states that Carter Observatory (in Wellington) was not opened until 1941, 31-years after the 1910 appearance of Halley's Comet. Thus, Wellington was somewhat slow in addressing this perceived deficit.

On 27 April, the *Otago Witness* published an article by Mr J.F. Morris (Senior) (Morris, 1910: 65), that provided an excellent summary of and prediction for 1P/Halley:

> When Halley's Comet was picked up by the camera in 1909 it was more than 300 millions of miles away from the sun. On December 1,1909, it was in opposition but was too faint to be seen with small telescopes, the wanderer being only equal to a star of about the twelfth magnitude. It continued to approach the earth until December 18, 1909, when its distance from us would be about 125 millions of miles. During January and February, 1910, it receded from us (its motion is retrograde), and increased its distance from the earth. In March it crossed the earth's orbit at a point which our earth will pass about the time of our Southern Vernal Equinox (September 23). Its distance then would be (March 24) about two radii of the orbit—that is, 180 million of miles, the earth being then at one end of the orbital diameter and the comet at the other. About the middle of April it will be in our eastern sky.
>
> On April 20 Halley's Comet will be in perihelion, according to Mr John Grigg, F.R.A.S., of the Thames, and, to quote his words, "will probably be an object grand, impressive, and brilliant." Then our visitor rushes along its parabolic course, still coming towards the earth. It passes between us and the sun about May 17, and a few days later comes very near us from a sidereal point of view. Although it is said to come very near us, it will be still 12 millions of miles distant. At this time the comet should be easily observable in the early evening, and, in all probability, still be an impressive and thought-promoting object. In a few days more our celestial visitor will double its distance from us; later still it will gradually diminish in apparent size and brightness till it fades from view altogether, not to appear again until 1987."

Professor H.W. Segar observed the comet from Auckland on Thursday 28 April and described the tail as being 5° long, as seen without optical aid. The waning Moon helped make clearer observations of the comet than previous mornings (Halley's Comet: Early Morning Views, 1910). Of note is that the general public were describing the tail as being "... eight or nine feet long …" with the tail pointing upwards in a NE direction (Local and General, 1910c). Others stated that it was "... four yards long …" (*Manawatu Standard*, 1910), and one newspaper reported it to be "... two and a'half inches [long] ..."! (Oamaru Mail, 1910c). Obviously non-astronomers did not know about distances being measured in degrees when subtended upon a sphere, although describing the tail in this way would have also been useful for non-astronomers because they did not know about using degrees to describe subtended angles. Others forewent description of length and opted for resemblance to animals, namely, that 1P/Halley's tail resembled a fox's tail (The Chronicle, 1910b). Mr A. Hansen, of Cape Palliser lighthouse, observed 1P/Halley with a refractor at 100× magnification and described the head as being yellowish and showing considerable condensation. The tail was described as being straight with a slight hint of a curve downwards (*Evening Post*, 1910b). Coupled with descripttions of 1P/Halley from many observers around New Zealand were astrophysical statements that comets were as plentiful as the fish in the sea and that the "... nucleus of a comet is a loosely associated mass of solid matter." (Evening Post, 1910c). The loosely associated mass could almost be likened to Fred Whipple's 'dirty snowball' hypothesis in the 1950s (see Whipple, 1950: 83).

By the end of April, the number of people seeing and reporting observations of 1P/Halley were so numerous that one article titled "They Say" suggested "That, for a change, the dailies ought to give us the names of people who have not seen the comet. The stories of those that have are becoming monotonous." (They Say, 1910). The *Manawatu Herald* (1910) confirmed this on 26 April, stating that "Halley's comet is the early morning attraction to hundreds of people throughout the Dominion." Indeed, J.A. Osbourne and his unnamed wife climbed Mount Eden (Auckland) for six mornings in a row in early May to view the comet, the newspaper concluding that: "J.A. Osborne's experiences are neither interesting nor entertaining. The "Star's" alleged news rarely is." (Pars About People, 1910).

In early May 1910, newspapers forewarned that 1P/Halley was slowly slipping towards the Sun (as seen from Earth) and was rising slightly later each morning until it would vacate the morning sky. It would reappear in the evening sky after transiting the Sun on 19 May 1910. During the transit, the comet's tail would "... brush across the face of Earth …", but there was no need to be concerned about impending disaster (*North Otago Times*, 1910a). The *Bush Advocate* (Halley's Comet, 1910q) published a table (Figure 32) of comet and Sun rise times, and the Earth-comet distance. The gradual reduction in rise times between the two illustrated the reduced elongation as May progressed.

On 4 May 1910, Australia's leading amateur astronomer of the time, John Tebbutt, also wrote of the impending swipe of Comet 1P/Halley's tail across Earth on 19 May. He asked people to look out for rare atmospheric, meteoric, and electrical effects as a result of the encounter between the two (Halley's Comet, 1910r). Writing about the Earth's encounter with the tail of the Great Comet of 1861, the *Marlborough Express* (The Comet of 1861, 1910) quoted an eye-witness account from England:

> In connection with this subject I may add that on Sunday evening [June 1861], while the comet was so conspicuous in the northern heavens, there was a peculiar phosphorescence or illumination of the sky, which I attributed at the time to an auroral glare; it was remarked by other persons as something unusual, and considering how near we must have been on that evening to the tail of the comet, it may perhaps be a point worthy of investigation whether such effect can be attributed to our proximity thereto.

Mr E.J. Lowe, of Highfield House Observatory, also wrote to *The Times* on 4 July 1861 (*ibid.*), with an interesting account of his observations:

> Sir, —In answer to Mr J. R. Hind's inquiry regarding the evening of Sunday, the 30th of June, I beg to say that the same peculiar appearance of the heavens was noticed here in a marked manner. The sky had a yellow, auroral glare-like look, and the sun, though shining, gave but feeble light. The comet was plainly visible at a quarter to 8 o'clock (during sunshine), while on subsequent evenings it was not seen till an hour later. In confirmation of this I may add that in our parish church the vicar had the pulpit candles lighted at 7 o'clock [June being summer for the Northern Hemisphere], a proof that a sensation of darkness was

> felt even with the sun shining. Without being aware that the comet's tail was surrounding us, yet being struck by the singularity of the appearance, I recorded in the daybook the following:—"A singular yellow phosphorescent glare, very like diffused Aurora Borealis, yet being daylight such Aurora would scarcely be noticeable." The comet itself had a much more hazy appearance than it has had since that evening.

Would people in New Zealand see a similar effect? Tebbutt of Australia certainly hoped so. Some even advocated capturing air in test tubes in the hope of capturing comet gas for analysis (*Southland Times*, 1910). Ward thought the tail would not reach Earth (Halley's Comet, 1910s), but Clement Wragge did, writing that there was nothing to fear and that the tail would have as much effect on Earth as an elephant (Earth) running through a spider's web (comet's tail) (*Woodville Examiner*, 1910). In two weeks they would find out.

As 1P/Halley raced towards Earth, the New Zealand newspaper hype was staggering, as illustrated in Figure 32. In the week prior to perigee (on 19 May 1910 NZST), there were 629 newspaper articles mentioning 1P/Halley. On the day of perigee (and transit across the Sun) and for the following two days there were 421 articles. During May 1910, the number of lectures on 1P/Halley peaked. Papers Past revealed that there were about 25 talks given across New Zealand in that month alone (see Table 1). Lantern slides showing the comet were often used for illustrative purposes, these frequently being taken from Meeanee Observatory near Napier (e.g. The Comet: Lecture by Mr G.V. Hudson, 1910).

During May 1910, the majority of newspapers were publishing reports not only about the brightness of 1P/Halley's head, but also about the extent of the comet's tail. On 7 May 1910, 1P/Halley was stated to be "... a splendid object, having vastly increased in magnitude and brightness …" The tail was described as curved, and with the conjunction of the Moon, Venus, and comet forming a right-angle triangle in the sky must have indeed been a splendid sight (Local and General, 1910d). Residents in Greymouth described the tail as narrow but that it was enormously extended (Halley's Comet: A Brilliant Spectacle, 1910). Even though the tail at that time was described as an impressive sight, J. Steele stated that the tail of the Great Comet of 1882 was better, writing that that comet was

> ... plainly visible in broad daylight. Its tail at one time exceeded 100,000,000 miles in length, while the tail of Halley's comet appears to be about 14,000,000 or 15,000,000 miles long. (Answers to Correspondents, 1910).

For a description of New Zealand observations of the Great Comet of 1881, C/1881 K1 (Tebbutt) see Drummond, 2023.

On 9 May the *Northern Advocate* wrote that astronomers (location unstated) noticed a twist in the tail (Here and There, 1910). This twist is not mentioned by observers in Kronk (2007: 151–152), but a dark rift running down the central length of the tail was noted. The photograph in Figure 33 taken from Lowell Observatory (USA) with a portrait lens on 13 May 1910 (UT)

**HALLEY'S COMET.**

| Date | | | Comet rises. | Sun rises | Distance of Comet from the earth. Millions of miles. |
|---|---|---|---|---|---|
| May | 3 | ... | 3.38 | 6.38 | 66 |
| May | 4 | ... | 3.37 | 6.39 | 62 |
| May | 5 | ... | 3.36 | 6.40 | 59 |
| May | 6 | ... | 3.34 | 6.41 | 56 |
| May | 7 | ... | 3.35 | 6.42 | 52 |
| May | 8 | ... | 3.36 | 6.43 | 48 |
| May | 9 | ... | 3.37 | 6.44 | 45 |
| May | 10 | ... | 3.40 | 6.45 | 41 |
| May | 11 | ... | 3.44 | 6.46 | 37 |
| May | 12 | ... | 3.48 | 6.47 | 34 |
| May | 13 | ... | 3.57 | 6.48 | 30 |
| May | 14 | ... | 4.9 | 6.49 | 27 |
| May | 15 | ... | 4.28 | 6.50 | 24 |
| May | 16 | ... | 4.48 | 6.51 | 21 |
| May | 17 | ... | 5.17 | 6.52 | 18 |
| May | 18 | ... | 5.58 | 6.53 | 16 |
| May | 19 | ... | 6.49 | 6.54 | 15 |
| | | | Sun sets. p.m. | Comet sets. p.m. | |
| May | 20 | ... | 4.39 | 5.48 | 14 |
| May | 21 | ... | 4.38 | 6.32 | 15 |
| May | 22 | ... | 4.37 | 7.33 | 16 |
| May | 23 | ... | 4.36 | 7.55 | 18 |
| May | 24 | ... | 4.36 | 9.27 | 21 |
| May | 25 | ... | 4.35 | 10.0 | 24 |
| May | 26 | ... | 4.35 | 10.19 | 27 |
| May | 27 | ... | 4.33 | 10.22 | 30 |
| May | 28 | ... | 4.33 | 10.33 | 34 |
| May | 29 | ... | 4.32 | 10.38 | 37 |
| May | 30 | ... | 4.32 | 10.44 | 41 |

Figure 32: The *Bush Advocate* (Halley's Comet, 1910q) published this table of Comet 1P/Halley rise and set times, as well as the Sun rise and set times, and the distance (in million miles) between the comet and Earth. The yellow highlighting is a result of the word-search in Papers Past (source: Papers Past).

shows the ion tail being altered by a solar magnetic disturbance (see Ip and Axford, 1982). Was this related to the 10 May (UT) observation? International observers in Kronk were reporting tail lengths of 20–25° in early May. The *Greymouth Evening Star* (1910a) reported a tail of 60° as seen from Perth (Australia). As an aside, the same paper summarised the total solar eclipse views from Australia, stating that one of the Tanzania expeditions had overcast skies and drizzle, whereas Parkes (New South Wales, Australia) saw the whole event. A poem by Chaz Johnson on 10 May in the *Hawera & Normanby Star* (1910) also mentioned the tail:

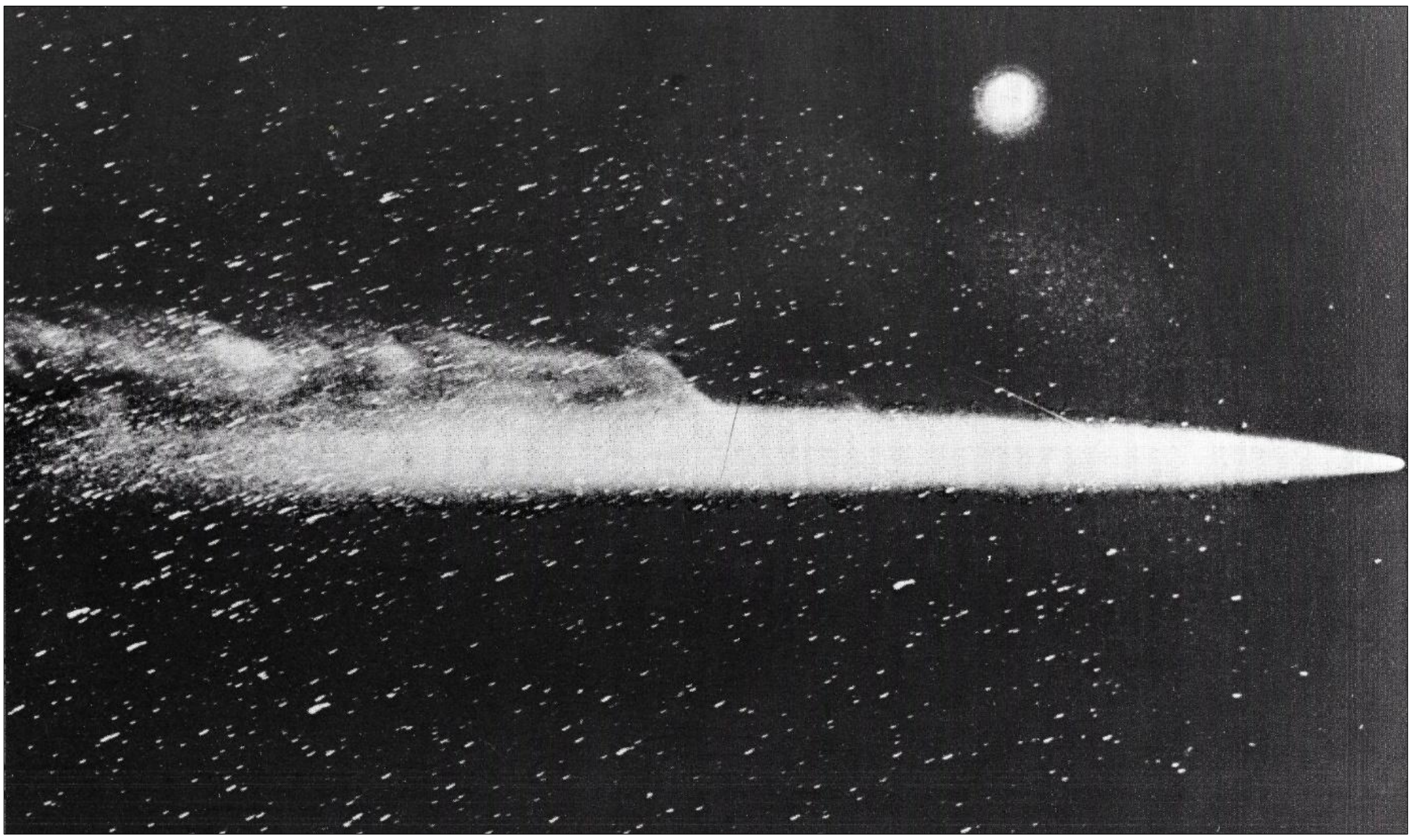

Figure 33: A truly remarkable (cropped) photograph of 1P/Halley as the head lay about 12° NNE of Venus on the local morning of 13 May 1910 UT. The thinly lined streak coming off the tail at 10 o'clock is a meteor. The uplifted section in the latter half of the tail is the ion tail being disturbed by the solar magnetic field. The lower main tail is the dust tail which was 35° long. Taken from Lowell Observatory, Flagstaff, Arizona (source: Lowell Observatory (Giclas/NASA).

> Now here's this comet up on high,
> Which travels fast and travels far;
> And just now seems at rest to lie,
> Tail up below the morning star.

The following provides a detailed description of the tail as seen from Auckland on 13 May 1910 (*Auckland Star*, 1910b):

> The past five or six mornings the weather has been so bad that the sky has been too overcast to permit of a view of Halley's comet, but this morning those who were fortunate to see it were rewarded with a truly magnificent sight of this restless traveller, which has been a source of recurring wonder to the world through the ages. Between two and three o'clock a shaft of greenish, silver light stole up from the eastern horizon, gradually lengthening with a northerly sweep, till the luminous ball which forms the head was clear of the earth's rim, and the phenomenon seemed to fill the whole of the eastern sky. It has grown remarkably in brightness since it was last visible, and the rapid change gives a faint idea of the terrific pace at which it is tearing through space in its orbit round the sun … The comet will be visible in the morning sky till Monday next, when its nearness to the sun will prevent it being seen, but towards the end of the week it will reappear in the evening as a glorious spectacle in the western sky.

On the same morning (13 May 1910), Captain Hewitt, "... the honourable astronomer …", stated that the tail was 20° long and that the comet was 60 million miles (97 million kilometres) from Earth and approaching at 3 million miles (4.8 million kilometres) a day (Halley's Comet, 1910t). He wondered if it would be viewable in the daytime sky on 18 May (Halley's Comet, 1910u). Adjectives from other NZ centres in relation to the tail of 1P/Halley at this stage were "enormous", "magnificent", "exceptionally luminous", "gorgeous", that the tail went "from here to yonder", "indescribably beautiful" and that it resembled the searchlight of a war vessel.

Around this period, 21 year old Charles Michie saw the comet from the dark skies of Kaitaia. Michie was fortunate to have six clear mornings in a row up until 19 May 1910 (NZST) in which to observe 1P/Halley. Being a keen amateur astronomer, Michie made comparative sketches of the evolving tail which can be seen in Figure 34. At its greatest length on 19 May (NZST) he described the tail as being a dull, orange colour and 150° in length. He said that it resembled a "... long search-light beam." (Walker and Blow, 1985: 24). Michie concluded

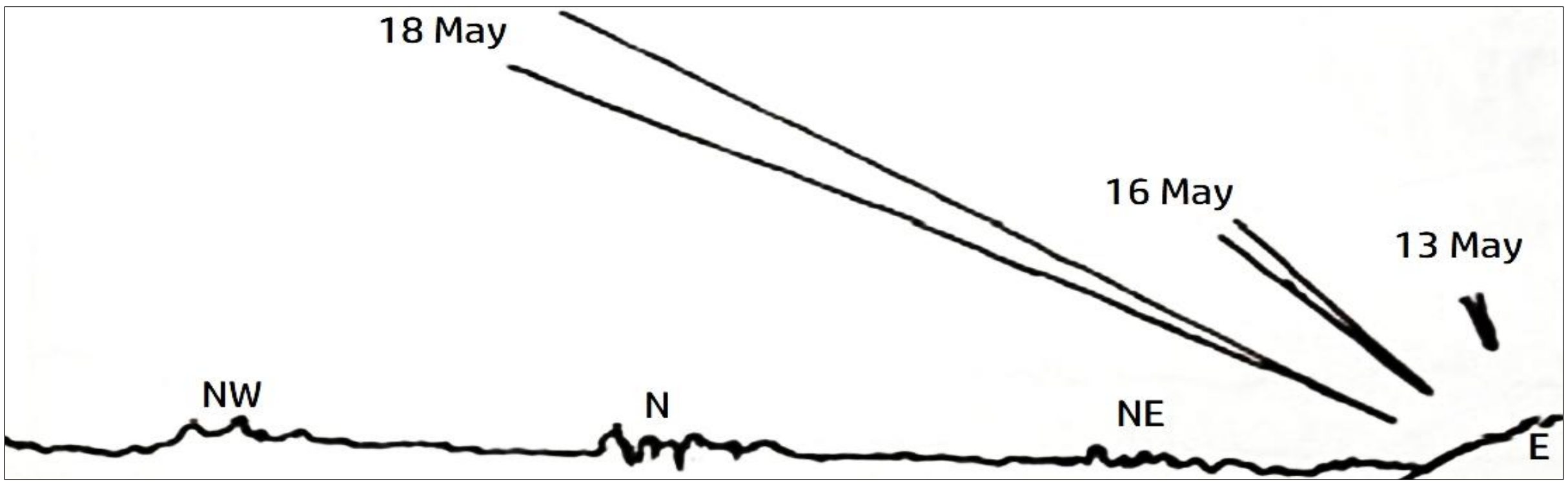

Figure 34: Sketch of the evolving tail of 1P/Halley made by Charles Michie of Kaitaia. Note how dramatically the tail increased in length from 13 May (short tail) to 18 May (long tail). The sketch (text) has been modified by the authors for clarity purposes (source: Canterbury Astronomical Society, *Halley's Comet 1986*).

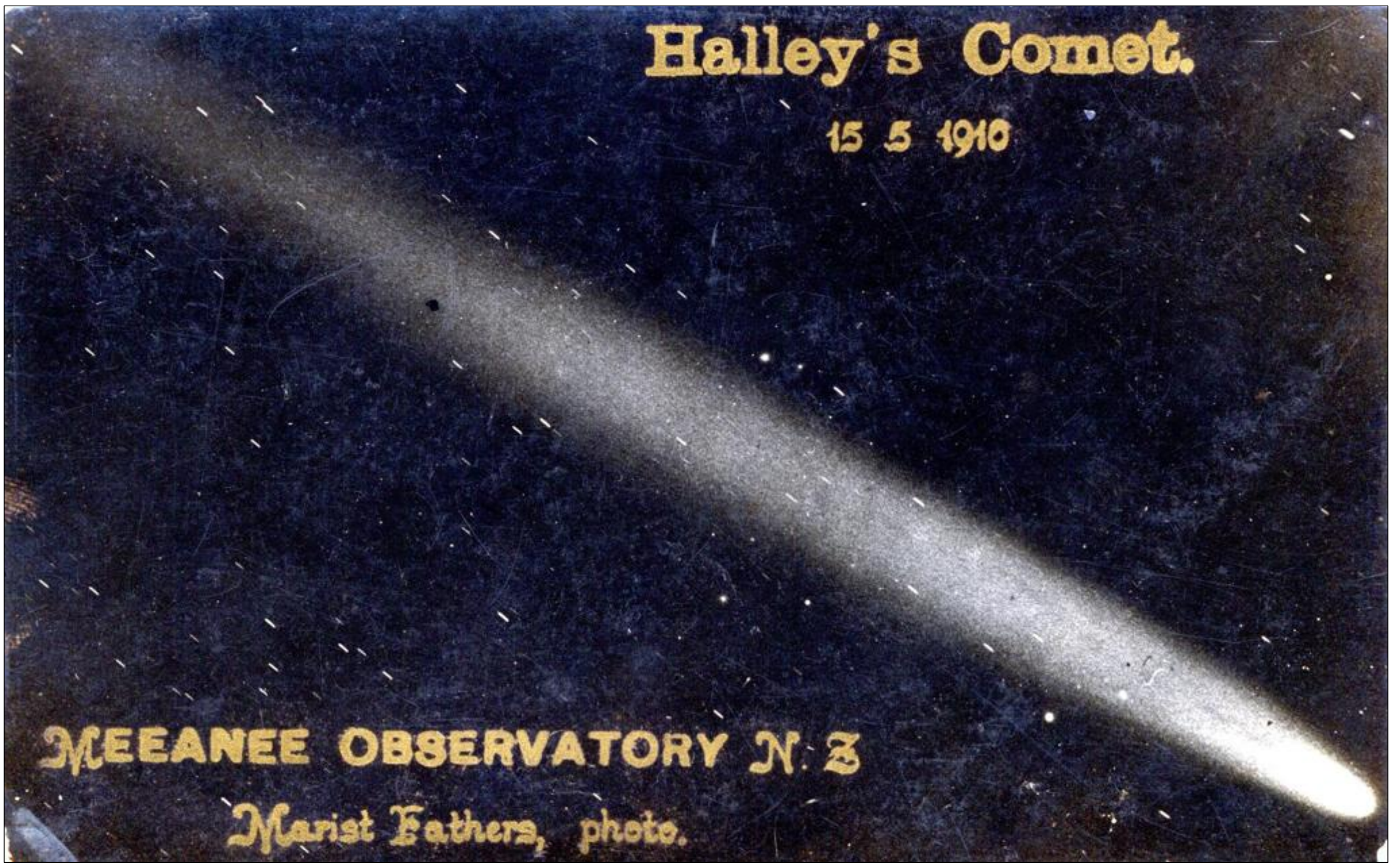

Figure 35: An example of a Meeanee Observatory photograph, taken on 15 May 1910 NZST. The tail must have been a real spectacle (courtesy: Marist Archives, Wellington).

that the sight of 1P/Halley on 18/19 May 1910 was "... the greatest spectacle of my life."

Measuring photographs of the comet taken from Meanee Observatory on 14 May 1910 (as an example see Figure 35), Father Allen estimated the 35° tail to be 18 million miles (~29 million kilometres) long and that it would reach Earth with a three million mile (approximately five million kilometre) overlap. Earth would take fourteen hours to transverse the tail on Thursday 19 May 1910, with the tail brushing Earth at 125,000 miles (201,000 kilometres) an hour. The coma was estimated to be 79 thousand miles (~127 thousand kilometres) in diameter (The Comet, 1910a). Four mornings later, the *Bush Advocate* published a report that the tail had vastly increased in length in the last few days and was ~80° long on 16 May (Local Observations, 1910). J.T. Stevenson stated that the 1910 tail was twice the length of the 1835 apparition (Halley's Comet, 1910w). On 16 May, Mr Marchant, the New Zealand surveyor, wrote that the coma had increased in size and that it was less sharply defined (Halley's Comet, 1910v). Kronk (2007: 153, 154) states international observers were reporting tail lengths of up to 140° on 19 May (UT). Using the Degree of Condensation chart (Figure 36), which visual comet observers use to determine the appearance of a comet's coma, it is possible, based on Marchant's observation, that the head changed approximately from a Degree of Condensation

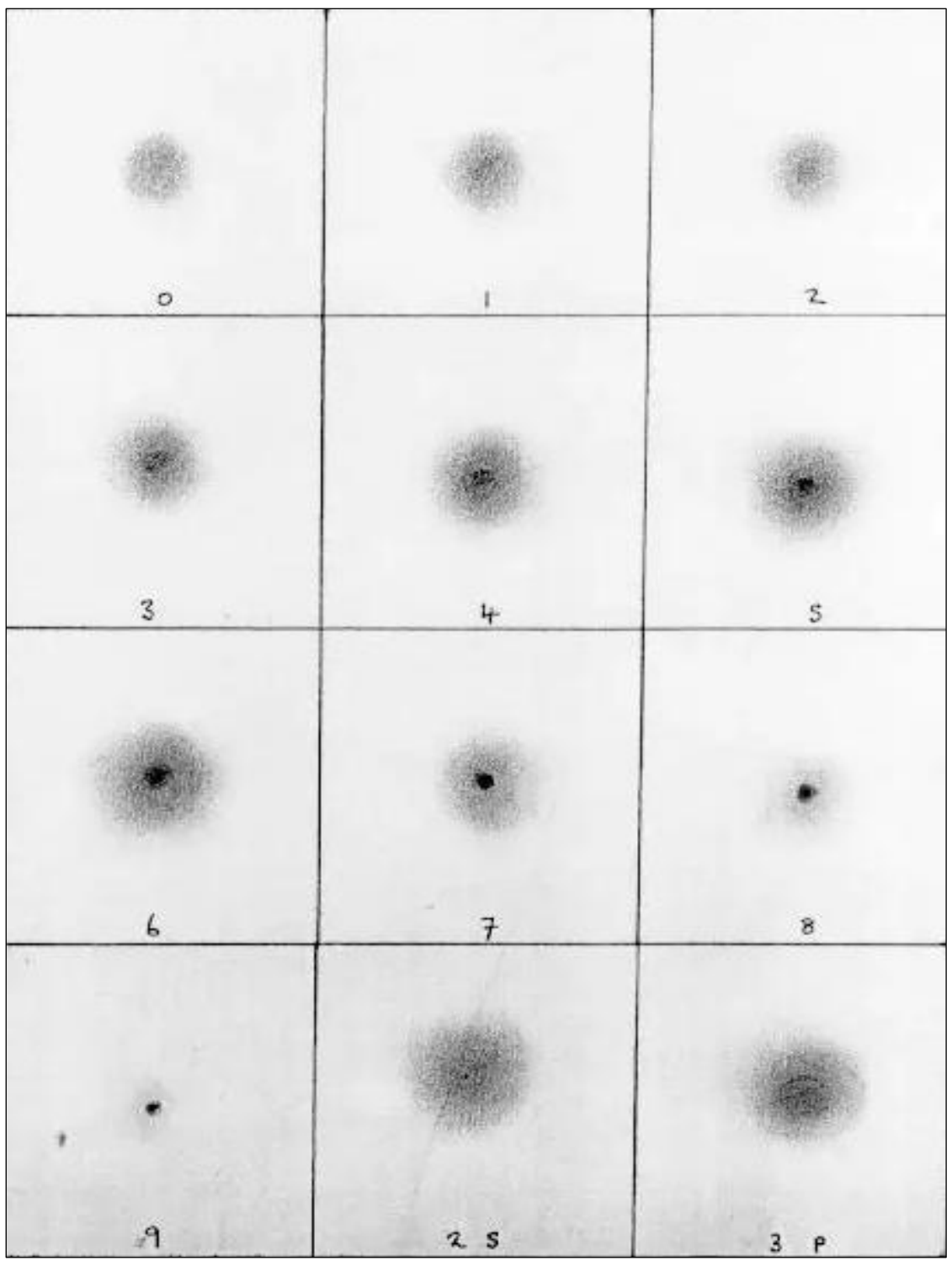

Figure 36: A degree of comet condensation chart. As the number increases the coma looks more star-like, those of 0–1–2 show hardly any central condensation while those of 8–9 display an almost star-like appearance (source: Wikicommons, 2020).

(DC) of 6–7 to a DC of 3–4. Of interest is a statement that

> Mr Marchant is convinced that what we have read of the waning of Halley's comet is true, and that what we see is the mere ghost of the splendid spectacle it once presented. (*ibid.*).

Basically, it was assumed that as the number of perihelion passages increased, the comet's potential for outgassing was diminishing. Of course, the position of Earth in relation to the perihelion passage of 1P/Halley played a big role in apparition appearances. Conversely, in his talk on Halley's Comet, Mr. G.V. Hudson, "... the well-known local observer …" stated that

> Astronomers at Home (*sic*) and in America classed Halley's as a comet of the second class, but this is beyond all shadow of doubt a first-class comet … (What to Expect on Thursday, 1910).

On 17 May 1910 one correspondent stated that Comet 1P/Halley:

> ... appeared to extend from the eastern horizon right across the zenith, and gave forth, in its wake, a brilliantly bright nebular streamer of great length and width, that was indescribably beautiful. It looked exactly like a huge and powerful searchlight playing on the heavens from the horizon. (Halley's Comet, 1910v: 2).

At this stage (17 May) 1P/Halley was rising at 5:17 am and the Sun (for Wellington) was rising at 6:52 am—approximately 1.5 hours before the Sun. On that date it was 18 million miles (~29 million kilometres) from Earth (Halley's Comet from Balloon, 1910). Refer to Table 4 for additional information.

On 18 May 1910, one day before 1P/Halley was closest to Earth, would transit the Sun, and the tail would brush Earth, Professor Segar stated that the head of the comet would rise at the same time as the Sun and probably not be visible due to the Sun's glare. He explained that the two million mile (~3.2 million kilometre) wide tail would appear broad since we would be looking along the tail towards the comet's head from within the tail (akin to looking up the skirt of a shuttlecock). He suggested that the tail could possibly be visible in both the morning and evening sky. Segar predicted that the comet would be a naked eye object for the next two to three weeks and then slowly fade (The Comet: Professor Segar Interviewed, 1910). On the same day (18 May) an unnamed observer reported seeing the tail rise at 1:15am, many hours before the Sun. They also reported seeing dozens of meteors from the vicinity of the tail (Bush Advocate, 1910). If this was related to 1P/Halley or circumstantial, is unknown. The number of people in towns and cities in New Zealand looking at the comet at this stage was substantial. The following quote (Otorohanga, 1910) could apply anywhere across the country, "... many of the residents—male and female—may be seen in their "fire escapes" at 5.30 a.m. admiring the long-tailed mystery in the heavens …" Artist's impressions of 1P/Halley around this time are seen in Figures 37 and 38.

## 5.3 1P/Halley Transits the Sun

Comet 1P/Halley was predicted to transit the Sun on 20 May 1910 UT (21 May 1910 NZST). However, predictions for this event varied. Even

Table 4: 1P/Halley and sunrise times, set for Wellington. Based on Halley's Comet from Balloon, 1910.

| Date | Comet Rise | Sunrise | Distance from Earth |
|---|---|---|---|
| 16 May | 4:48 am | 6:51 am | 21 million miles (34 million km) |
| 17 May | 5:17 am | 6:52 am | 18 million miles (29 million km) |
| 18 May | 5:58 am | 6:53 am | 16 million miles (26 million km) |
| 19 May | 6:49 am | 6:54 am | 15 million miles (24 million km) |

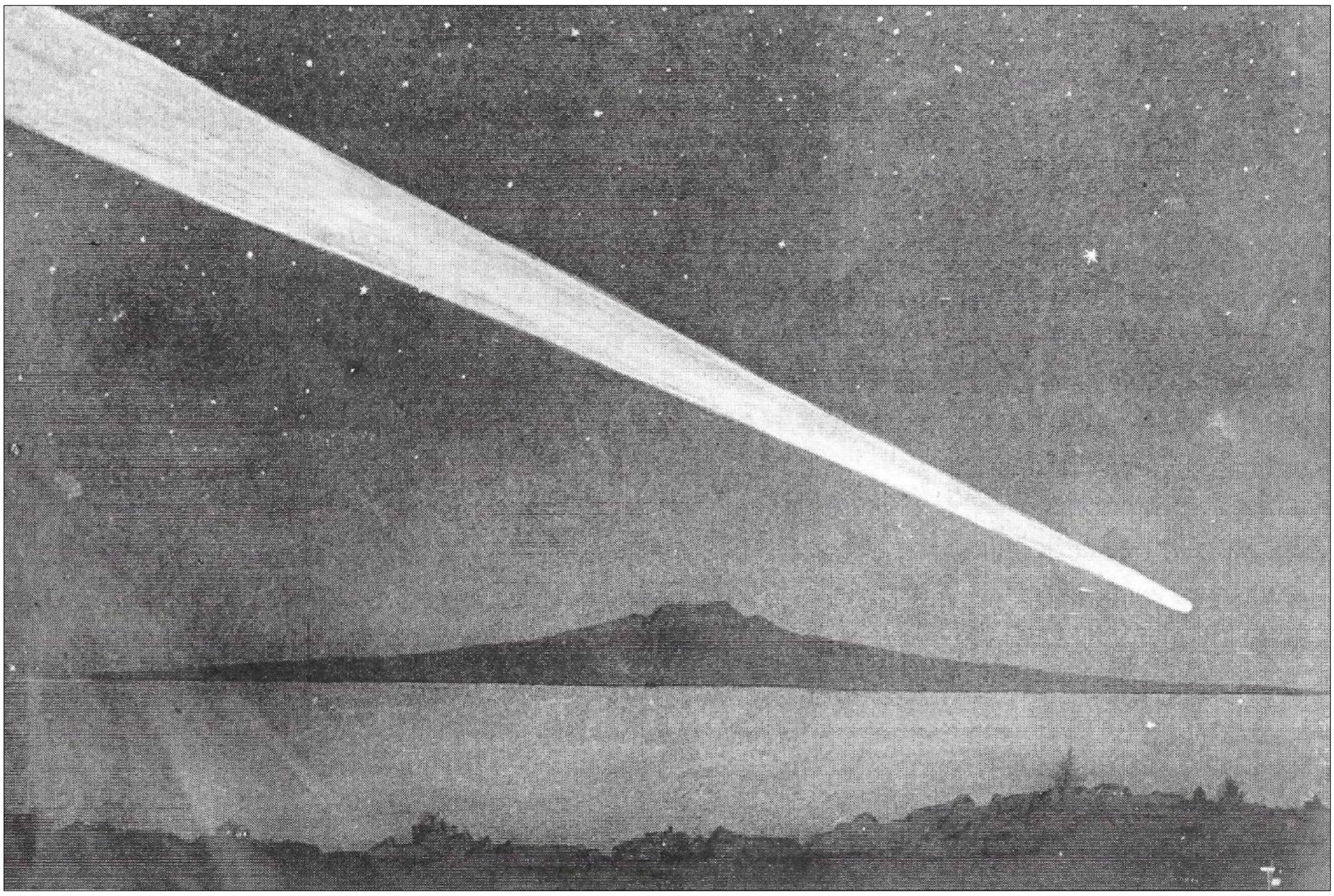

Figure 37: From Walker and Blow (1985: 1). The inscription reads: "A magnificent spectacle in the eastern sky: Halley's Comet as viewed from Auckland on the morning of May 17, 1910." At this stage Comet 1P/Halley was rising 1 hour and forty minutes before the Sun in the ENE morning sky. On 17 May, magnitude 0.5 Saturn was above the head of the comet and could perhaps be the 'bright' star above and to the left of 1P/Halley's head in the drawing. Observers were reporting a tail length of approximately 70° on this date (Kronk, 2007: 153).

after the transit, astronomers were issuing varying transit dates and times. An example is in August 1910, A.O. Leuschner, based on updated astrometric positions and refined calculations suggested that it had occurred on 18 May between 15h 38m and 16h 37m UT—that is 19 May between 3:38 am and 4:37 am NZST (Leuschner, 1910: 135–136). Based on modern orbital calculations using the global observations of 1P/Halley during the 1910 apparition, the comet actually passed directly between the Earth and Sun on 19.15 May UT (Kronk, 2007: 153)—that is 19 May at 3:36 pm NZST, with the alignment so precise that the nucleus of the comet would have transited the disk of the Sun. As stated, a variety of predictions were made that erroneously calculated the timing of the

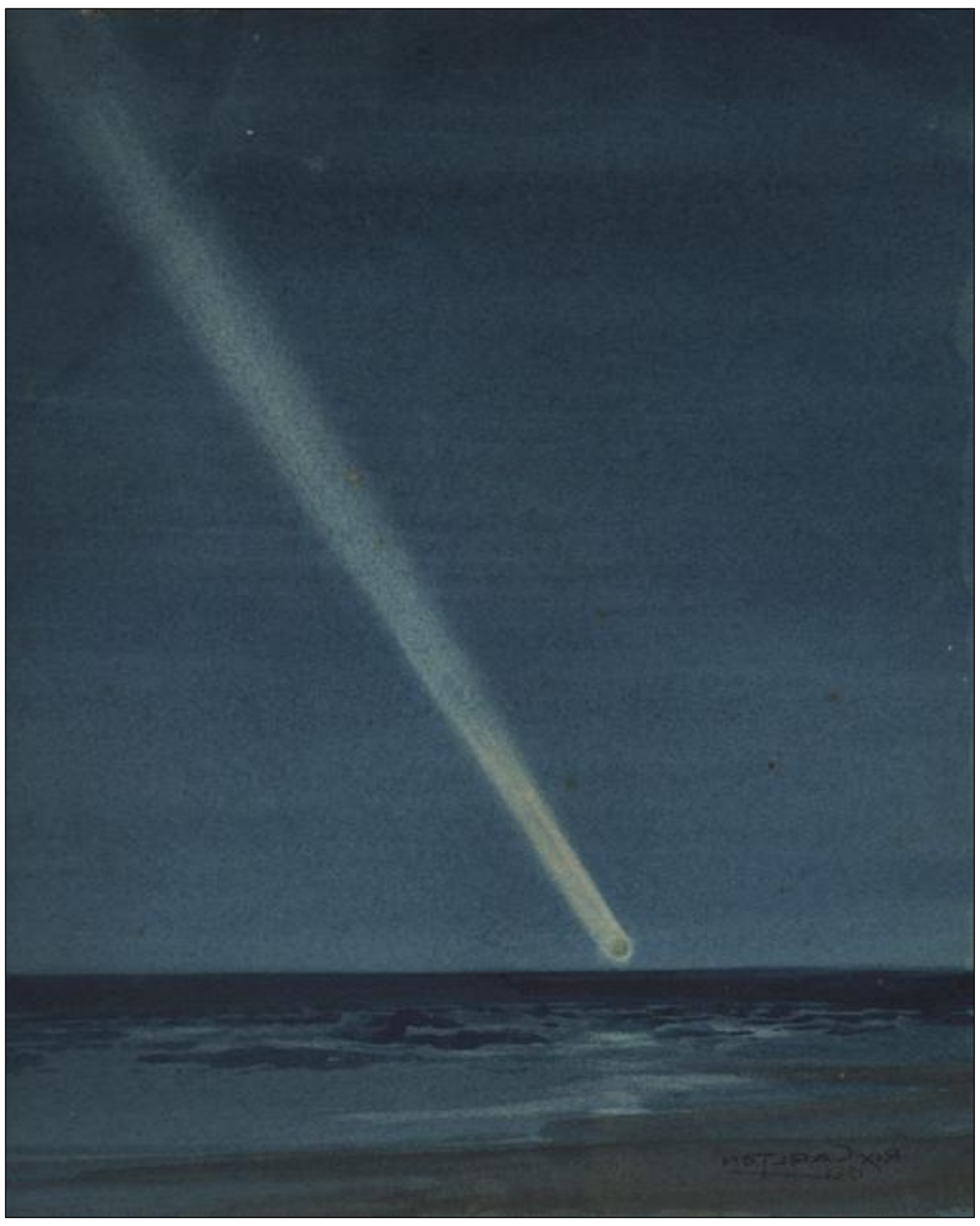

Figure 38 (right): Comet 1P/Halley over Saint Kilda's Beach, Dunedin, on 17 May 1910 at approximately 5:15am. Painted by Herbert Cole (Aka Rix Carlton) (1890–1962) in 1954, possibly based on a photograph of the comet or from memory. The painting is reversed (right=left, left=right) to the actual appearance that 1P/Halley would have had in the sky. Auckland astronomer Jennie McCormick, who is the guardian of the painting, is unsure why, but speculates that perhaps Cole painted it from a lantern slide that was reversed in the projector. Cole/Carlton was a surgical procedure artist and was appointed the Honorary Artist of the Auckland Astronomical Society in 1954. The image has been reversed by the authors to more correctly represent how the comet would have appeared in the sky (source: Jennie McCormick/Auckland Astronomical Society).

transit—such as that made by Leuschner.

Several attempts were made to observe the comet's passage across the bright luminary. The Marist Brothers at Meeanee Observatory photographed the Sun at five minute intervals during the transit but, not surprisingly, nothing was conclusively recorded (Halley's Comet, 1910ah). The image of the Sun on the glass plates was 5.5-inches in diameter. John Grigg of Thames Observatory (the discoverer of three comets—see Drummond et al, 2026; Orchiston, 2016) visually and photographically examined the Sun from 16 to 21 May, he only observed the usual movement of sunspots revealing the rotation of the Sun (Astronomical Notes for June, 1910: 41). Observers at the Canterbury University College Observatory also reported a negative visual observation (Halley's Comet, 1910x); Mr H.F. Skey at the same Observatory did record a small magnetic anomaly but concluded that it was not as a result of the Earth's passage through 1P/Halley's tail (Through the Comet's Tail, 1910). Astronomers, one of which was the New Zealand Surveyor-General, at the Observatory in the Botanical Gardens in Wellington also had negative visual and photographic transit observations (The Comet, 1910b), as did Clement Wragge who travelled to Tasmania to observe it (Through the Tail of the Comet, 1910), and Joseph Ward using the 9.5-inch telescope at Wanganui Observatory (Transit of the Comet, 1910). Surprisingly, an article in the *Evening Post* stated that astronomers in Munich caught a glimpse of the comet as it transited the Sun (The Comet, 1910b), however, this seems improbable.

Indeed, an ADS journal search of approximately a dozen articles relating to the transit of 1P/Halley in 1910 all reported a negative result. Examples are A.B. Chatwood (1910: 614) who used an 8-inch (20-cm) Watson refractor and W.E. Cooke at Perth Observatory who used a 10-inch (25-cm) telescope. Cooke's resolution was clear enough to see granules on the Sun. Also in Australia, astronomers at Sydney Observatory used an 11.5-inch (29-cm) Schroeder refractor without success. Walter Gale, R.P. Sellors, and W.P. Raymond also observed the Sun for an hour from Newcastle but also failed to see the comet (Raymond, 1910: 653). The same results were repeated in all articles.

It is no surprise that attempts to see Halley's nucleus transit the Sun were unsuccessful. If we assume that the smallest feature that telescopes would be able to resolve on the Sun's disk in 1910 would have a diameter of 10″, then it is simple to calculate the minimum size of a comet's nucleus that would be required for such a detection to be possible. At its closest approach to Earth 1P/Halley was approximately 23 million kilometers from our planet. At that distance, to subtend an angle of 10″, an object would have to have a diameter of approximately 1100 km! By contrast, observations made by the Giotto and Vega flybys of Halley's nucleus in 1986 revealed a length, on the long axis, of approximately 15 km. In other words, 1P/Halley's nucleus is approximately two orders of magnitude too small to have possibly been detected during the 1910 transit across the Sun. Or, to put it another way, the comet would have had to approach within two hundred thousand kilometers in order to render the nucleus detectable during the transit.

Similar to the observer mentioned above, Auckland's J.T. Stephenson reported numerous meteors from the realm of the tail on Friday 20 May. He noted that the tail rose from the eastern horizon to the zenith and was fainter than the Milky Way (Halley's Comet, 1910y), although Yerkes Observatory astronomer E.E. Barnard reported that it was brighter than the Milky Way and 107° in length (Halley's Comet, 1910x). Stevenson wondered if, due to the curve of the tail, the tail passed over Earth at a different time than the head's transit times, possibly in the morning hours of Saturday 21 May 1910 NZST (Halley in the West, 1910). Some suggested this tail swipe occurred in the evening hours of Thursday 19th and early morning hours of Friday 20th May 1910 NZST (Through the Comet's Tail, 1910).

## 5.4 Did Earth Encounter 1P/Halley's Tail?

Some people, including Stevenson, reported seeing the head of the comet in the evening sky on 22 May NZST (one day after the transit) and a 75° tail in the morning sky on 21 May NZST (The Comet Scare, 1910). Stevenson explained that, after the transit, the head had moved into an eastern elongation (evening) but the curved tail was still west of the Sun (in the morning sky) (Halley's Comet, 1910y). An article in *Nature* (The Transit and Tail of Halley's Comet, 1910) states:

> The narrowness of the tail (8°) on May 18 [UT] and the increased breadth next morning suggest that it was bent back in the orbit, and probably did not begin to sweep past the earth before 12h. on May 20 [UT]. At this time the earth was some four million miles [6.5 million kilometres] south of the comet's orbit plane, and consequently the tail probably passed well to the north of the earth, for the Helwan observations, during May, suggest that it was not nearly wide enough to envelop the earth at that distance. They also

> show that its length was well over twenty million miles [32 million kilometres], and would therefore have enveloped the earth had the planes coincided.

Those watching the sky for atmospheric effects mostly reported no change during the day or night, when the 12-day old Moon ruled the night (Halley's Comet, 1910y). The *Bush Advocate* also reported no unusual phenomena from overseas (Halley's Comet, 1910x). The following conclusion was published in *The Press* (Through the Comet's Tail, 1910) on 21 May:

> There was no glowing illumination of the heavens, or any dazzling shower of meteors. The sky did not do any honour to this phenomenon, one of the most wonderful of Nature's side-shows. The night was as hundreds of nights have been - a mixture of starry blue sky, fleecy clouds, attenuated haze, and a smiling moon. If the comet's tail was about, it retired into shamed obscurity before the radiance of the Queen of the Night. The tail gave the earth a swish, but it was such an apologetic and innocuous brush that the earth was not sure when it happened. The absence of any display whatsoever causes almost a feeling of disappointment, especially when the popular mind had conjured up so many fanciful and disastrous possibilities.

However, observers in Greytown stated,

> The comet caused much interest in Greytown on Thursday [19 May], and through a telescope, with a red lens, the sun was noticed to change its colour and become a peculiar yellowish hue. (The Comet: 1910c).

They also observed spots on the Sun and initially thought that they were the comet but later concluded they were sunspots. Perhaps the observers in Munich succumbed to this same wrong initial interpretation. A farmer in the Amuri district wrote to the *Lyttelton Times* (Halley's Comet, 1910z) stating that astronomers had their calculations wrong for the transit of 1P/Halley across the Sun (and presumably the tail-swipe), he wrote that at 7:25 pm on Thursday 19 May he and others saw:

> ... peculiar rays projected from the sun and five minutes later something appeared in front of the sun, which flickered and blazed. A blinding light then settled over the sun in the form of a white opaque shade. The air grew cold and there were peculiar rainbow lights on the hills.

This report was published in fourteen newspapers across New Zealand. Was this atmospheric phenomena experienced overseas? In the Blue Mountains, Australia, Mrs Bessie Hamond sent a report to Mr. W.E. Raymond, Director of Sydney Observatory, stating that she (and others) saw with the naked eye and binoculars what looked like meteors emanating from just below the obscured Sun on 19 May. This occurred around the time of the transit (the authors of this article base this on Kronk's 19.15 May UT transit time), possibly when the comet's tail brushed Earth (Halley's Comet, 1910aa). The day before, on 18 May, the surveyor Mr E. Du Faur in northern Sydney, blocked the Sun with a tree and using binoculars noticed a "... bright scintillation …" that looked like an insect swarm below and to the east of the Sun. Closer inspection revealed it was not insects. Du Faur concluded that he had seen the lower part of Halley's Comet—or its tail (*ibid.*). However, GUIDE software suggests that at the time of observation, the head of the comet was approximately 12° west of the Sun, with the tail pointing towards the west. Both Australian corresponddents made their observations when the Sun was culminating around noon.

## 5.5 1P/Halley Enters the Evening Sky

By 21 May 1910 (NZST) observations started being submitted to newspapers stating that 1P/Halley was seen in the western sky after sunset, however, due to the influence of the 12 day old Moon, its brightness was "... greatly diminished …" (Star, 1910). At this stage it lay in northern Orion with a declination of approximately +18°. On 22 May, J.T. Stevenson observed it at 5:30 pm local time, the Sun setting just after 5pm. He also remarked that it was not as bright as the morning views several days before. Stevenson, and those observing with him, were amazed at the apparent speed of the comet. NASA/JPL suggests that at that stage it was travelling at nearly 0.5° per hour towards the east. The estimated distance to the comet was 16 million miles (25.8 million kilometres) (Halley's Comet, 1910ab). With telescopic aid, he thought that the "... comet's nucleus appeared very small …" He also observed the dark band in the tail (The Comet Scare, 1910), most probably similar to the photographs in Figures 39 and 40. NASA states in the 11 November 2024 *Astronomy Picture of the Day* (showing Comet Tsuchinshan-Atlas) that

> The reason for the dark streak is currently unclear and a topic of some debate. Possible reasons include a plume of dark dust, different parts of the bright tail being unusually superposed, and a shadow of a dense part of the coma on smaller dust particles.

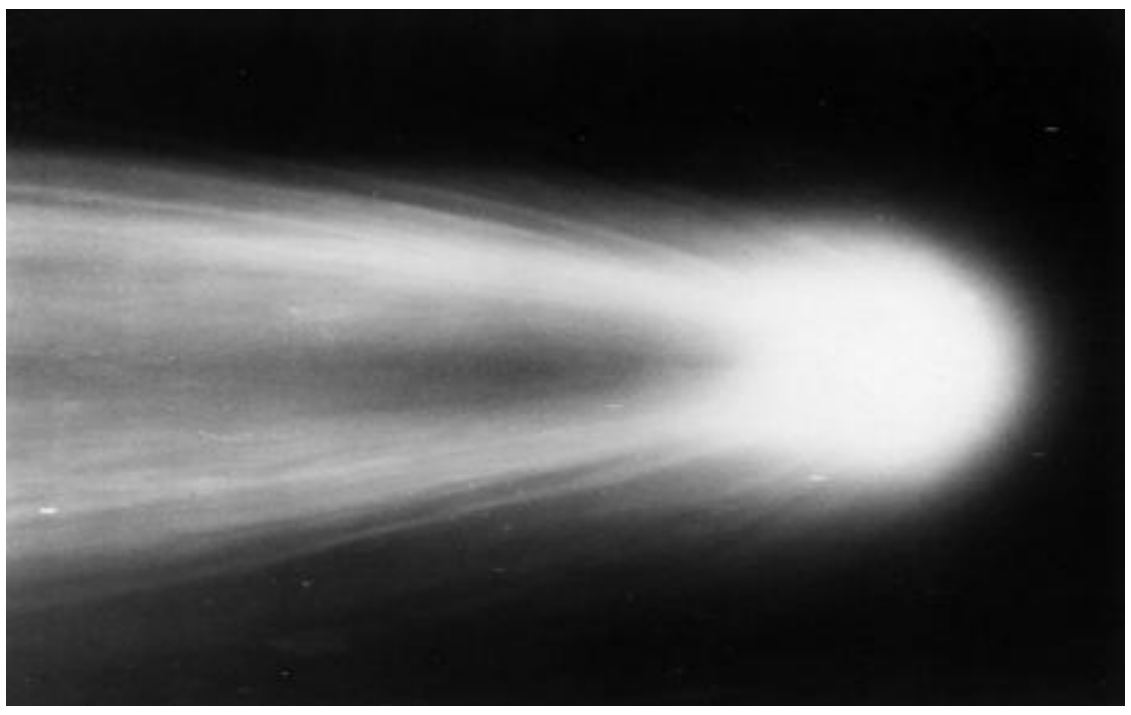

Figure 39: An incredible photograph of the head and nearby tail of Comet 1P/Halley taken on 8 May 1910 by Dr G.W. Ritchey at Mount Wilson Observatory, California. Note the dark central region to the left of the head. The possible reason for this is is not clear yet, but astronomers believe that it could be a "... plume of dark dust, different parts of the bright tail being unusually superposed, and [or] a shadow of a dense part of the coma on smaller dust particles." APOD 11 November 2024 (after Walker and Blow, 1985: 51).

Stevenson determined the tail to be about 30° in length, but this was with moonlight interference. The *King County Chronicle* (Local and General News, 1910b) wrote of two tails on 22 May: "On Sunday night the comet plainly showed two tails, one stretching in a straight line [gas], and the other shaped like half a rainbow [dust]."

Many hoped that with the total lunar eclipse on 24 May 1910, when the comet was in the evening sky, this would permit a clearer view of the comet (Halley in the West, 1910). Stevenson (Halley's Comet, 1910ac) published comet set times for Nelson (in NZST), as listed in Table 5.

On 22 May, observers in Dunedin determined a tail length of 20°, stating that it was not as impressive as a week or two before (*Otago Daily Times*, 1910). Several days later, spectators were suggesting that 1P/Halley was fading (*Ashburton Guardian*, 1910). J.T. Ward of Wanganui Observatory agreed, stating that the comet was getting smaller and fainter due to the increasing distance between Earth and the comet (*ibid.*). On 24 May, the night of the total lunar eclipse, many viewed the comet. One report stated that the

> ... the face of the moon [was] of a dark coppery hue … [and that] the moon was lost to view entirely, but it emerged from this obscurity into sight again a minute or two before a quarter-past five. (Astronomical Phenomena, 1910).

From this description, it sounds like the Moon may have been 0 on the Danjon Scale of lunar eclipse brightness, that is, a "… very dark eclipse, Moon almost invisible, especially at mid-totality." (Espenak, 2005). Thanks to this darkened sky, numerous reports were published indicating there were great views of the comet. One stated that:

> ... the comet was seen to the best advantage, and as a spectacle easily surpassed all its other evening displays. The people who were most disappointed were those who had seen the comet in all the splendour of its morning glory; others were content to admire without criticising. As the shadow gradually moved off the moon the comet shed its illumination, and by seven o'clock it had again become a comparatively insignificant ob-ject. (Astronomical Phenomena, 1910).

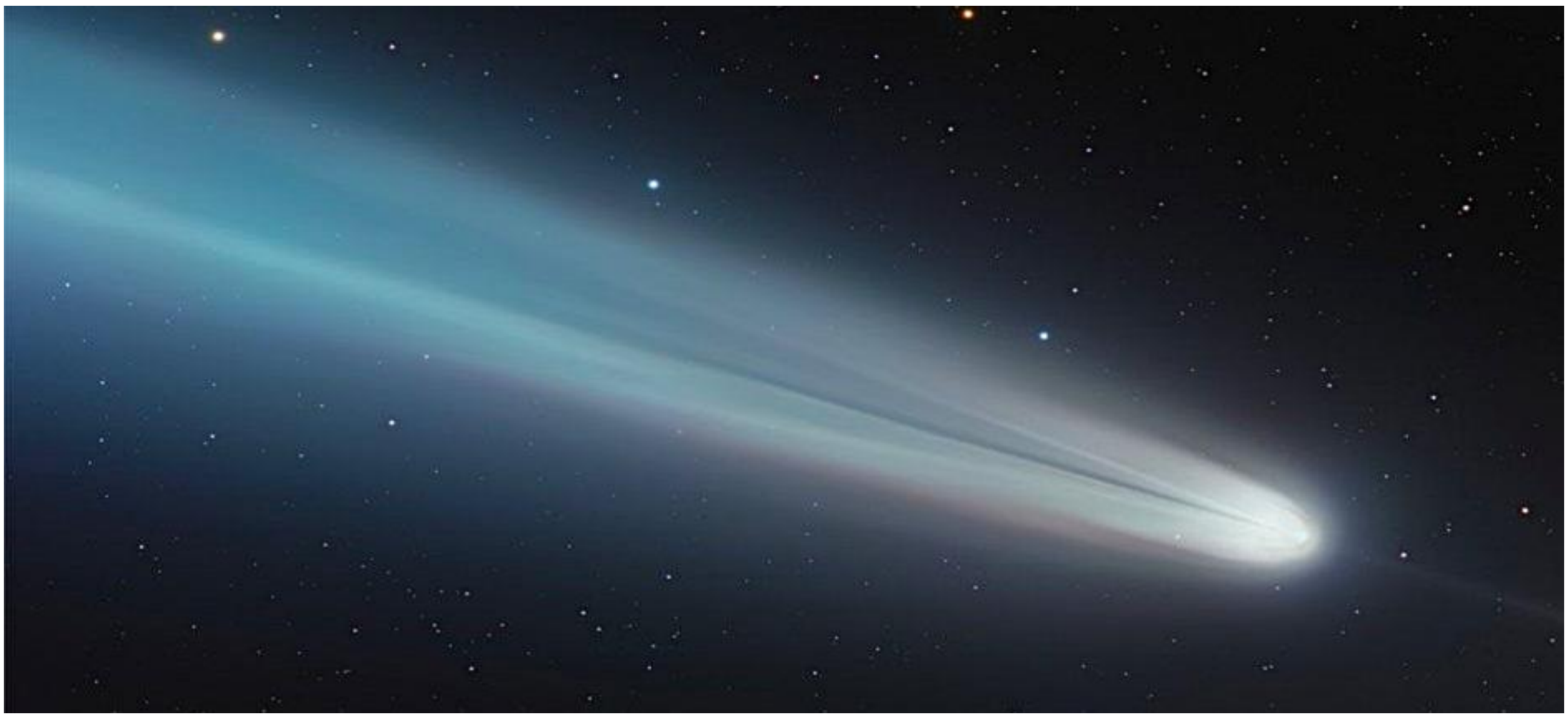

Figure 40: A beautifully detailed image of Comet Tsuchinshan-Atlas (C/2023 A3) taken on 14 October 2024 by Bray Falls from Texas, USA. The dark spinal feature is clearly evident, as it was in the photograph of 1P/Halley in 1910 (see Figure 39). This image of C/2023 A3 appeared in the 11 November 2024 Astronomy Picture of the Day (https://apod.nasa.gov/apod/ap241111.html).

Table 5: Comet set times provided by J.T. Stephenson (after Halley's Comet, 1910ac).

| Date (1910) | Sunset | Comet Set | Moonrise | Comet Distance | Notes |
|---|---|---|---|---|---|
| 23 May | 5:11 pm | 7:55pm | 4:33 pm | 18 million miles | 13 day old Moon |
| 24 May | 5:09 pm | 9:27pm | 5:01 pm | 21 million miles | Lunar eclipse (Full Moon) |
| 25 May | 5:08 pm | 10:00pm | 5:36 pm | 24 million miles | 15 day old Moon |
| 26 May | 5:08 pm | 10:19pm | 6:17 pm | 27 million miles | 16 day old Moon |
| 27 May | 5:07 pm | 10:22pm | 7:06 pm | 30 million miles | 17 day old Moon |
| 28 May | 5:06 pm | 10:33pm | 8:04 pm | 34 million miles | 18 day old Moon |
| 29 May | 5:06 pm | 10:38pm | 9:08 pm | 37 million miles | 19 day old Moon |
| 30 May | 5:05 pm | 10:44pm | 10:16 pm | 41 million miles | 20 day old Moon |

Immediately following the above quote, the reporter wrote that "Halley's comet had, however, made an impressive farewell—one that will long be remembered by those who witnessed it." The fading of the comet was now being widely noted, with reports from Wellington on the night of the eclipse stating that "... reduced though it is in magnitude, [it] was a most beautiful object, its tail sweeping over an arc of many degrees, and its nucleus as bright as, if not brighter, than ever." (Evening Post, 1910d). However, in Nelson, the conclusion was that:

> To those who saw the comet during the last few mornings of its appearance in the eastern sky, however, the view has been a disappointment, as the visitor has not been anything like as imposing in the western sky. As a matter of fact, those who missed the morning view have lost their only chance of beholding the comet at its best. (Halley's Comet, 1910ad).

Indeed, an observer writing to the *Press* stated that they saw 1P/Halley on 29 and 30 May 1910 and concluded:

> The nucleus is comparatively brilliant, while the tail is well defined, and can be traced for quite a long space. It is quite evident, however, that Halley's comet is saying farewell, and that in a few more nights it will be lost in infinity. (News of the Day, 1910c).

Joseph Ward believed that he was witnessing the comet break apart on 25 May. He wrote that

> The comet's appearance this evening is of a most striking form. There are two distinct nuclei … It really looks as if the comet might, as did Biela's, break into two separate bodies. (Mackrell, 1985: 137).

Figure 41 shows three photographs taken from Meeanee Observatory. The difference between the 27 May, 29 May and 31 May appearance of the comet's brightness and tail length is evident. Note the in-photo comment that clouds affected the photograph on 27 May. On 30 May (when the Moon was 21-days old and rising at approximately 11 pm) an unnamed observer described the coma as faint and the tail being 10° in length (Local and General, 1910e). By 7 June (New Moon) an unnamed observer stated in the *Evening Post* (1910e):

> The clear heavens last night afforded exceptional facilities for observation, and high in the heavens, still in advance of the planet Jupiter, the nucleus of the comet, with a considerable length of tail, might be [seen] by the naked eye. But the light was so extremely faint that very careful search was necessary to detect the receding comet. In fact, it is really invisible to all save those possessed of good sight.

On 1 June 1910 Wanganui Observatory's Joseph Ward, published his 'Astronomical Notes for June' (Ward, 1910e) and stated:

> Halley's Comet has made the greatest appearance seen by man during the last half-dozen times it has re-visited this earth. Many, unfortunately, despite frequent warnings in these notes, waited to see it in the evening skies after its transit, and were greatly disappointed. Those who saw it on the morning of the 22nd, 23rd, and 24th, carry in their memory something well worth recalling. A great white riband [sic] of light, stretching from the north-eastern horizon right up into the zenith, and still more below the horizon, is a greater manifestation of this interesting visitor than any of us had reason to expect. Nevertheless, Halley's Comet has been a disappointment to the average sightseer in the telescope, its comparative faintness standing very little magnifying, excepting on the evenings of the 24th and 25th [May 1910?] when a sudden rise in activity and brilliance in the nucleus, owing to the extrusion of a great jet of gas towards the Sun, and its gradual folding back into the tail, broke the normal appearance of a dull nebulosity. Still when we view it, even now, at the last of the month, and its distance from the earth has increased to something near forty millions of miles, we must con-

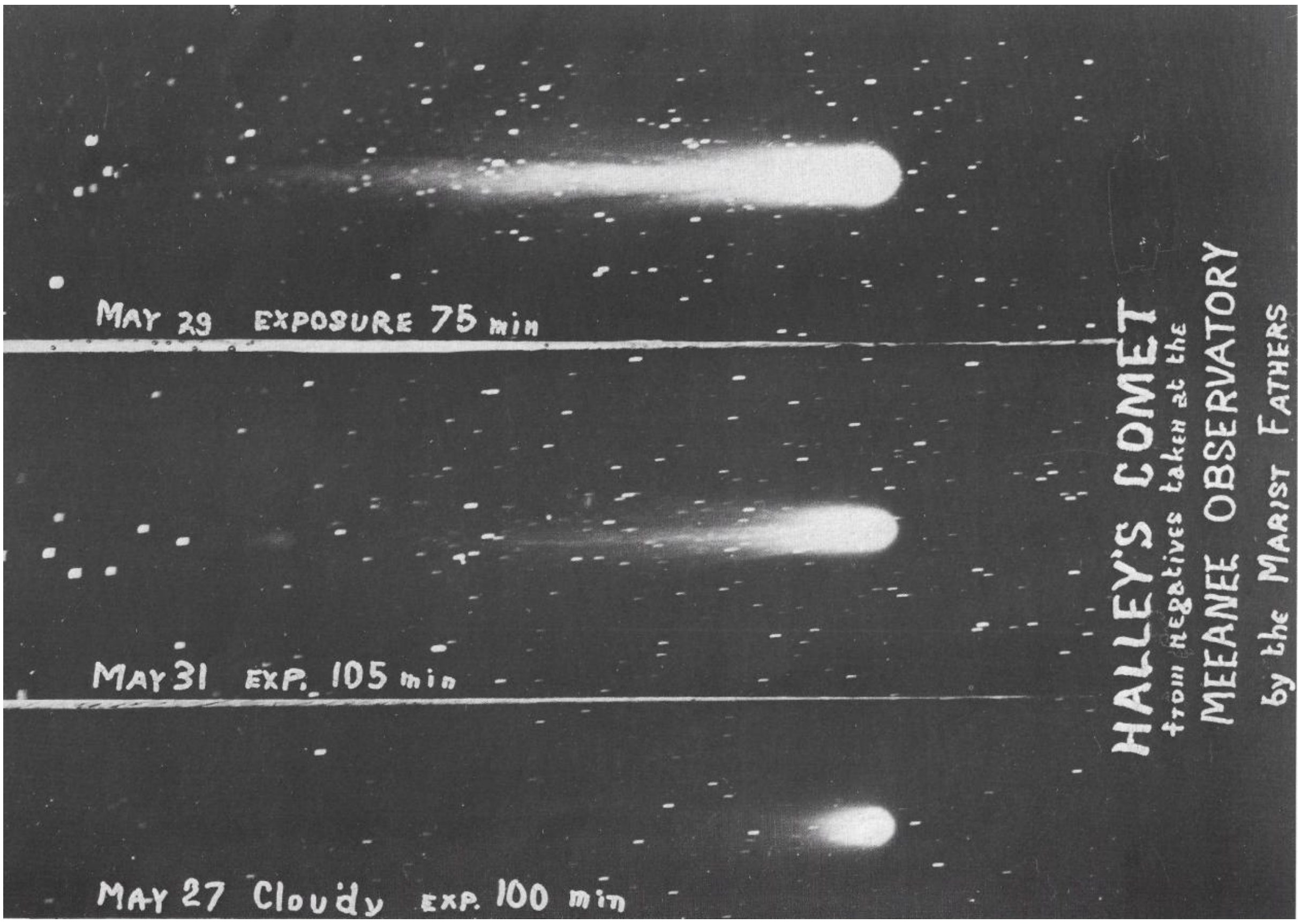

Figure 41: Photographs of 1P/Halley as imaged from Meeanee Observatory, Napier, at the end of May 1910. The comet was just days past its brightest appearance (courtesy: Marist Archives, Wellington).

> fess to being favoured by a fine sight. Its length was between 30 and 40 degrees on the 29th (Sunday), when on a beautyfully clear sky it made a fine appearance. And we have it on record that more than one noted astronomer was dubious as to whether we should see it at all, owing to the poor appearance it made in 1835, and the uncertainty as to the behaviour of these bodies.

From early June 1910, the number of newspaper articles in New Zealand about Halley's Comet declined rapidly (as can be seen in Figure 30). An example being a statement made in the *Otago Witness* (1910) on 8 June that "Oamaru" "… sends an exhaustive account of Halley's comet. I am afraid we have had enough of comets for a time …"

Observations continued in June 1910. On 10 June an unnamed observer in the *Dominion* newspaper merely stated that it was "... distinctly a comet …" (Local and General, 1910f). On 12 June an observer at Queen's Wharf in Auckland saw the comet with the naked eye and wrote that it was "... a small object, yet quite discernible." (Comet Still Visible, 1910). On 22 June, an observer with the *nom de plume* "Curly Nut" wrote "Halley's comet is just about out of sight now; it is very dim to-night." (Our Young Folks Column, 1910). At the end of June, astronomers were noting that 1P/Halley required binoculars to see it and that it was already outside of the orbit of Mars (Astronomical Notes for July, 1910).

Comet 1P/Halley was too faint to see with the naked eye by late June 1910. It provided a spectacular display when close to the Earth around 18 May. Those that followed its progress concluded that it was best in the few mornings before its transit across the Sun. The forward scattering of light, when close to the Sun (as in a line of sight) also undoubtedly aided in the comet's brightness. The observations above were primarily visual ones—be they with telescopes, binoculars, or naked eye. However, the use of the fairly recent technological development of photography saw many photographs of the comet taken from across New Zealand.

## 5.4 Photography of Comet 1P/Halley from New Zealand in 1910

In 1910, photography was very primitive compared to today's technology that utilises Charge-Coupled-Devices (CCDs) and Complementary Metal-Oxide-Semiconductors (CMOS) sensors. Before digital camera sensors, most photographers used photographic negative film that was then reverse printed onto photographic paper to produce photographic prints, this being first developed by Eastman Kodak in August 1888 (Gernsheim, 1965: 48). Prior to film, F.S. Archer invented wet collodion glass plates in 1851 (Gernsheim, 1965: 32–36). This was a wet collodion process where glass plates coated with potassium iodine and collodion (AKA Gun Cotton), were placed in baths of silver nitrate and exposed when wet. A bright terrestrial scene could be recorded in approximately 10 seconds. Following this, Dr R.L. Maddox invented Gelatin dry-plates (Albumen) in the late 1870s (Gernsheim, 1965: 46–47). These plates were more sensitive than wet-plates; a bright terrestrial scene could be recorded in ~1⁄15 second exposure time. The first deep sky photograph was taken by Henry Draper (1880: 583), who photographed Messier 42 (the Great Orion Nebula) in September 1880 with a 51-minute gelatin dry-plate exposure.

Surprisingly, glass plates continued to be used in astrophotography into the 1970s and beyond (David Malin, pers. comm). For convenience, most amateur astronomers used Single Lens Reflex cameras and negative film for astrophotography, whilst the more serious amateur and professional astronomers used glass plates. Prior to Eastman Kodak's invention in 1888, Walker and Blow (1985: 14) mention S.J. Lambert who photographed the transit of Venus in 1882 from Auckland and Josiah Martin who exhibited a "... photograph of the Sun obtained using a combination of bromide of silver and iodide of silver …" in 1882.

Through the course of its apparition, Comet 1P/Halley was imaged across New Zealand by observers using a variety of techniques. In the following subsections, we discuss the 'key players' in these photographic endeavours.

### 5.4.1 Meeanee Observatory

Searches through Papers Past for articles on 1P/Halley regularly highlighted one key New Zealand observatory as the most frequent photographer of the comet, namely Meeanee Observatory near Napier, on the East Coast of the North Island. Orchiston describes the 9-inch Cooke refractor used at Meeanee Observatory in great detail in papers, posters and books (e.g. Orchiston, 2016: 337–367; Mackrell, 1985; and Orchiston and Drummond, 2022e). This instrument, which was made by Thomas Cooke & Sons in 1867, is now housed at Carter Observatory in Wellington, New Zealand, was upgraded to become a 9.75-inch refractor by the late Gary Nankervill in about 2001. The driving force behind the initial purchase of the telescope from England and its installation in New Zealand was the Reverend Dr David Kennedy (1864–1936; Mackrell, 1985), the Superior at the Catholic Society of Mary, a Marist order of priests and lay brothers at Meeanee. Royalties from the sales of his very popular school text book, '*Natural Philosophy for Junior Students*' (1891) helped purchase the telescope (Orchiston, 2022). Additional fund-raising and donations helped build the observatory that this telescope was housed in.

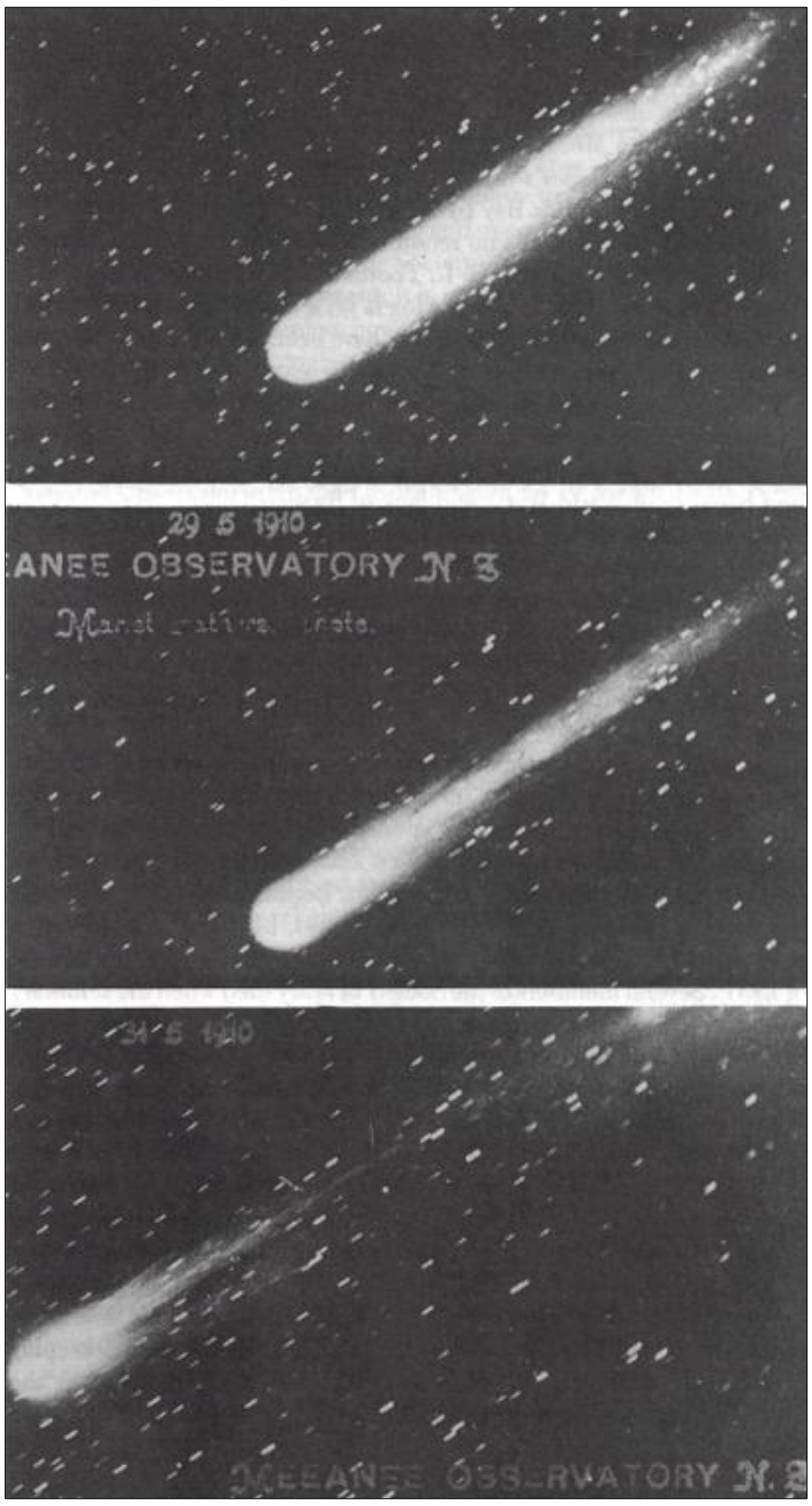

Figure 42: Photographs of 1P/Halley taken by the Marist Brothers at Meeanee Observatory in May 1910 (after Mackrell, 1985: 143).

Figure 42 shows examples of Comet 1P/Halley as photographed from Meeanee Observatory, and the photographic plate-holder is in Figure 43.

The first Meeanee Observatory was established in 1892 by Kennedy with a 6-inch (15.2-cm) reflector (Mackrell, 1985: 80). The larger observatory, with the 9-inch (23-cm) Cooke refractor, was built later, and officially opened in July 1907 (Mackrell, 1985: 88). It was located at at the seminary on the junction of Meeanee and Powdrell Roads. The metal dome was three metres in diameter and weighed one tonne and sat on walls 6 metres long—not in an observatory 6 metres square as Mackrell (1985: 84) states! The instrument had photographic and spectroscopic capabilities. For wider angle field of views, a 3.5-inch (8.9-cm) Dallmeyer portrait lens with gelatin photographic plates could be attached to the top of the main telescope and take 30–60 minute exposures as the main telescope was manually guided to keep a reference star in the crosshairs.

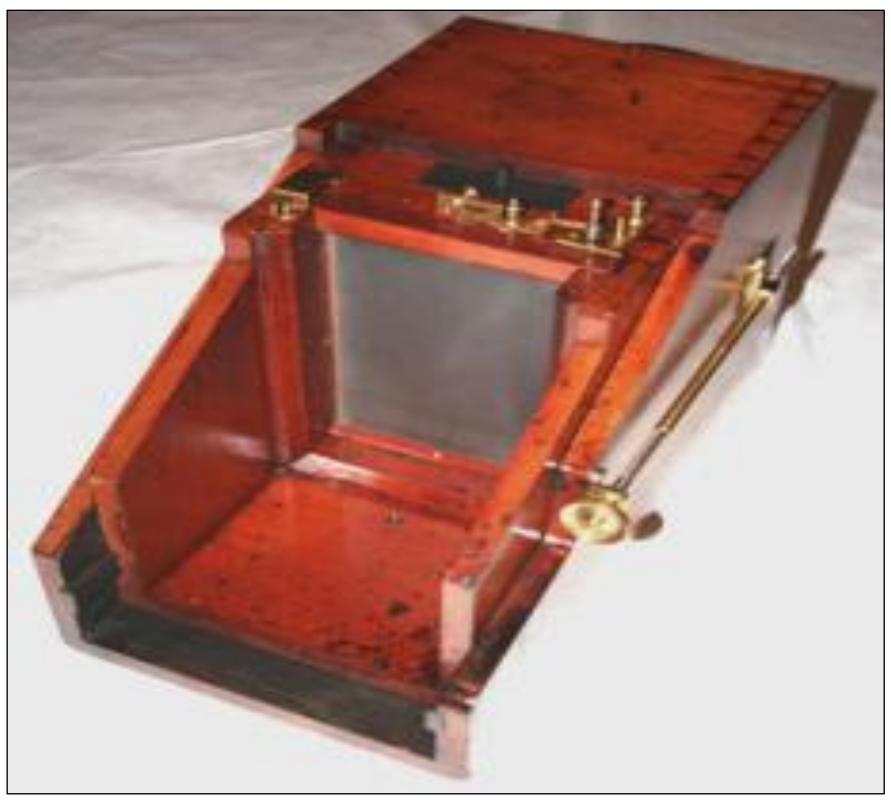

Figure 43: The wooden plate-holder used by the Meeanee Observatory astronomers (photograph: Orchiston Collection).

Unfortunately, as stated earlier, Kennedy was transferred to Wellington to become the Rector of Saint Patrick's College in May 1909, one year before 1P/Halley would be at its brightest for the 1910 apparition (Orchiston, 2016: 348). Kennedy's two observing colleagues at Meeanee, Ignatius von Gottfried (1864–1951) (Figure 45a) and Joseph John Cullen (1886–1963) (Figure 45b) continued their departed leader's role in observatory duties. Both would produce world-class photographs of 1P/Halley in the following year on glass plates, some of which grace the pages of this paper. In late April 1910, a newspaper article stated that it was the intention of the Meeanee astronomers to ".... take a photo every morning for some time." (*Waipawa Mail*, 1910). Given Napier's position on the east coast of the North Island and that the prevailing wind in New Zealand is westerly, there was a reasonable chance that they could quite successfully attain their goal. Indeed, photographs of Comet 1P/Halley were subsequently sent from Meeanee Observatory to the Royal Astronomical Society and the British Astronomical Association (Mackrell, 1985: 140–141; Orchiston, 2016: 350). Many of these 1P/Halley photographs taken from Meeanee were used in public lectures (e.g. Romance of Science: 1910) around the country and also sold as postcards.

Meeanee Observatory sadly fell into disrepair in the years following the apparition of Comet 1P/Halley and Cullen and von Gottfried were transferred elsewhere in 1910 and 1911,

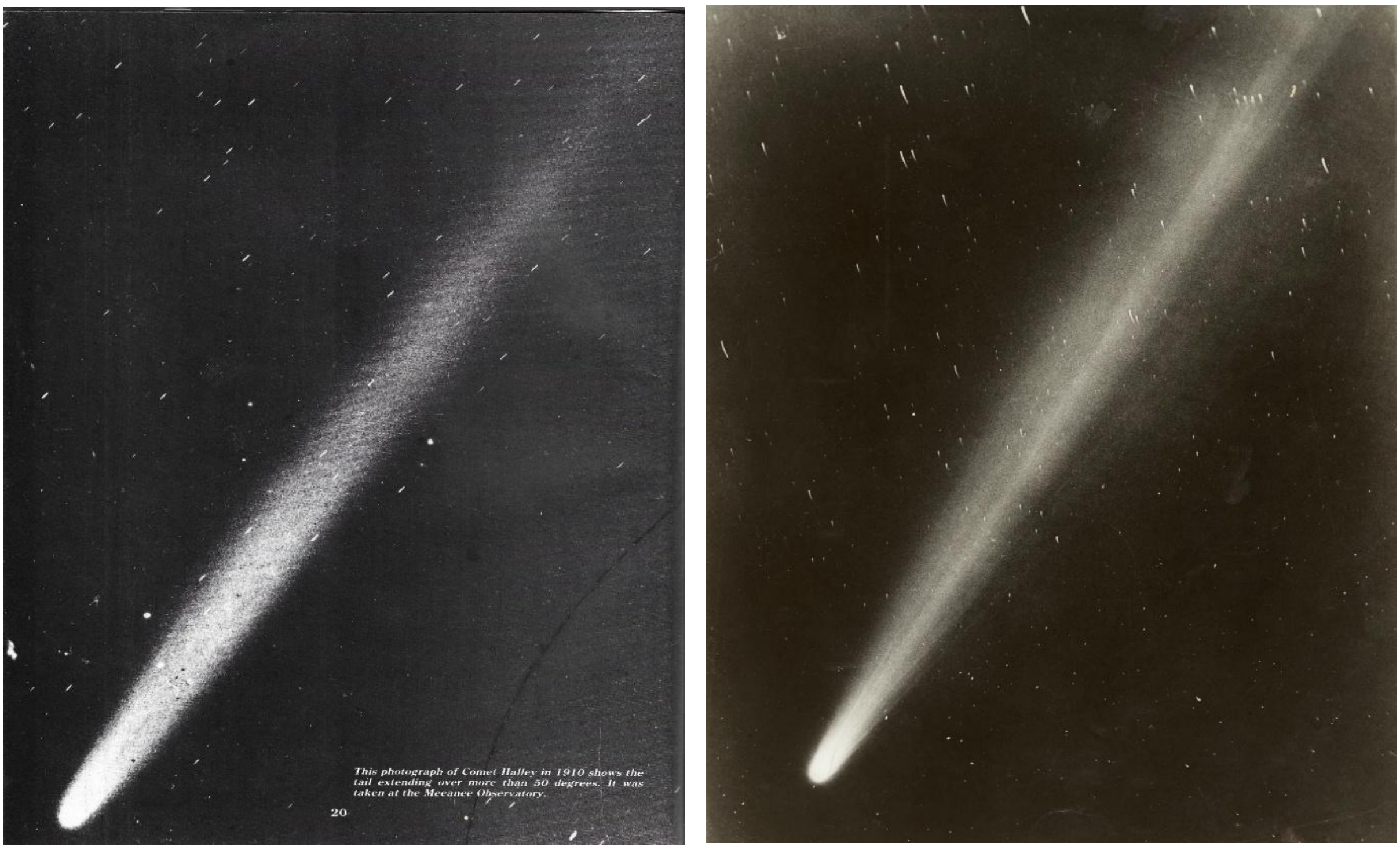

Figure 44: A comparison of photographs of 1P/Halley taken at Meeanee Observatory (left) and Yerkes Observatory (right). The photographs were taken around the same time, in early May 1910. Considering that Meeanee Observatory was of amateur status, it compares quite favourably with that of Yerkes Observatory's professional standing (source, left: Walker and Blow, 1985: 20; right: https://lampiaum.tumblr.com/ (2015)).

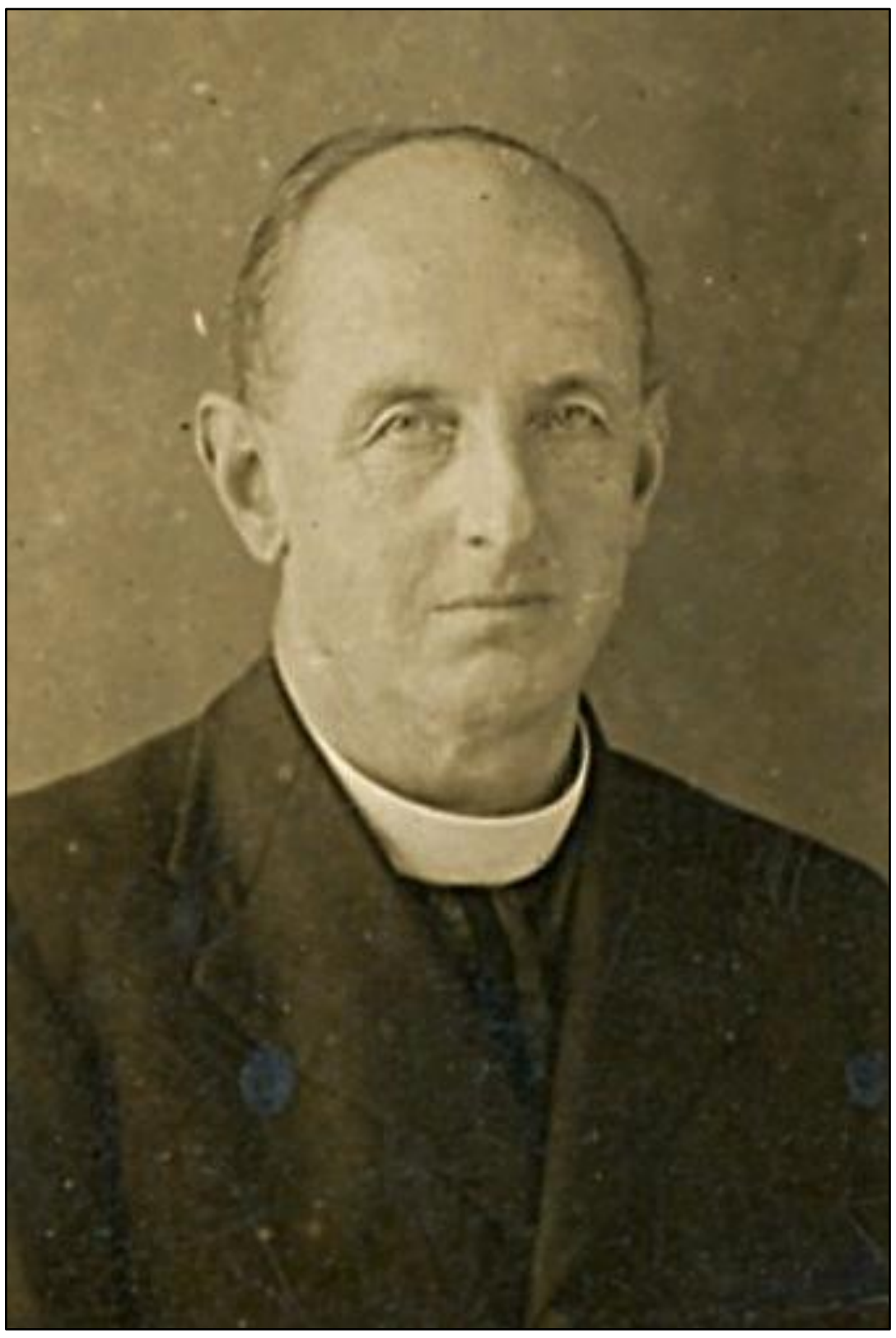

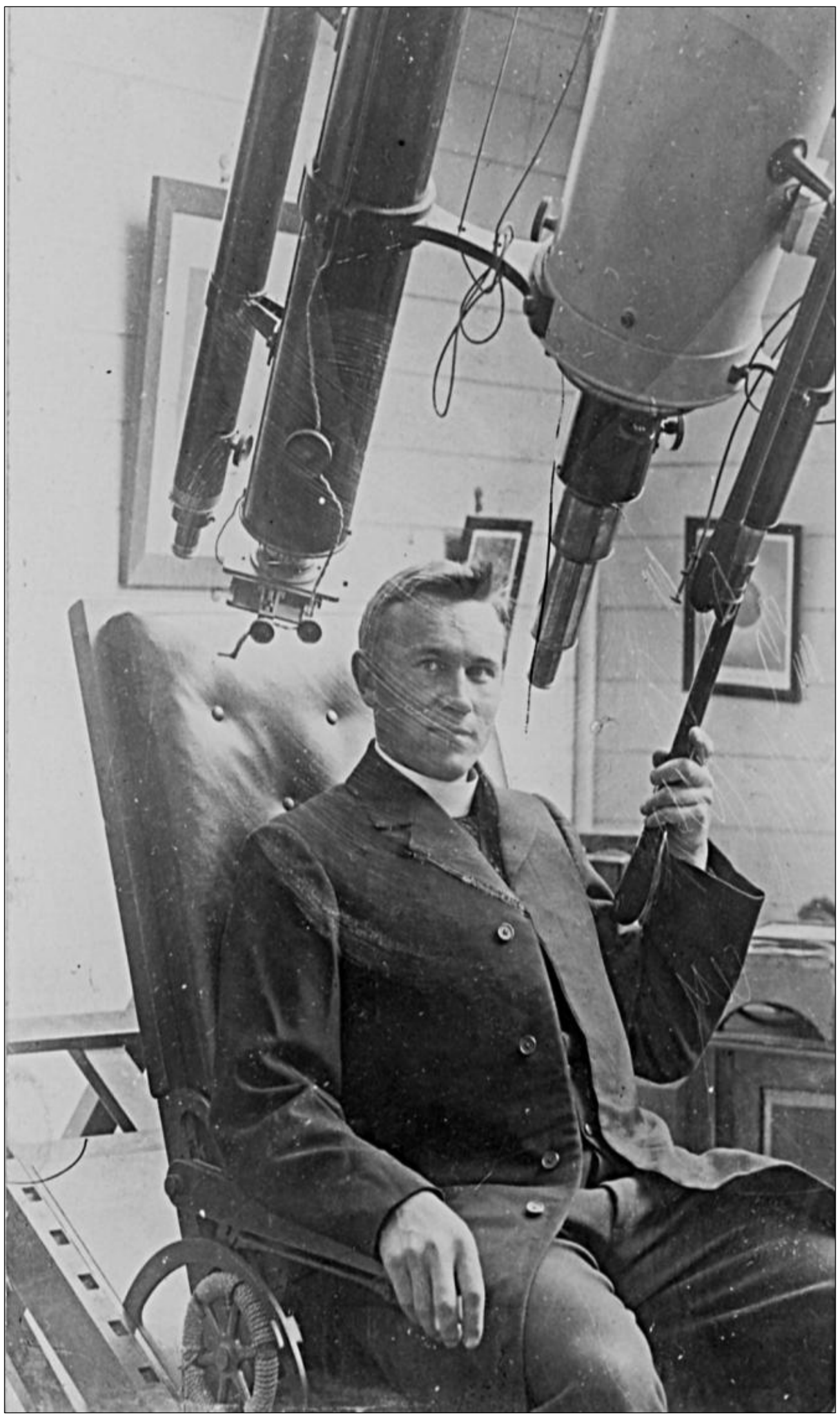

Figure 45a (above): Father Ignatius Gottfried in 1917; Figure 45b (right): Father Cullen at the 9-inch Cooke refractor in circa 1908.. Both men continued the astronomical and cometary photographic work after Reverend Kennedy was transferred to Wellington in 1909. Mackrell (1985: 112) states that Cullen and Gottfried rose at 3am for the two weeks prior to 1P/Halley entering the evening sky (photographs courtesy Marist Archives, Wellington).

respectively, leaving no-one at the Meeanee Seminary with the will or interest to maintain the facility. Then, in 1911 the Seminary was transferred from flood-prone Meeanee to a hill-side at nearby Greenmeadows. On 26 October 1912 a major storm roared through the Hawkes Bay region (Storm in the South, 1912), ripping the heavy dome off the observatory and hurling it down the hillside. The fate of Meeanee Observatory was sealed. The telescope, instruments and eyepieces were hastily placed in storage. Eleven years later, Reverend Father Aubrey notified Dr. C.E. Adams, the Government Astronomer, that the telescope was for sale. The Wellington City Corporation (later the Wellington City Council) decided to purchase it for £500 (Orchiston, 2016: 351), and it was housed in a roll-off roof observatory in the upper reaches of the Wellington Botanic Garden, near the cable car terminus. Known as the Wellington City Observatory, it was officially opened on 26 August 1924, and the telescope was used by local amateur and professional astronomers for research and public outreach. Later, Carter Observatory was built at this same site, and the 9-inch Cooke refractor was housed there (Figure 46). Murray Geddes was the first Director, and the official opening occurred on 20 December 1941 (Drummond et al, 2026: 305–312; Orchiston, 2016: 356). Interestingly, the telescope was used later for the 1985–1886 apparition of Comet 1P/Halley by Carter Observatory staff, Wellington Astronomical Society members, and members of the general public.

### 5.4.2 *Charles J. Westland*

Another keen astronomical photographer of 1P/Halley was Charles J. Westland (1875–1950; Figure 47; Orchiston, 2016: 509–522), a North Canterbury farmer who for a time worked as a professional astronomer. Mackrell (1985: 89–90) points out that Westland's photographic equipment was "... primitive homemade, but effective, equipment." He used a 4- or 5-inch Ross portrait lens (Mackrell, 1985: 90. Orchiston, 2016: 519) as the primary objective at the end of a wooden box that housed the photographic plate. Mounted on this box was a small refracting guide scope with which he manually guided during his exposures. This assembly was on a simple English equatorial mounting (Figure 48). Westland became assistant to Dr Charles Edward Adams (the first Government Astronomer) after his return from serving in the 1914–1918 war. Orchiston (2016: 519) states that Douglas Charles Berry, the then Director of the Comet Section of the Royal Astronomical Society of New Zealand, identified "... Westland as a pioneering New Zealand astronomical photographer, and he joins the likes of Thames' John Grigg and Meeanee's Reverend Dr David Kennedy, both of whom also photographed comets …" Westland produced some nice photographs of 1P/Halley as evidenced by Figure 49.

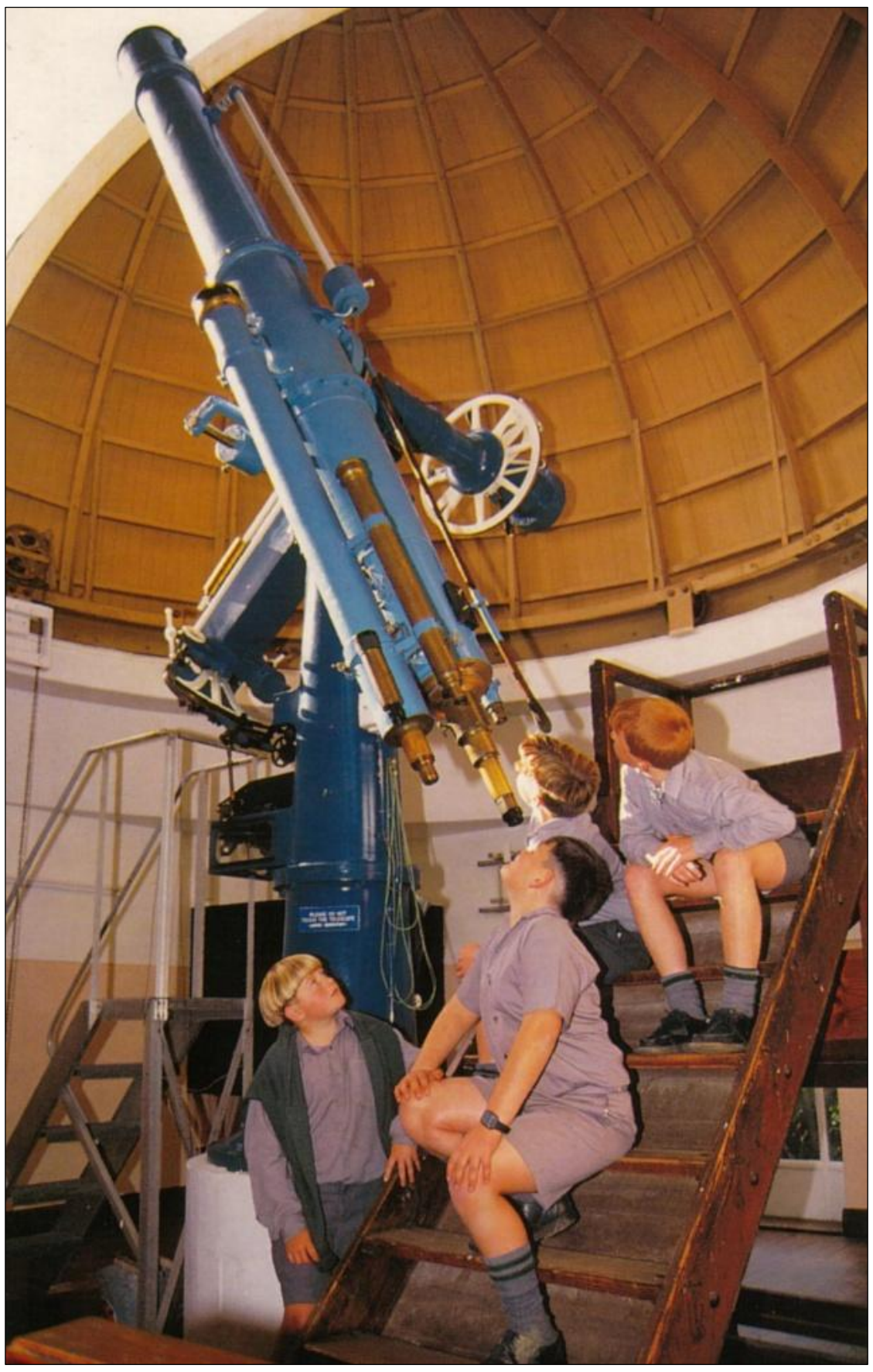

Figure 46: The old 9-inch (22.9-cm) Cooke photovisual refractor at Carter Observatory in the 1990s. When he designed the photovisual triplet for Cooke, Dennis Taylor was unaware that the glass selected for one of the three objectives would deteriorate over time. This also happened to the Carter Observatory telescope and in 2001 Gary Nankivell was forced to fabricate replacement optics. He chose to install a 2-element aplanatic objective with an increased diameter of 9.75 inches (24.77-cm) (photograph: Frank Andrews).

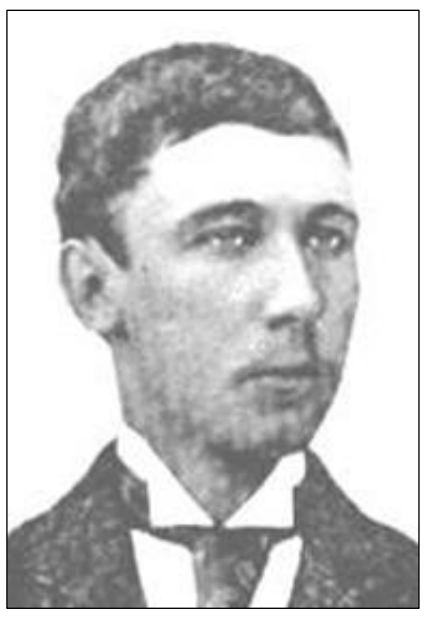

Figure 47: Indian-born Charles James Westland, photographed Comet 1P/Halley and in 1914 independently discovered a comet. After working as Adam's Assistant at the Dominion Observatory and as Director of Apia Observatory he retired and returned to his North Canterbury farm (after The Cyclopedia ..., 1903: 567).

Mackrell correctly writes, "Considering the primitive equipment and low-powered telescope Westland used, these pictures are outstanding work." To place it in context, the right-hand photograph in Figure 49 was taken when the comet was in Sextans, and α Sextantis is the bright star on the extreme left (magnitude 4.5). The bright star just to the right of the 1P/Halley's coma is β Sextantis (magnitude 5.1). The field of view is approximately 10° × 10°. The comet's tail is 6° long in the photograph. The faintest star is approximately magnitude 8.5 to 9. Note the dimming of the comet between 29 May and 8 June—although this could be due to different exposure times (which were not stated in Mackrell's book; however, Mackrell (1985: 115) does publish one of Westland's photographs and states that it was a 32-minute exposure, so an indicative time is given). Newspaper articles do support this dimming of the comet though between the two dates.

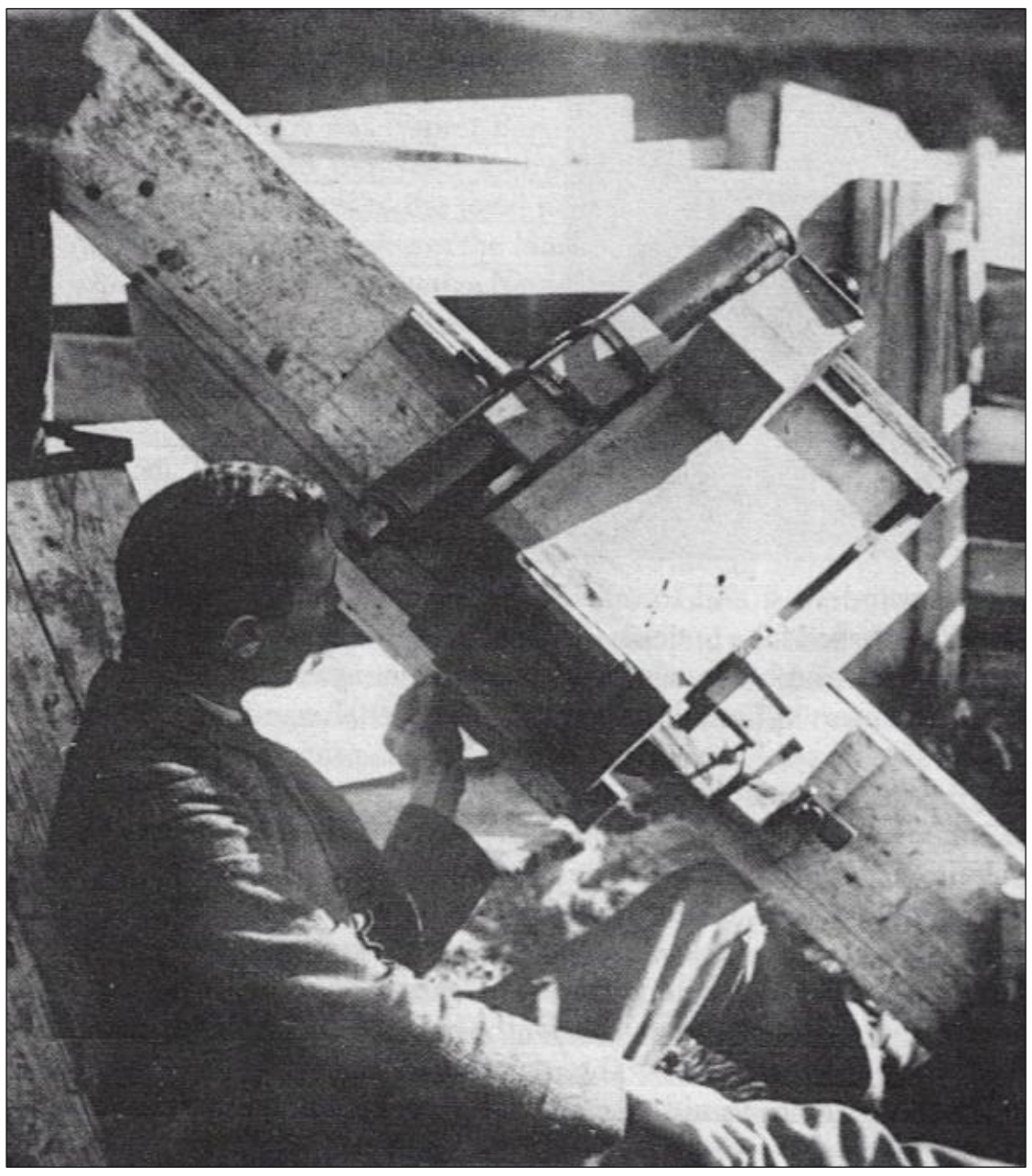

Figure 48: Charles Westland with his rustic home-made camera and guiding telescope around 1909 (after Mackrell, 1985: 90).

After 1P/Halley, Westland also took a succession of photographs of Comet C/1914 S1 (Campbell), which are discussed in Orchiston (1983; 2016; 515–517).

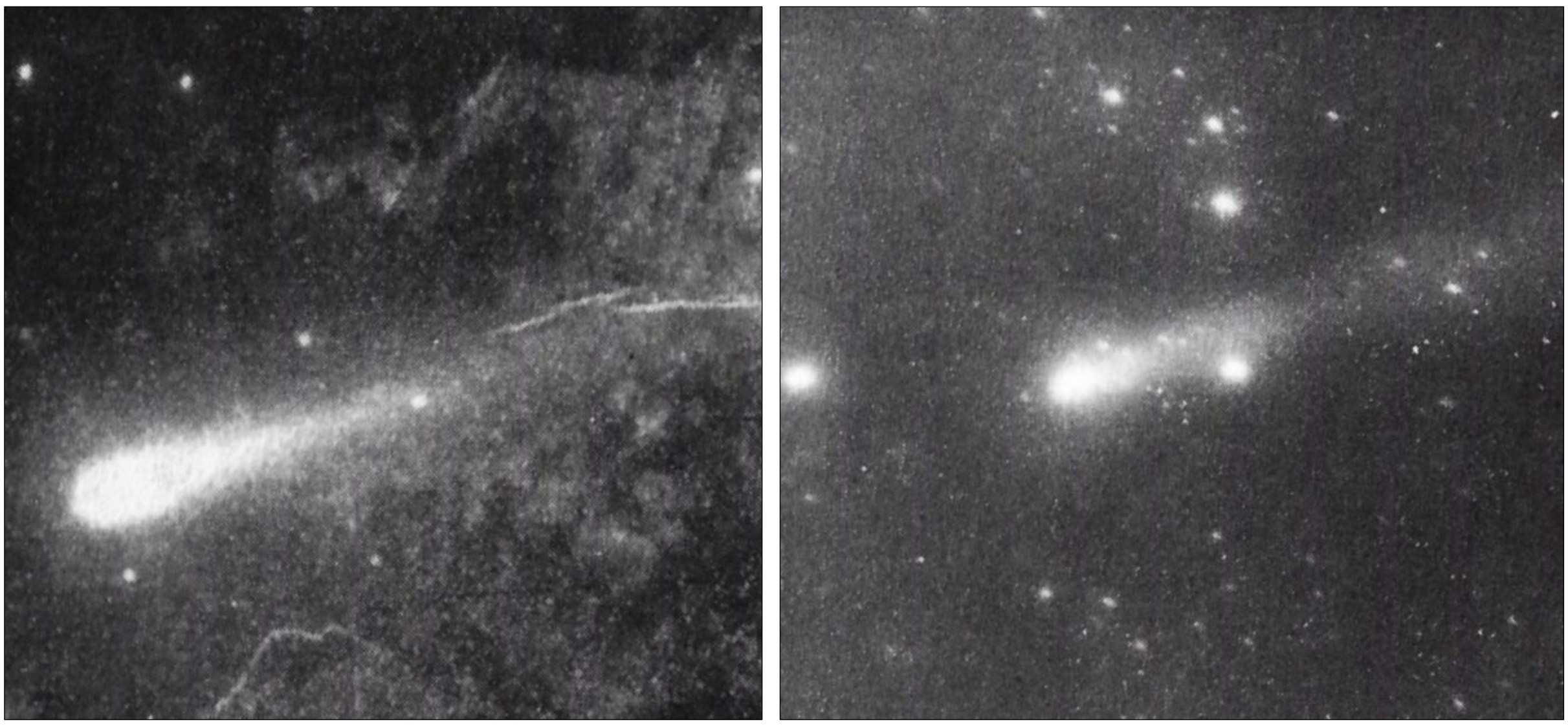

Figure 49: 1P/Halley as photographed by C.J. Westland with his 4 or 5-inch lens and using glass plates. The left photograph was taken on 29 May 1910 and the right one on 8 June 1910. Comparing the photograph taken on 8 June with the GUIDE planetarium software package (REF) reveals that the field of view was approximately 10° × 10° and the limiting magnitude was about 8.5–9. These two images were published in Mackrell (1985: 139). Mackrell, states that the photographs were provided by Mrs Edna Westland (C.J. Westland's widow).

### *5.4.3 John Grigg*

John Grigg of Thames was also an enthusiastic photographer of comets (Drummond, et al., 2026: 290–303; Orchiston, 2016). Orchiston (2016: 597–601, 604–606) discusses the photographic equipment that Grigg used to take his cometary photographs. A newspaper article (Astronomical Photographs, 1896) states that:

> ... Mr Grigg … has for many years been grappling with the difficulties attendant upon the reproduction by means of photography of views of the heavens–difficulties that are even puzzling to a professor at a large observatory … We believe that he is the only astronomer in New Zealand who has so far advanced, and it is in that connection we have written this notice … and it is only right that his townsman should know what he has accomplished.

On the evening of 12 May 1901 (NZST), Grigg photographed the Great Comet of 1901 (C/1901 G1) with his astrocamera (Figure 50). His photograph captured both the gas and dust tails extending to a distance of at least 9° from the comet's head. This was the first successful photograph of a comet taken from New Zealand (Mackrell, 1985: 77–78; Orchiston, 2016: 613). An earlier photograph, claiming to be a photograph of Comet C/18821 R1 over Mount Taranaki was shown by Orchiston et al. (2020b: 633–636) to be a painted-on addition to a daytime photograph of the mountain.

On the basis of Grigg (1970: 27), Orchiston (2016: 606), states that Grigg took several photo-graphs of 1P/Halley on 13 May 1910 when it was in the predawn sky. One such image is shown in Figure 51. Orchiston (2016: 606) suggests that 1P/Halley was the last comet photographed by Grigg.

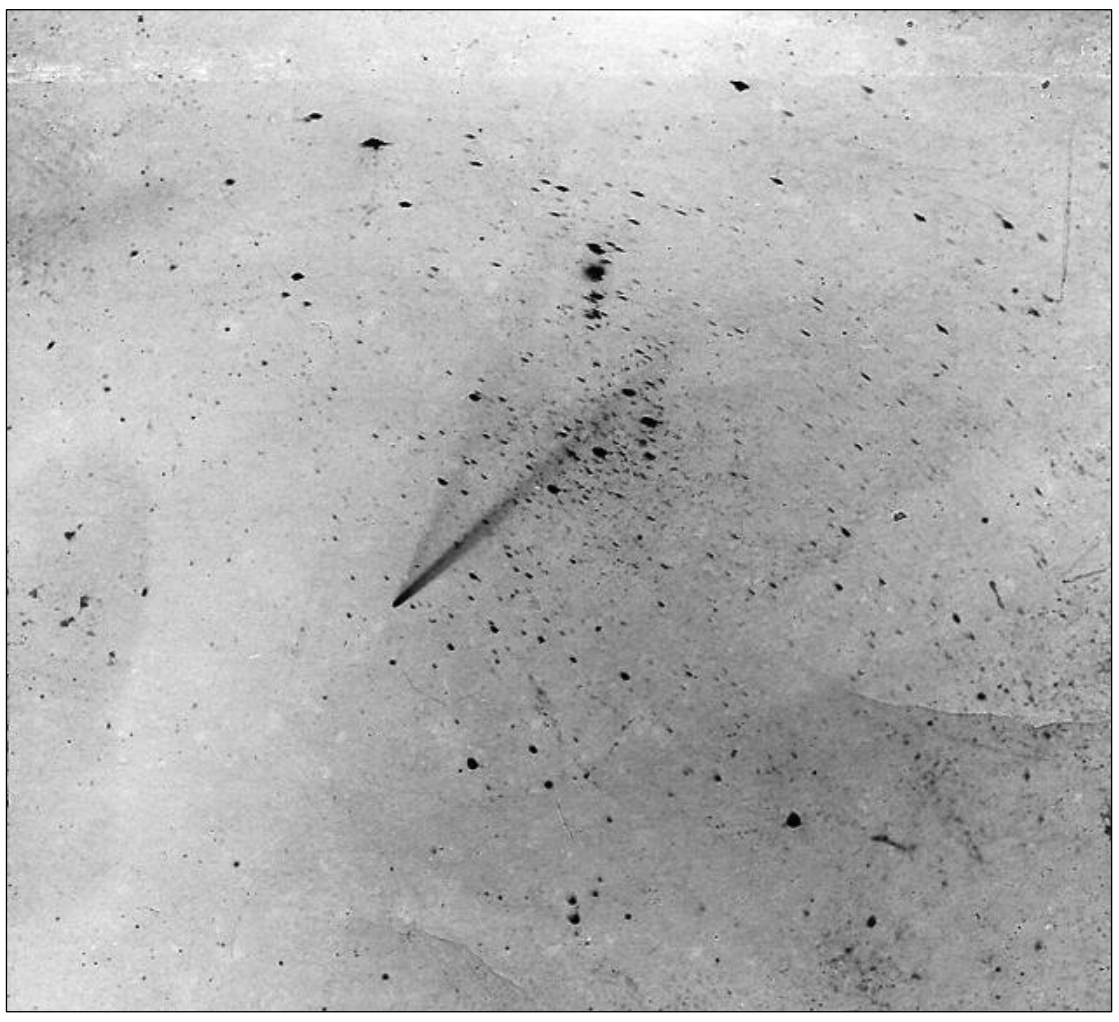

Figure 50: Comet C/1901 G1 as photographed by John Grigg on the evening of 12 May 1901. Around the date of Grigg's photograph, C/1901 G1's nucleus was determined to be magnitude 5.4 by Nijland, and both tails were observed by several astronomers, with the dust tail being much brighter at about 15° long (Kronk, 2007: 12; Vsekhsvyatskii, 1964: 352). To enhance the visibility of the tails, the authors have converted the image into a negative. When photographed, the comet was in Orion and the head was 4.2° northwest of the western-most star in Orion's belt (Mintaka). The brighter tail is the dust tail, which is approximately 9° long in this photograph. The tail pointing towards M42 (the Great Orion Nebula) is the gas (ion) tail, and is a similar length. If uncropped by Grigg, the field of view of this camera and lens combination was 27° (vertical) × 25° (horizontal). The faintest star recorded (near the comet's head) is about magnitude 8.5 to 9. Note the spherical aberration in the corners of the photograph, caused by the lens aperture being fully opened creating coma (courtesy: Mitchell Library, Sydney).

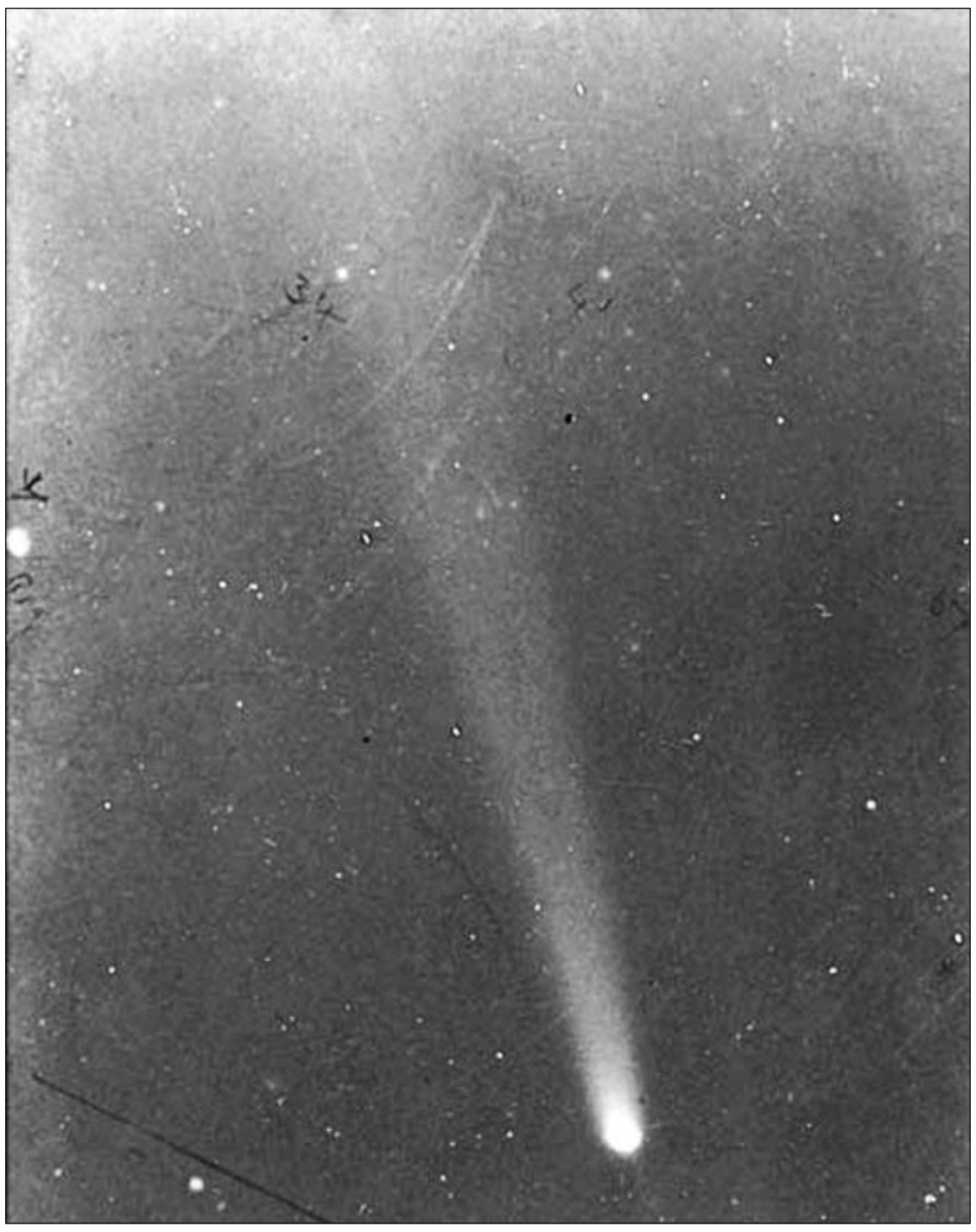

Figure 51: Comet 1P/Halley as photographed by John Grigg of Thames on 13 May 1910. Based on the ink inscriptions on the photograph and GUIDE software, the 'bright' star near the middle at the extreme left is Gamma Pegasi (magnitude 2.8), the star with 34 written beside it is 34 Piscium (magnitude 5.5), the star at 41 is 41 Piscium (magnitude 5.4). The star to the left of 1P's head may be 47 Piscium (magnitude 5.1). The tail is about 11 degrees long in the photograph and pointing approximately to the SSW. The lighter region at the top is possibly deterioration of the print or negative. This fairly wide field image was probably made with a portrait lens and was possibly cropped so as to enhance the viewer's attention just on the comet (photograph: Orchiston Collection).

### *5.4.4 Other New Zealand Comet Photographers*

The Meeanee Marist Brothers, Westland, and Grigg were the key New Zealand astronomers leading the photography of comets, including 1P/Halley in 1910. However, newspaper articles suggest that some members of the general public were also interested in photographing this 'once in a lifetime' event. A correspondent wrote that on 19 May 1910 that many members of the public in Wellington climbed Mount Victoria and the Tinakori Hills to view the comet, and that "... others took cameras with them …" (Halley's Comet, 1910ae). On 21 May the *Marlborough Express* (Local and General News: 1910c) stated:

> Mr M.C.H. Cheek showed a representative of The Express this morning an undeveloped proof of a photograph taken of Halleys comet on Sunday morning last. A good view of the comet and tail was secured; in fact, it appears to be a better picture than any which have been exhibited locally.

The Wanganui Observatory also attempted photography of 1P/Halley, albeit, not always with success. On 26 May 1910, the *Wanganui Chronicle* reported:

> The comet again presented a fine appearance in the western sky last evening, although it is fast getting smaller, and its brightness suffers owing to the light of the moon. The night was also cloudy, but those who patronised the Observatory had good views of the comet and other heavenly bodies. The photo taken on Tuesday evening was not altogether a success. The exposure was for 15 minutes at intervals, but in the closing of the dark slide the camera was, unfortunately, moved and the comet developed in triplicate. The plate shows a large view of the comet, with two diminutive comets alongside. Another effort will be made tomorrow night, when the moon, rising later, will enable the comet to be seen on a darker sky. (Halley's Comet, 1910af).

The photographer is not mentioned, but it was probably Joseph Ward, the Observatory Director. It would seem that the Observatory had photographic capabilities. The instrument used is unclear. A clue suggests that it was taken through the 9.5-inch Cooke refractor, the developed photograph showed "... a large view of the comet …" This probably being due to the long focal length of the photographic instrument.

Other New Zealanders also attempted to photograph 1P/Halley. Some of these are undoubtedly not mentioned in newspaper articles, while others are. An example is Joseph Taylor of Nelson who stated:

> The comet showed a definite nucleus nearly as large and bright as Venus herself; while her brilliant tail streamed upwards in a north-easterly direction about three degrees long and one degree wide at the termination (when it was bifurcated). I secured two photographs, one at 5.30 and another at 5.40; but the comet was then rapidly fading in the light of the dawn, and by 6 o'clock it was lost even to an opera glass. (Halley's Comet, 1910m).

### 5.4.5 New Zealand Paintings and Drawings of Comet 1P/Halley in 1910

In addition to photographs of 1P/Halley, some images of this comet were sketched or painted. As an example see Figure 37 from Walkers and Blows' book (1985: 1). A similar painting can be seen in Figure 38. It was painted by Herbert Cole (aka Rix Carlton) and shows 1P/Halley over Saint Kilda's Beach, Dunedin, on 17 May 1910. Other painters, such as Nelson's Edwin Harris and his daughter Emily (Leggott and Field-Dodgson, 2025) had previously painted the Great Comets of 1882 and 1901, as well as the solar eclipse of September 1885 (Local and General, 1910g; Orchiston et al., 2024). The authors have not been able to locate any paintings that they made of Comet 1P/Halley.

## 6 THE HUMAN ELEMENT OF COMET 1P/HALLEY: ANXIETY, NON-SCIENTIFIC CLAIMS, AMUSING STORIES, AND THE EFFECT ON ADVERTISING.

### 6.1 Fears Associated with Comet 1P/Halley

Finally, the impact that 1P/Halley in 1910 had on humanity is worth noting for it revealed the thinking that was prevalent at the time in relation to comets. As example see Figures 52 and 53. Many books and articles about comets open with discussions of the fact that comets are seen as portents of doom, harbingers of disaster, precursors to the death of princes, and so on. Such statements were valid for people prior to the early 1900s who often associated the appearance of a brilliant comet in the sky with an arching tail as something to be feared. In 1910, many articles were written by eminent astronomers about the malignity of comets, and pockets of humanity still cowered in suspense as to what disaster a comet such as 1P/Halley might bring.

The *Manawatu Standard* (Halley's Comet: How the Earth Will Pass Through the Tail of Halley's Comet, 1910) concluded in April 1910, "No meteor has ever excited so much terror or so much interest as Halley's comet." Many articles were written highlighting the negative events in human history associated with Comet 1P/Halley, such as the fall of Jerusalem in 66 CE (Jaroff, 1985), and the fall of England to William the Conqueror in 1066 CE (*ibid.*). One sad example of comet-induced fear was a Hungarian man who committed suicide due to his terror of 1P/Halley, stating that he would rather take his own life that it being taken by a star (News Tit-Bits, 1910a). In Bergenburg, Germany, a woman threw herself and her infant into a well for fear of the comet (News Tit-Bits, 1910b). One hopes that they were rescued. The following brief accounts are stories of the fear that 1P/Halley produced, as well as scientific misconceptions and humorous observations that this, the most famous of all comets, produced.

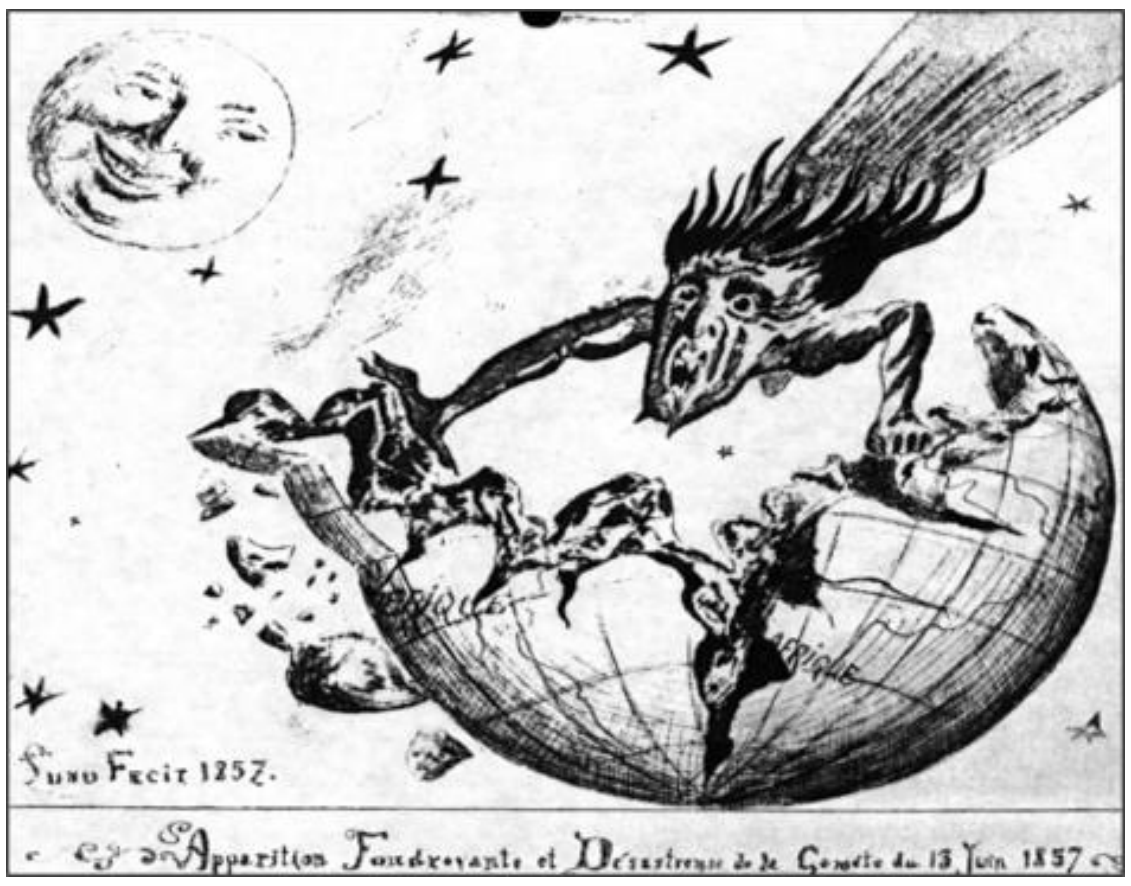

Figure 52: A cartoon from 1857 showing a comet (not 1P/Halley) destroying Earth. Today's attitude towards comets is not as fearful as earlier generations, however, we still fear Earth being hit by a comet or asteroid, this being a key reason for today's comet and asteroid search surveys (source: ifa, 2019).

A Hungarian village in the Theiss Valley feared that Halley's Comet might smash into Earth and atomise it (Town and Country, 1910). At midnight in early April 1910, the night watchman saw the glow of a neighbouring village on fire and assumed that it was the end of the world being caused by 1P/Halley, so he alerted the local population with a horn and shouted "The last day has come!" Residents ran half clothed into the town square to die; men trembled, women screamed, and children cried. They then decided to party until the end and ate and drank all the town's provisions. It was stated that "... hasty prayers were made between bites for the salvation of their souls." On a similar note of drunkenness, an eyewitness to a crime who was set to testify in court for the prosecution in late April 1910 was warned not to drink on the day of her court appearance, however, in her words, the comet upset her and she needed to take something for it – alcohol. Needless to say she was too inebriated to testify (Australian News, 1910).

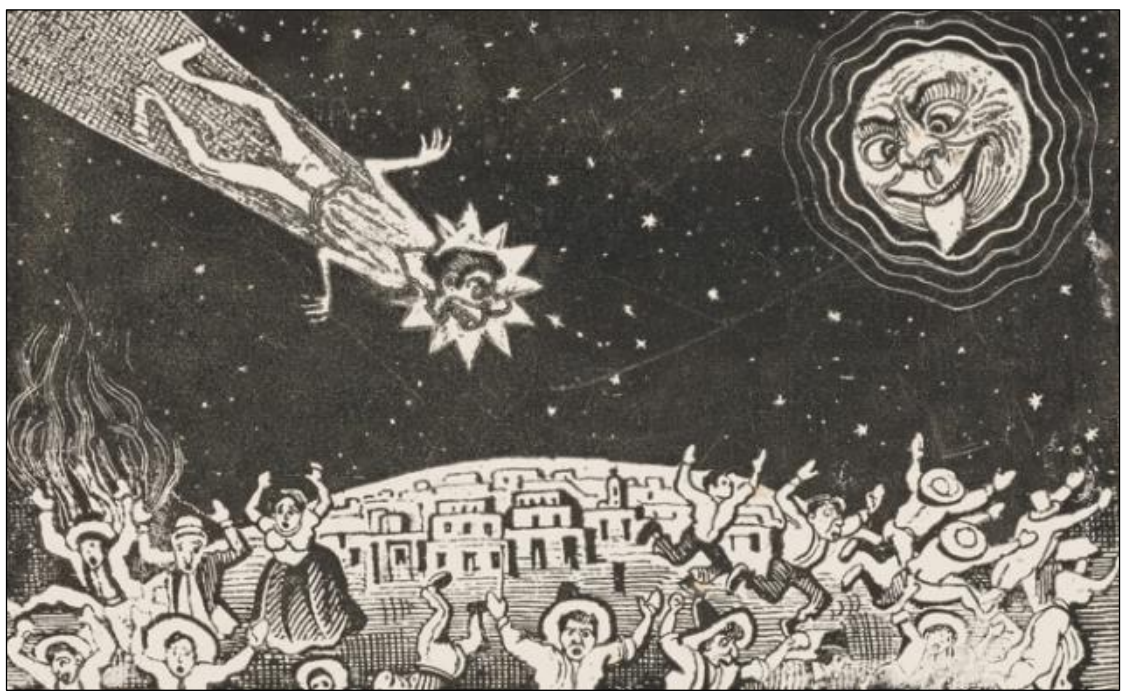

Figure 53: A cartoon from 1910 showing the fear of Comet 1P/Halley that some people possibly had (source: Science History Institute, 2025).

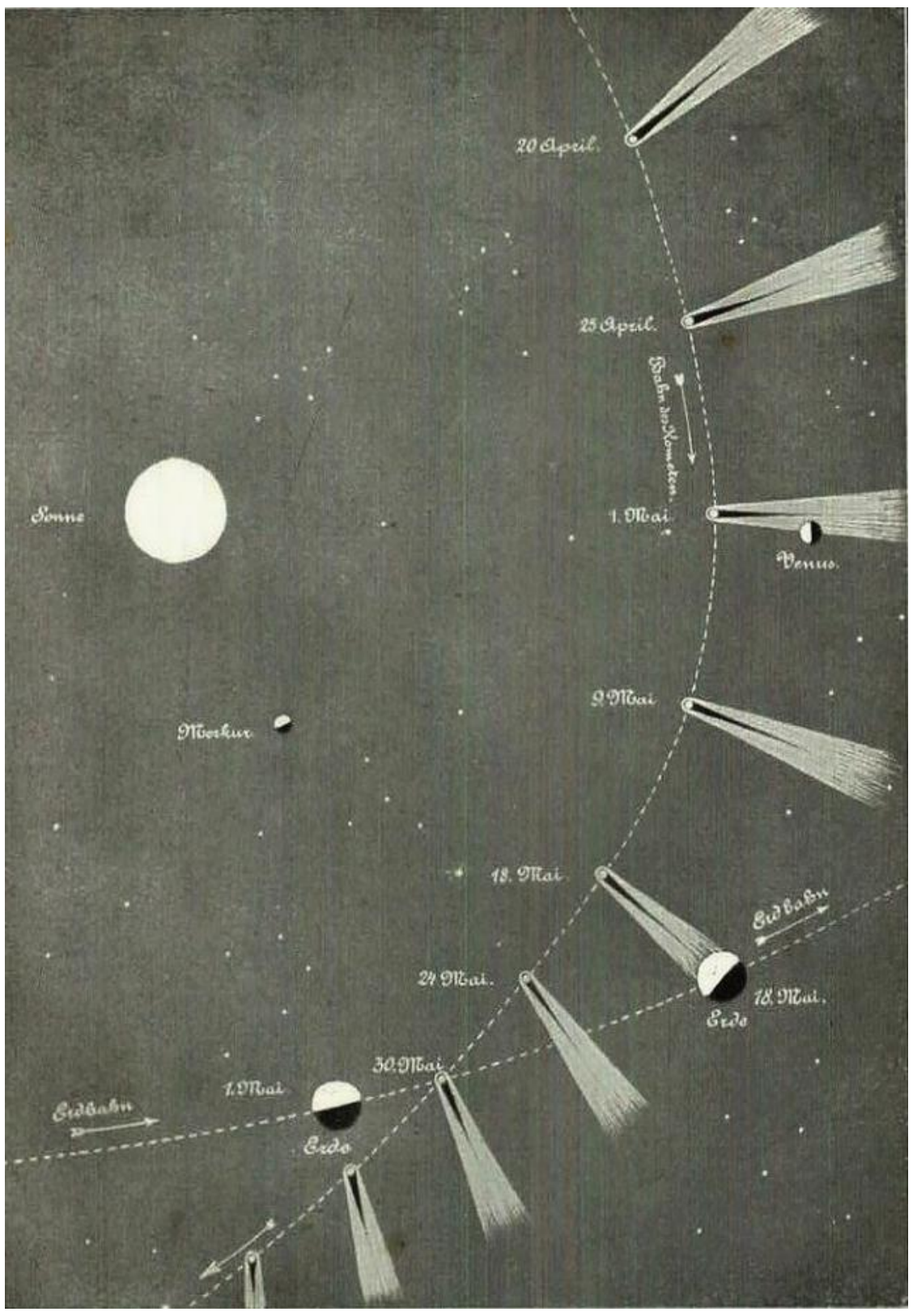

Figure 54: A diagram showing the paths of 1P/Halley and Earth in 1910. Note how the tail of the comet appears to brush the Earth on ~18 May (source: https://i.pinimg.com/736x/1f/69/cc/1f69cc00e8993438710a281c2ecf4cfc.jpg).

Many believed that when the tail of Comet 1P/Halley brushed the Earth on ~19 May 1910 (Figure 54), poisonous gases would infiltrate the atmosphere and kill many—or the comet would outright hit Earth and destroy it. Some people in New Zealand decided not to work that day, preferring to be at home with the family when the end of the world came. Others stated that they would rather not be working on the day when they would be burned alive (Auckland Star, 1910c). Some believed that the comet was actually an inhabited planet, and that the tail was caused by the inhabitants signalling to Earth with giant mirrors (By the Way, 1910). There is perhaps an eerie link to Comet C/1995 O1 (Hale-Bopp) and the Heaven's Gate cult here. Some entrepreneurs made a killing by selling 'comet pills' to protect people from such a worrisome end, or, as was stated, the pills were "… an elixir for escaping the wrath of the heavens." (Powell, 2022). One boy, hearing stories that the tail of 1P/Halley would wipe the hair off the head of those who had not cemented it in place with glue, smothered his head with golden syrup, much to the consternation of his teacher (*North Otago Times*, 1910b). In order to calm people's emotions, Clement Wragge (and other eminent New Zealand and international astronomers) wrote the following (The Comet's Tail, 1910):

> Sir,—-Calculations show that the earth should enter the upper part of the tail of Halley's comet at 1.52 p.m. on May 19 and emerge therefrom at 2.52 p.m., New Zealand standard time. Watch sun and star effects. Now people must not be afraid. Nothing very serious is likely to happen, for the tail is so thin and tenuous that this earth will tackle it like an elephant running amuck in a spider's web.

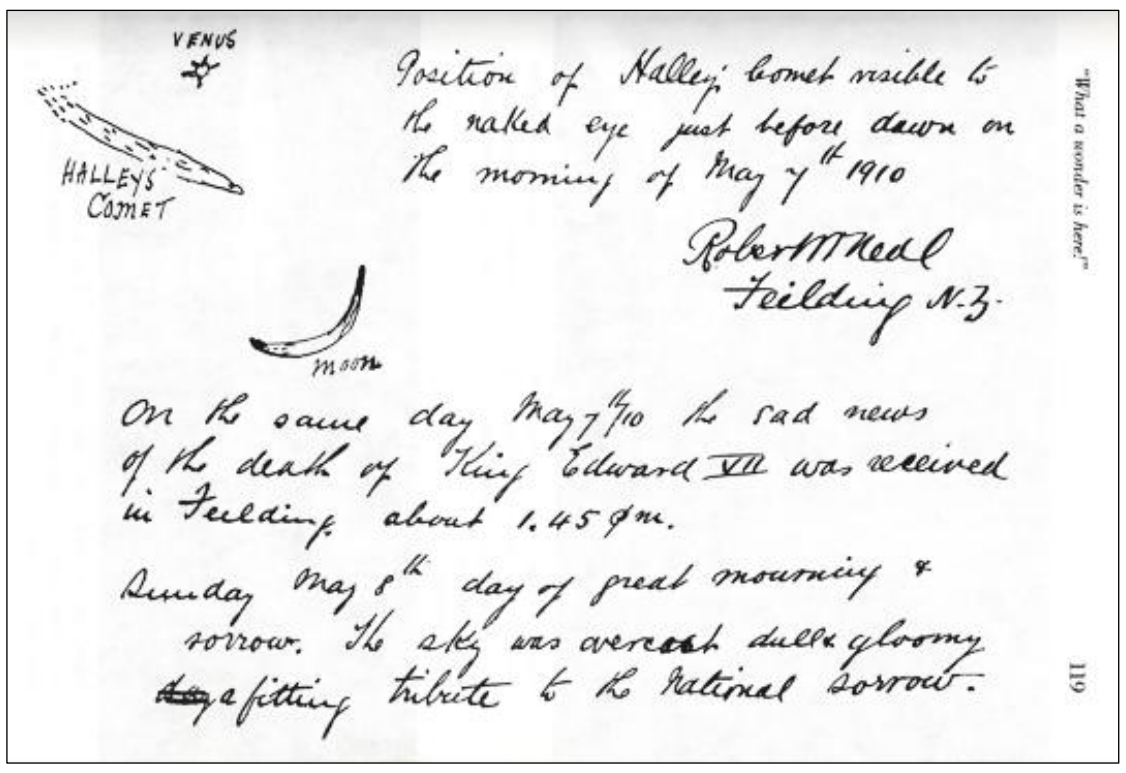

Figure 55: A sketch of where Comet 1P/Halley was located in relation to Venus and the crescent Moon on 7 May 1910, the day that King Edward VII died (after Mackrell 1985: 119).

On 6 May 1910, two weeks before perigee, England's King Edward VII died at the age of 68 from bronchitis complications that led to a heart attack (Wikipedia, 2026). Some associated Comet 1P/Halley with his death (The Late King, 1910). Figure 55 displays a sketch of the comet, the Moon and Venus by Robert W. Neal on the day the King died.

One Wellington pundit was so convinced that the comet would destroy Earth at 2:30 pm on Thursday 26 May 1910 that he bet £5 that it would. Of course, he ended up being wrong and paid his debt at 3pm! (It is Town Talk, 1910). Many predictors of comet-related doom would similarly later quietly disappear into anonymity whilst eating a serving of humble pie. Several months before, in mid-February 1910, during a fireworks display a mock Earth was blown up to represent 1P/Halley hitting Earth. This would surely provide comfort to the fearful (Amusements, 1910)!

Just north of the first author's city, Gisborne, the Māori, under the advice of a local elder, believed that Comet 1P/Halley was a presage to a huge tidal wave that would "... sweep the pakehas (white settlers) into the seas, and so dispose once and for all of the land question, the noxious weeds inspectors, and other barnacles of civilisation." (Local and General, 1910h).

## 6.2 Beliefs that Comet 1P/Halley Would Affect the Weather and World Events

Some believed that 1P/Halley's approach would affect the weather and cause earthquakes. In early May in Nelson, Joseph Taylor stated that the comet was "... one and a half times larger than the earth, and weighed fifteen thousand million tons …" He believed that as it approached Earth it would cause heavy rain and wind storms in July or August and cause heat resulting in famine and pestilence in tropical countries (Greymouth Argus and Blackball News, 1910). An earthquake in Vienna on 11 May lasting four minutes was also blamed on the comet (Earthquake, 1910). In the *NZ Truth*, under the heading The Comet as an Omen (1910) an unnamed writer produced the following list (Table 6) of 1P/Halley appearances and apparent associated disasters (tabulated by the authors).

Of course, most years in human history have had their fair share of wars and natural disasters, so attempting to link them to appearances of Comet 1P/Halley is a rather futile effort. Unless a comet hits the Earth, it has no influence on the affairs on Earth, although there is a small minority of humans who still view comets with superstitious fear.

## 6.3 Amusing Stories Relating to Comet 1P/Halley in 1910

On a humorous note are the number of dance parties held in 1P/Halley's honour. The *Evening Post* called them 'Comet Parties' (Women in Print, 1910). The attendees partied all night until the comet rose in the small hours before dawn. In late April 1910, at Rakaunui, one party was so crowded that people found it difficult to dance in the town hall (Local and General, 1910j). The article concluded: "It was not until the comet was blazing in the sky that one of the most enjoyable evenings ever spent in Rakaunui was brought to a successful conclusion." In early May, when P/Halley was rising at 4am, Mr and Mrs Park celebrated their daughter's 21st birthday with over 100 guests with games, competitions, dance and musical items until the arrival of the comet (Women in Print, 1910). Others, preferring an early evening's retirement, would wake very early and meet friends at designated viewing sights, look at 1P/Halley, and then go to someone's house for breakfast (*ibid.*).

In Fielding, two bank clerks got up early to observe the comet clad in "picturesque pyjamas". Afterwards, they realised to their horror when trying to go back to bed that they had locked themselves out of their house. 1P/Halley also caused a night of broken sleep for a correspondent on 19 April when he also attempted to view it (Halley's Comet, 1910k):

Table 6: On 14 May 1910, an unnamed correspondent using the title "The Comet as an Omen" wrote to the *NZ Truth* attempting to link the following 'catastrophes' on the Earth with appearances of Comet 1P/Halley in the sky.

| Stated Year | Correct Year | Kronk Reference | Events – As Listed in The Newspaper Article |
|---|---|---|---|
| 240 BC | 239 BC | 1999: 5 | Romans defeated Carthanians, concluding a 23 year long war. |
| 163 BC | 163 BC | 1999 | Judas Maccabaeus defeated the Syrians and occupied Jerusalem. |
| 87 BC | 86 BC | 1999: 17 | Rome was besieged and taken by four armies. |
| 12 BC | 11 BC | 1999: 24 | Drusus Senior invades Germany, establishing Roman supremacy there. |
| 66 AD | 66 AD | 1999 | Jerusalem was destroyed by the Romans in 67 AD. 1P/Halley appeared the year before. |
| 375 AD | 374 AD | 1999: 66 | The Huns invaded the Roman Empire and drove out the Goths. |
| 452 AD | 451 AD | 1999: 81 | Attila invaded Italy and Gaul. |
| 531 AD | 530 AD | 1999: 86 | Persia visited by the plague. |
| 610 AD | 607 AD | 1999: 97 | Hahomet(*sic*) received his first 'revelation' and began to preach in Mecca. |
| 1066 AD | 1066 AD | 1999 | William the Conqueror invaded and defeated England – portrayed on the Bayeux Tapestry |
| 1146 AD | 1145 AD | 1999: 200 | The second crusade began. Conrad III's army destroyed by the Greeks. |
| 1223 AD | 1222 AD | 1999: 209 | Khorassan and Persia conquered by Genghis Khan. |
| 1456 AD | 1456 AD | 1999 | Constantinople taken by the Turks (the comet was again unusually bright, as in 1066). |
| 1531 AD | 1531 AD | 1999 | Holland flooded due to dyke failure, 400,000 drowned. Great earthquake in Lisbon, 30,000 buried. |
| 1607 AD | 1607 CE | 1999 | The Dutch destroyed the Spanish fleet at Gibraltar. |
| 1758 AD | 1758 CE | 1999 | The English defeated the French fleet in the East Indies. The Russians and Austrians invaded Berlin. |
| 1835 AD | 1835 CE | 2003 | Chile experienced a devastating earthquake. |

I vowed to feast my eyes on that great celestial visitor this morning. Of course when I went to wind up the alarm clock I found that the youngest pet of the family had recently been using the coil for some wireless telephone experiments, and I went to bed trusting to memory to waken me. After having provided myself with one of Gaustad's best glasses [binoculars], with a 2½-inch objective (double), and having arranged the necessary clothing handy to my bed, I went off to sleep at 10.55 in calm confidence that I should awaken at the appointed hour. I slept blissfully at peace with all the world, and finally rose with a start. Having carefully attired myself as a preventive against the chilly air [May is late autumn in New Zealand], I looked at my watch. It was exactly five minutes past twelve! As I knew the comet was not due until 4 o'clock I laid down once more on the bed, and fell into a troubled sleep, in which I saw the whole heavens covered with a veil of gossamer, as delicate as a spider's web. Behind this veil I could see comets by the thousand—all sailing away in different directions, some making noises like the screech of a shell. The Southern Cross was spinning round on its axis like a Catherine wheel at Hopkins' fireworks, while Venus and Mars were bobbing up and down like floats on a stream. Many other scenes came into my line of vision, and I was eventually awakened by seeing what I took to be Halley's Comet sweeping right into the sun! I roused myself into action, and seizing my opera glasses, found the time was exactly 10 past four. Making my way outwards I saw Venus shining brightly, and searched in the vicinity for the comet. I tried everything with a nucleus, but although I made every allowance for the comet being head on, I could see nothing with the semblance of a tail. At last I got the glasses on what seemed a blur, and satisfied that it was the comet I returned to my bunk feeling disgusted that I had wasted so much valuable sleep to such little purpose. When I eventually made my appearance, somewhat tired and distrait after the broken night, I looked over the timetable of the comet which had appeared a few days previously in the Advocate, and found that I had concluded my final inspection of the heavens and gone back to bunk at least ten minutes before the comet was timed to appear above the horizon. No more comets for me!

Another New Zealand resident had an alarming night when he looked out of his window in Geraldine to view the comet and saw his neighbour disappearing into the gloom with a sack of his coal (Omnium Gatherum, 1910). Papers Past revealed numerous humorous stories concerning people's experiences with Comet 1P/Halley. Not only that, but the name 'Halley's Comet' was given to many racehorses and racing dogs across New Zealand at the time (Dunedin Coursing Club, 1910).

### 6.4 Advertisements in New Zealand Newspapers using Comet 1P/Halley to Promote Sales

Coupled with the onslaught of comet observations, omens, and amusing stories in early- to mid-1910 about 1P/Halley, there were numerous newspaper articles that used the comet in a bid to bolster sales.

As the comet's brightness increased, so did the number of adverts. A range of retailers cashed in on the popularity of 1P/Halley, such as the 'Planet' sewing machine, with obvious astronomical links between 1P/Halley and planets. Watchhorn's Clothing was said to eclipse Halley's Comet as the talk of the town. Tie's Cheap Meat was purported to have more far-reaching consequences for people than Halley's Comet did. Jewellers were also making Halley's Comet brooches. 1P/Halley was also used for advertising photographic studios, Lipton Tea, and movies such as Cleopatra (1910).

In addition, Frank Shaw Chemist said that they would help with disrupted sleep from observing Halley's Comet and conversely Milners Alarm Clock would wake you in time to view the comet.

The comet was also used to flagrantly promote shoe repair, motor cycles, furniture, cocoa, bridal photos, Red Cross ointment, shoe polish, real estate, Halley's Comet postcards, 'comet soup' and much more.

More related to 1P/Halley were binocular and telescope advertisements (e.g. see Figure 56). Also pictured is a Milner's Alarm Clock advertisement.

## 7 COMET 1/P HALLEY'S INFLUENCE IN THE FORMATION OF ASTRONOMICAL SOCIETIES IN NEW ZEALAND

On 3 May 1910, the *Greymouth Evening Star* (Notes and Comments, 1910) reported:

> The boom in astronomy occasioned by the reappearance of Halleys (*sic*) comet, continues with unabated vigor and the

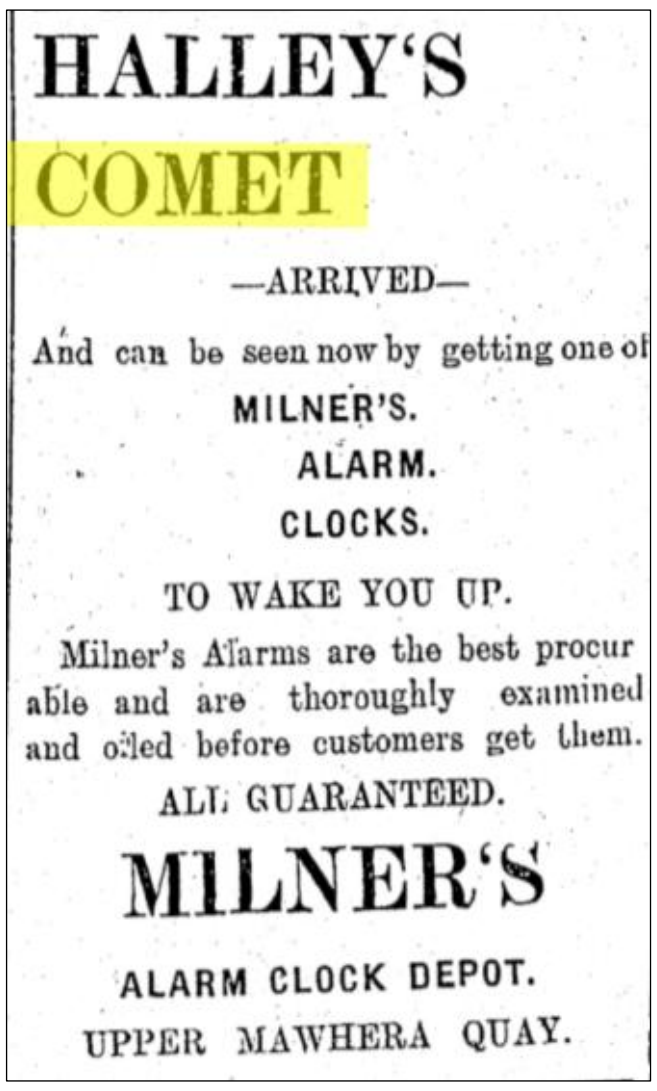

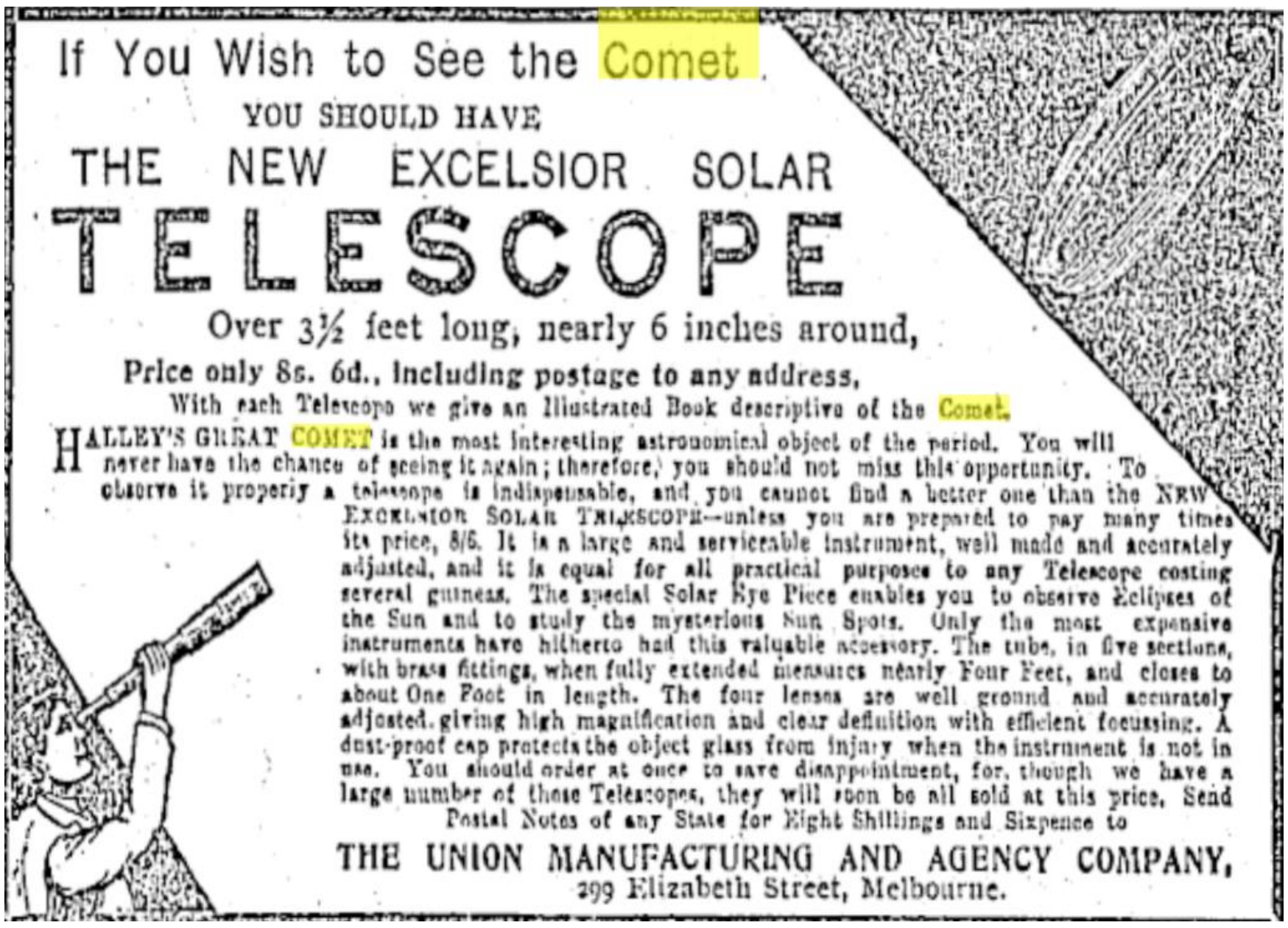

Figure 56: Examples of two advertisements in 1910 that readily incorporated Halley's Comet into their advertising campaigns. The yellow highlight is a result of the keyword searches in Papers Past.

> fact that this month we are to experience eclipses of both the sun and moon should lend added interest. Indeed we might go so far and say that four interesting events are predicted to occur during May., viz., a total eclipse of the sun, a total eclipse of the moon, a transit of Halley's comet across the sun, with probable passage of the earth through the tail, and the appearance of the comet as an evening object at its greatest brightness.

With the upsurge of interest in the heavens due to 1P/Halley's appearance in 1910, a reasonable question is whether this interest translated into New Zealanders forming local astronomical societies. We utilised Papers Past searches as well as contacting local astronomical societies across New Zealand for historical input. One problematic issue was that there may have been societies that formed as a result of the comet but that no longer exist today. This will be addressed in a future paper on the history of New Zealand astronomical societies. Through contact with existing astronomical societies, it was found that the vast majority formed well after the 1910 appearance of 1P/Halley, but the authors found that possibly three societies were founded around this period possibly as a result of the comet.

## 7.1 Wanganui Astronomical Society

In 1901, a public lecture on comets (including 1P/Halley) by Professor McLaurin in Wanganui was intended to act as a catalyst to the proposal that that city form an astronomical society (The Winter Lectures, 1901). However, due to the lateness of the hour, the question was postponed until a month later with pleasing results. The Wanganui Astronomical Society formed in 1901. Having the driving force of Joseph Ward in the city undoubtedly helped with the formation of the Society. Several years later Ward secured the 9.5-inch Cooke refractor, and he was elected Observatory Director.

## 7.2 The Palmerston North Astronomical Society

The Palmerston North Astronomical Society had its roots in the Manawatu Philosophical Society which purchased Revered Kennedy's British-made 6.5-inch (17-cm) reflecting telescope and installed it in an observatory in 'The Square' in the centre of town in the early 1900s (Seymour, 1985). See Figure 57. Captain Dudley Hewitt then offered to "... occupy the position of honorary astronomer, and in many an interesting little talk explained the mysteries of the heavens to visitors to the observatory." (Seymour and Orchiston, 2021). He did this successfully until his passing in 1913.

## 7.3 The Wellington Astronomical Society

Heritage New Zealand (2026) states that, "In 1910 members of the Wellington Philosophical Society formed the Astronomical Section in order to build a public astronomical observatory, a facility the city was lacking." As stated earlier, the lack of an observatory in Wellington was something of an embarrassment to the city and the astronomers that resided within. Two years later, in 1912, after 1P/Halley had made its historic visit, the Thomas King Observatory (formerly the King Edward VII Memorial Observatory

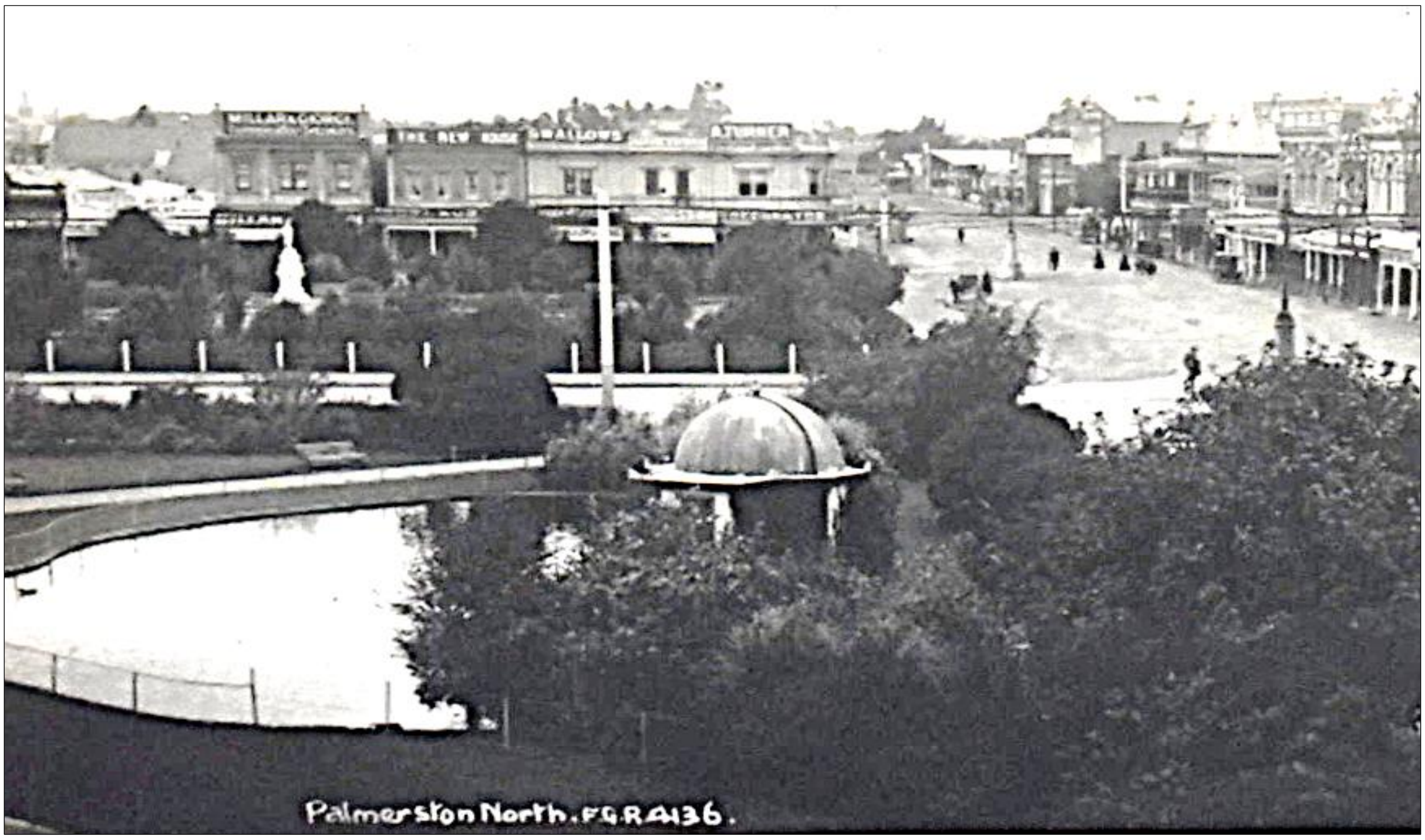

Figure 57: The Manawatu Philosophical Society's observatory in The Square in Palmerston North in the early 1900s. It housed the 6.5-inch reflector that previously belonged to Reverend Kennedy (see Seymour, 1985). In Palmerston North it was widely used to observe 1P/Halley (photograph: Seymour Collection).

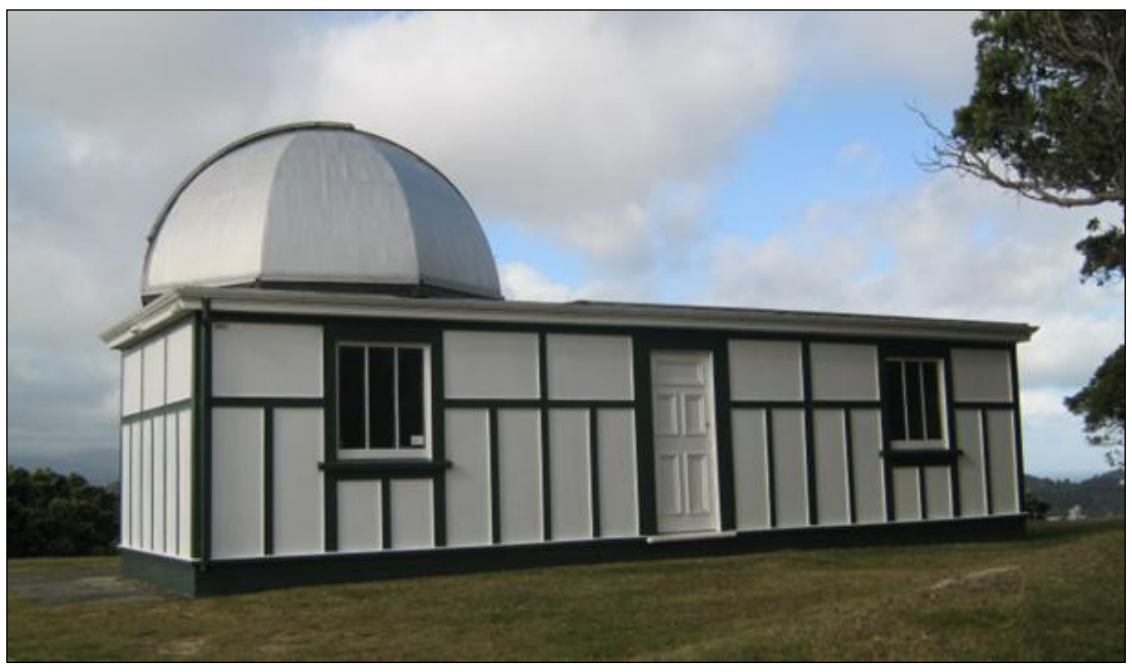

Figure 58 (left): The historic Thomas King Observatory was built in 1912 by the Astronomical Section of the Wellington Philosophical Society. Photographed in 2012, after it had been refurbished by Carter Observatory (photograph: Wayne Orchiston).

until 1918) was constructed in the Wellington Botanic Garden (Figure 58). The Observatory's telescope was a 5-inch (13-cm) Cooke refractor which was replaced in 1916 by a 5.5-inch (14-cm) Grubb telescope which belonged to Thomas King until his death that year (Seymour, 1995). As stated, the 9-inch refractor used at Meeanee was purchased by the Wellington City Corporation and housed in the 'Green Tin Shed' Observatory not far from the Thomas King Observatory (Figure 59). It was subsequently installed in the newly built Carter Observatory in 1941, which was sited just metres from the 'Green Tin Shed' Observatory and near to the Thomas King Observatory (see Figure 60). The Society maintained close links with Carter Observatory after it opened.

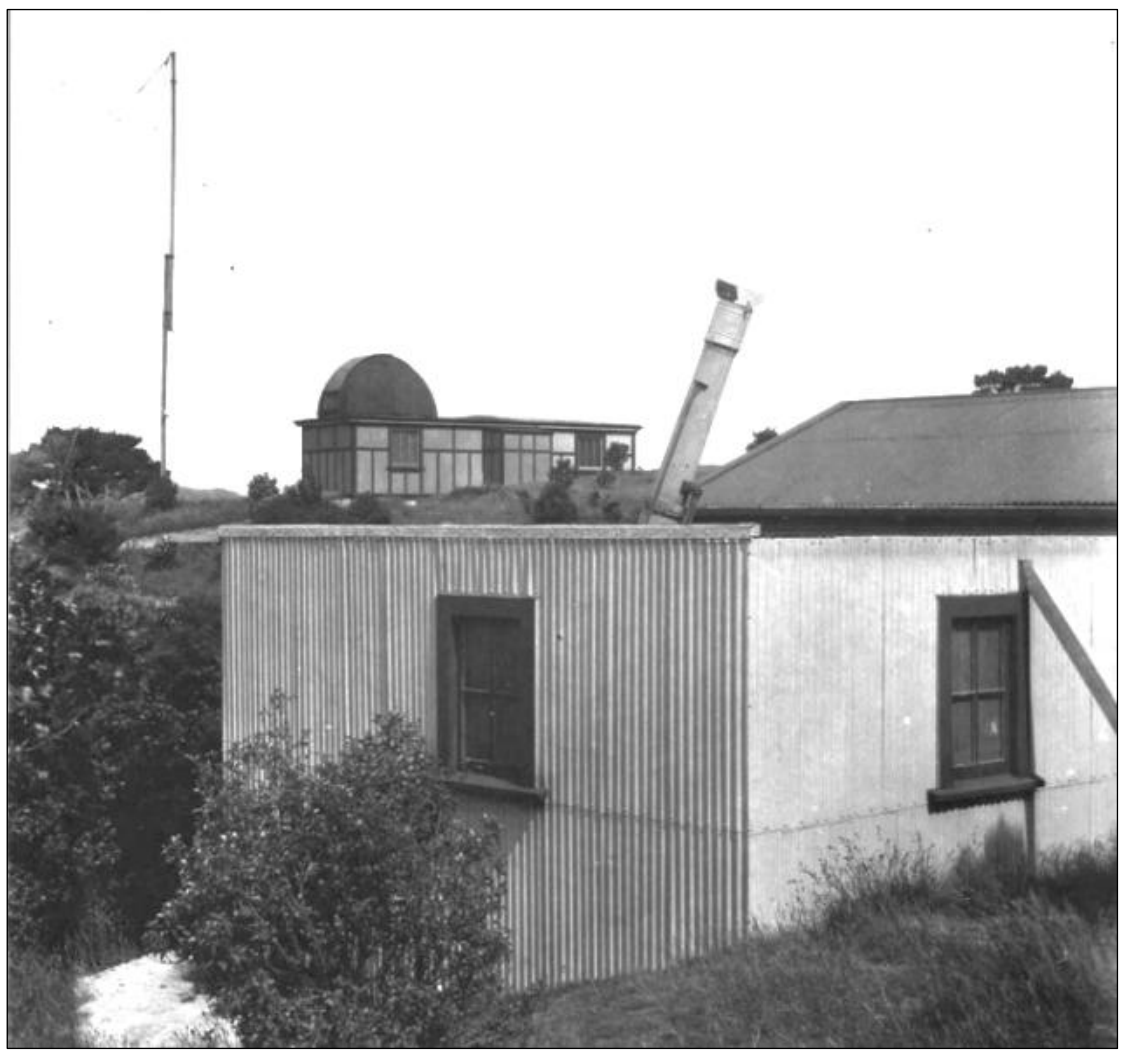

Figure 59 (left): The 9-inch Cooke refractor poking out the top of the 'Green Tin Shed' Observatory. The Thomas King Observatory is on the hill in the background (source: Orchiston Collection).

## 7.4 The Dunedin Astronomical Society

On 18 June 1910, Robert Gilkinson wrote a letter to the editor of the *Evening Star* under the heading, 'Wanted, an Observatory for Dunedin' (Wanted, an Observatory for Dunedin, 1910). In the lengthy article the first sentence alluded to the recent visit of Halley's Comet and how it generated interest in astronomy with local citizens. The question was asked as to why, with the city's Otago University and the large number of students interested in the topic, an astronomical society and observatory with suitably equipped telescope was not in existence. Gilkinson then used the formation of the Wanganui Astronomical Society in 1901 as an example of how it could be done. He also mentioned the purchase of the 9.5-inch refractor for £450 and self-funding by members for building of the £350 observatory. Joseph Ward was described as the President and Honorary Director of the Wanganui Observatory. The question was directed at the University council and interested parties.

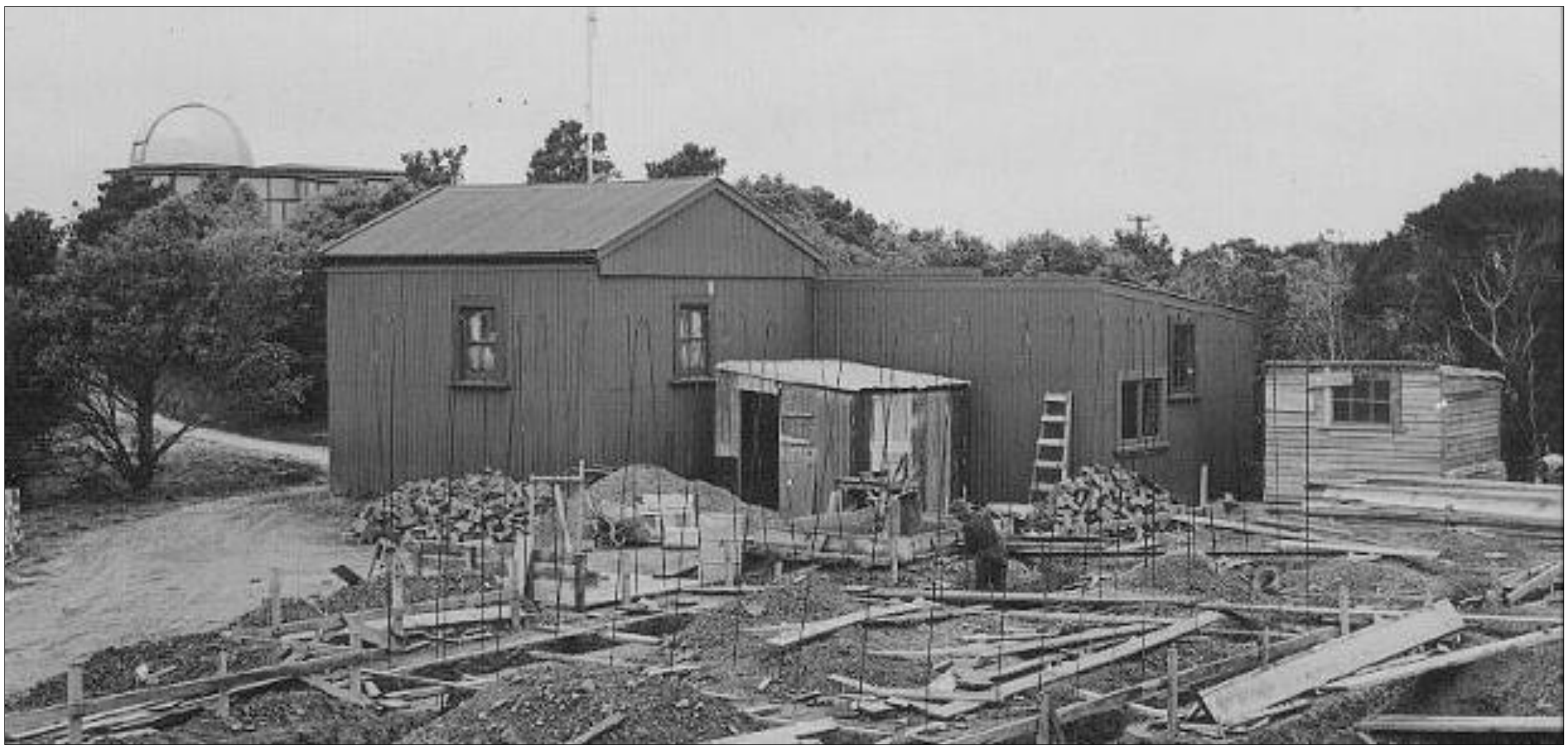

Figure 60: A photograph taken during excavation of the Carter Observatory foundations (in the foreground), showing the adjacent Wellington City Observatory (aka 'Green Tin Shed'), and on the hill behind the Thomas King Observatory (Orchiston Collection).

Two months later, on Thursday 18 August 1910, an advertisement for a lecture by the Reverend P.W. Fairclough on astronomical discoveries was placed in the *Evening Star*. It was intended that at the conclusion of the meeting, a proposal would be brought forward to form an astronomical society (Lectures, 1910).

On 19 August 1910, the *Evening Star* published an article (Astronomy, 1910) which reported that approximately 100 people attended the previous night's meeting at the Art Gallery. A thorough recount of the lecture by Fairclough was given and at the end of the meeting Gilkinson moved "That it is desirable that an Astronomical Society be formed in Dunedin." Mr A. Meggett seconded the motion which was carried unanimously. About 30 people were appointed as a Provisional Committee and to draw up a constitution which would be reported on at a future meeting. At the time, the University already had a 2.5-inch (6.4-cm) transit telescope, an astronomical clock, and a 3-inch (7.5-cm) refractor bequeathed by Mr Beverly. A small observatory with a dome and "... moderately sized …" telescope was planned for the future (*ibid.*).

Table 7 following, provides a list of current astronomical societies around New Zealand and when they were formed.[3] Most of the data were collected from individual societies and from earlier research carried out by the second author of this paper.

## 8 COMET 1P/HALLEY: GONE FOR 1910, BUT RETURNING IN 1985–1986

Probably the key finding when comparing the apparitions of 1P/Halley in 1835, 1910, and 1986 is the evolutionary change in technology utilised to observe it. In 1835 it was with pencil on paper, in 1910 it was photographic plates and film, in 1986 it was (amongst other things) digital photographs of the comet transmitted across space from space probes to astronomers on Earth eager to finally view close-up images of a comet's nucleus (Figure 61). These images would be something that the Marist Brothers, Westland and Grigg would have only dreamed of. In this paper we focused on the 1910 visit by 1P/Halley; exploring both visual and photographic observations. These observations were disseminated mainly through submitted reports to newspapers. In 1835, few New Zealand newspapers existed, so reporting sightings was a challenge. In 1985–1986, observations were distributed throughout New Zealand via newspaper reports, as well as electronic communication (fax, mobile phone, television, etc). This advancement in technology allowed more observations to be submitted to international journals. Who knows what it will be like

Table 7: A list of current New Zealand astronomical societies and when they were founded. This list is based mainly on the Royal Astronomical Society of New Zealand's 2026 list of astronomical societies, and focuses on societies rather than observatories.

| Year | Astronomical Society | City | Island | Notes |
|---|---|---|---|---|
| 1901 | Wanganui AS | Wanganui | North | Associated with 1P/Halley |
| 1900s | Palmerston North AS | Palmerston North | North | Part of Philosophical Society |
| 1910 | Wellington AS | Wellington | North | Associated with 1P/Halley |
| 1910 | Dunedin AS | Dunedin | South | Associated with 1P/Halley |
| 1919 | New Plymouth AS | New Plymouth | North | Started by Rev. Oscar Blundell |
| 1923 | Auckland AS | Auckland | North | Started by Prof. Percy Burbidge |
| 1926 | Hawera AS | Hawera | North | Started by G. Townsend |
| 1933 | Hamilton AS | Hamilton | North | Started by Allan Bryce |
| 1948 | Canterbury AS | Christchurch | South | Started by Robert Francis Joyce |
| 1956 | Nelson AS | Nelson | South | Started by Thomas Cole |
| 1960 | Whakatane AS | Whakatane | North | Started by Clive Kingsley-Smith |
| 1961 | Southland AS | Invercargill | South | Started by G. Hall-Jones & R. Beck |
| 1962 | Gisborne AS | Gisborne | North | Group initiative |
| 1986 | Rotorua AS | Rotorua | North | |
| 1980s | Hawkes Bay AS | Hawkes Bay | North | |
| 1990 | Horowhenua AS | Horowhenua | North | |
| 1997 | Phoenix AS | Carterton | North | Started by R. Hall and K. Leather |
| 2001 | Tauranga AS | Tauranga | North | Started by Denis Twiss |
| 2007 | Skydome Observers Dargaville | Dargaville | North | |
| 2012 | Aoraki Mackenzie Int.D.Sky Res. | Tekapo | South | Group initiative |
| 2017 | Wairarapa Dark Sky Ass. | Wairarapa | | Started by D. Kershaw, B. Lilley & L. Mauger |
| 2018 | Hibiscus Coast AS | Hibiscus Coast | North | Started by James Smith |
| 2020 | Kaikoura Dark Sky Trust | Kaikoura | South | Started by Nicky McArthur |
| 2020 | Omaka Obs Charity Trust | Marlborough | South | Started by Lee Harper |
| 2024 | Oxford Dark Sky | Oxford | South | |
| 2025? | Central Otago Dark Skies Soc | Otago | South | |
| Unsure of Establishment Date | | | | |
| | North Otago AS | Otago | South | |
| | Northland AS | Whangarei | North | |

when 1P/Halley approaches Earth next in 2061?

At least three astronomical societies were formed in New Zealand around 1910 as a result of the visit by 1P/Halley. Those who saw the comet in 1910 (or the Great January Comet earlier that year) passed the wonderment of the comet on to eager ears, especially as it drew close to its 1985–1986 visit. The media was also naturally enthusiastic to spread the news about the comet—and what the public may see. Expectantly, there was an increased public attendance at observatories as the comet brightened. There was also a definite increase in societal telescopic instrumentation to show the public views of the comet. These visits resulted in increased membership numbers for many astronomical societies. Sadly, the 1985–1986 visit was quite a disappointment and nothing like the 1910 visit. A paper on the 1985–1986 apparition of 1P/Halley and the influence that it had on New Zealand astronomical societies, observatories, and the general public is long overdue.

## 8 CONCLUDING REMARKS

1P/Halley put on a spectacular display in its 1835 apparition. The superlatives from those who witnessed it made the next generation of observers eager to view this once-in-a-lifetime comet—although, as we revealed in this paper, some were fortunate enough to see it twice. Excitement was raised through newspaper articles, books and public lectures. When Comet 1P/Halley did return in 1910, rather surprisingly it was superseded by another brilliant comet, namely the Great January Comet of 1910 (C/1910 A1). Not many New Zealanders viewed this comet that appeared in southern skies and brightened to naked-eye visibility. This may have been due to it appearance in January, summer in the Southern Hemisphere. Before Daylight Saving was introduced in New Zealand, not many people were awake before the early sunrise to view the comet. Those who did witness this spectacle often incorrectly assumed that it was 1P/Halley (Drummond et al., 2025). We examined observational reports of Comet 1P/Halley from New Zealand during its 1910 apparition and found that the vast majority were not submitted to international observatories or journals. For this reason, observations from New Zealand are not found in such classics as *Physical Characteristics of Comets* (Vsekhsvyatskii, 1964) or *Cometography*, *Vol-*

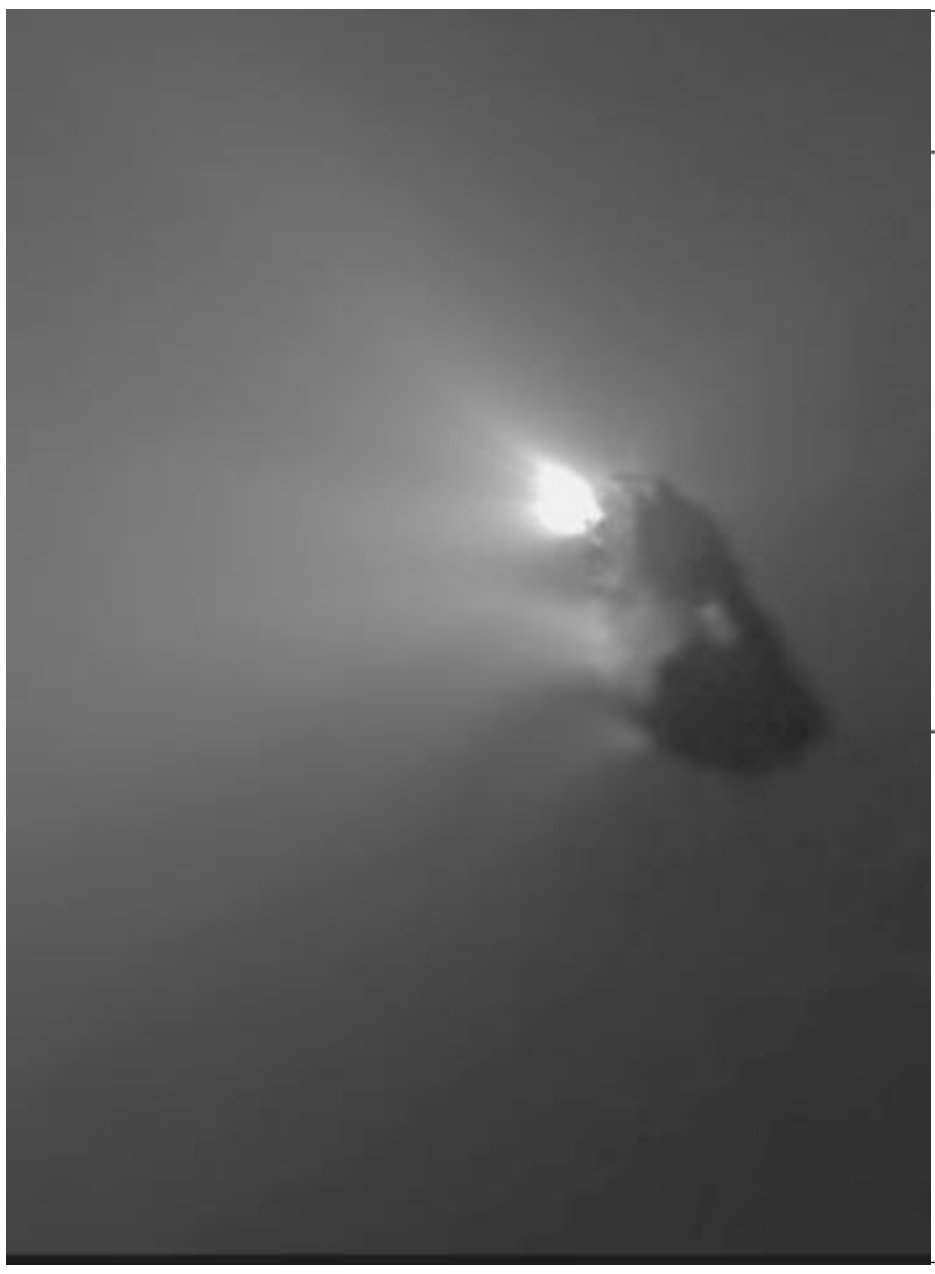

**COMET HALLEY: HISTORICAL AND PHYSICAL DATA**

| Historical Data: | |
|---|---|
| Earliest probable recorded apparition | 240 B.C. |
| Number of recorded apparitions From 240 B.C. to 1910 A.D., only the 164 B.C. apparition was not recorded | 28 |
| Shortest period between returns to perihelion | 74.42 years (1835-1910) |
| Longest period between returns to perihelion | 79.25 years (451 —530) |
| Closest approach to the Earth | 0.04 Astronomical Units (6 million kilometres) on April 11, 837 |
| Longest angular tail length recorded | 93° (mid-April, 837) |
| Brightest apparent magnitude recorded | —3.5 (April 11, 837) |
| **Physical Data:** | |
| Estimated diameter of nucleus | 5 km |
| Estimated density of nucleus | 1 gm $cm^3$ |
| Estimated rotation period of nucleus | 10.3 hours |
| Observed spectra in 1910 | CH,CN,C2,C3,Na D,CO+,N2+ |
| Observed tails | Type I ion and Type II dust |
| Associated meteor streams | Eta Aquariid (early May) & Orionid (late October) |
| Orbital velocity at perihelion | 54.55 km/sec |
| Orbital velocity at aphelion | 0.91 km/sec |

Figure 61a (left): The nucleus of 1P/Halley as photographed by the Giotto space probe on 14 March 1986 (credit: https://www.esa.int/ESA_Multimedia/Images/2012/11/Comet_Halley_close_up); Figure 61b (right): some of the findings from the space probe missions to 1P/Halley in 1986 (NASA).

*ume 3 (1900 - 1932)* (Kronk, 2007). In order to distribute observations throughout the country, most New Zealand observers published them in local or national newspapers. Utilising the on-line search tool, Papers Past, we found 1,157 articles relating to Halley's Comet between January 1908 and December 1912 (inclusive). Most of these were observations from profess-sional and amateur astronomers, as well as the public. Many of these matched favourably when compared to international observations of the same time. We also found a gradual rise in the number of articles published about 1P/Halley as the comet approached. There was a sudden upward trend when Max Wolf photographically recovered it from Germany in September 1909. Professors Burnham and Barnard visually detected it with the 40-inch Yerkes refractor a few days later. Several New Zealand astronomers tried to view it as it approached the inner Solar System with the first being the Marist Brothers at Meeanee Observatory near Napier. This was done photographically with a 3.5-inch (8.9-cm) Dallmeyer portrait lens on 14 October 1909. The first to visually see it from New Zealand was Joseph Ward of Wanganui Observatory who saw it on 3 January 1910. It was described as a "... very faint nebulosity."

As time progressed and the comet came closer to the Sun and Earth, the brightness began to increase dramatically. It became a naked-eye target in early to mid April 1910. In the ensuing months it was a spectacular comet, and reached naked-eye brightness during the daytime. The length and brightness of the tail just prior to perigee was measured to be over 100°. The comet had a profound effect on the public. Thousands of New Zealanders (and international viewers) woke before dawn to etch the view into their memories. We found that as a result of the appearance of 1P/Halley, there was an upsurge of interest in astronomy. At least three astronomical societies formed around New Zealand during the 1910 epoch as people began to meet on a regular basis to ponder the heavens. 1P/Halley also became the 'toast' of many as people revelled all night until the comet rose prior to dawn. Not only were people using their eyes to view the comet, but those fortunate enough to possess photographic equipment took photographs of the comet, some of which almost rivalled those obtained at professional international observatories. The Marist Brothers at Meeanee Observatory led this charge. Other prominent New Zealand astronomers who studied the comet were Charles Westland, John Grigg and Joseph Ward.[4]

When 1P/Halley transited the Sun on 21 May 1910, many attempted to view the nucleus against the bright disk. No one detected it due to the smallness of the nucleus and the distance it was from the Earth. This was a common occurrence around the world. As the comet came between the Sun and Earth, many feared the long tail and the gases contained therein. Numerous stories were uncovered that revealed the anxiety that the comet produced among the New Zealand public. Some believed that the tail of the comet would cause atmospheric, meteoric and magnetic phenomena and eagerly looked to see if these eventuated. We highlighted how some believed that they witnessed such events, but most people saw nothing unusual on that May day. After the transit and perigee, the comet swept into the evening sky in late May. Those who saw it in the morning sky several days before commented on how it appeared to be fading quite rapidly, this being due to the increasing distance between it and the Sun and Earth. By 22 June, it was described as being "... just about out of sight now; it is very dim ...." (Our Young Folks Column, 1910). Seventy-five years would pass before it was seen again with the naked-eye. Once again, many New Zealanders were eager to witness the 1985–1986 apparition. We hope that the observations and recollections of these people will be the focus of a future paper by the authors.

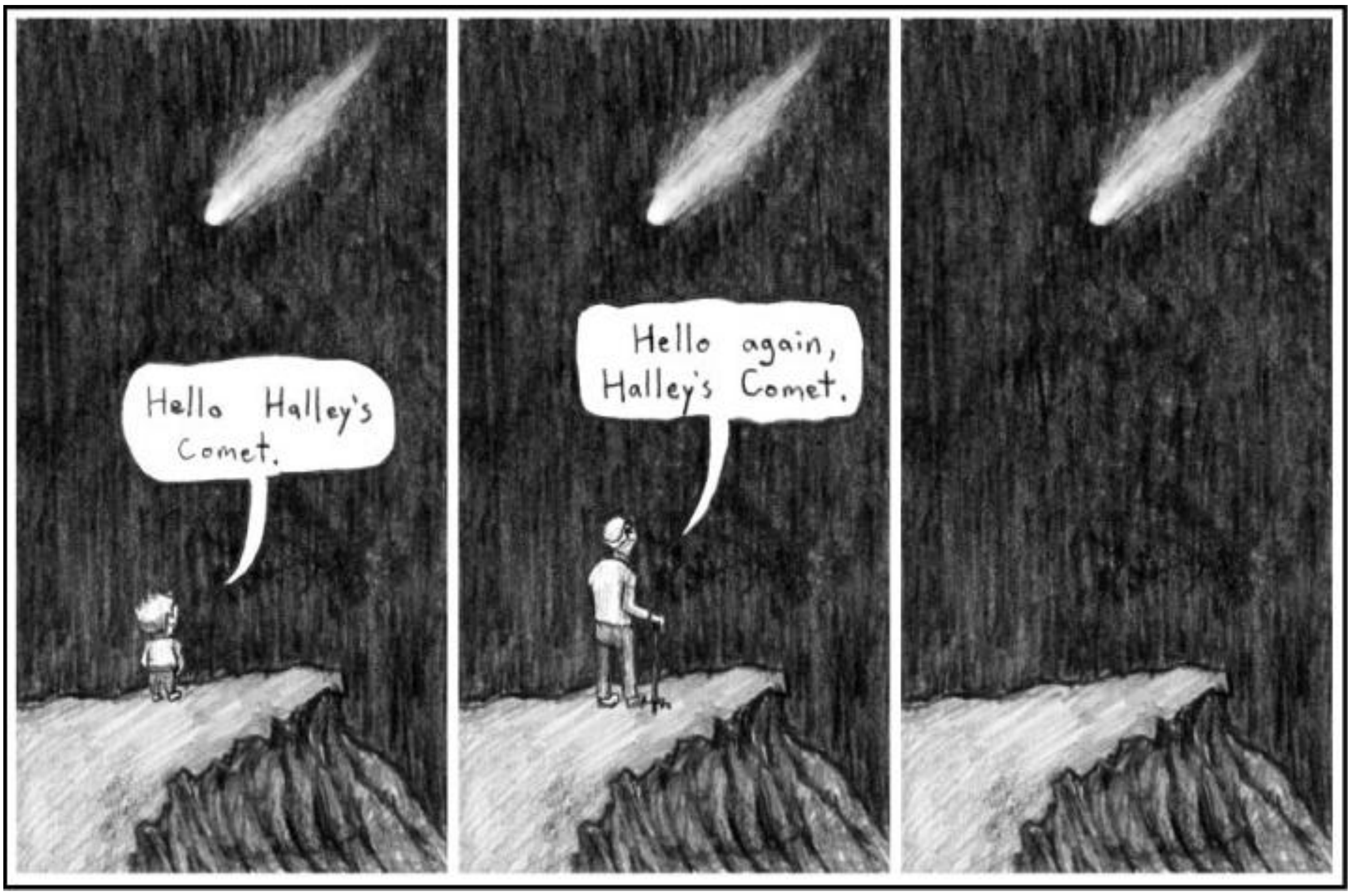

Figure 62: Sad, but true (source: https://www.todayifoundout.com/wp-content/uploads/2013/10/halleyscomet.jpg).

Meanwhile what does the future hold for those of us who are reading this paper right now? How many of us will see Comet 1P/Halley in 2061? I guess for most of us Figure 62 is all too true!

## 9 NOTES

1. In other related studies by the authors they have reviewed NZ cometary astronomy over the past 750 years (Drummond and Orchiston, 2021) and examined comets and Māori cometary astronomy (Orchiston, 2016: 55–63; Orchiston and Drummond, 2019), plus NZ's six comet discoverers (Drummond et al., 2026b), NZ observations of various Great Comets of the nineteenth and early twentyeth centuries (Drummond, 2023; Drummond et al., 2025; Orchiston, 1997a; 2016: 565–583; Orchiston and Drummond, 2022a; 2022b; 2022c; 2024b; Orchiston and Orchiston, 2024; 2026; Orchiston et al., 2020b), other NZ cometary observations (Kronk et al., 2026; Orchiston, 1983; 2016: 509–522; Orchiston and Drummond, 2024a; Orchiston et al., 2020a), NZ paintings of comets (Orchiston et al., 2024c), and NZ meteors (Baggaley and Orchiston, 2024; Orchiston et al., 2021; Taibi et al., 2026), and meteorites (Evans and Orchiston, 2023; Orchiston, 1997b; Orchiston and Drummond, 2022d; Orchiston et al., 2024b; 2024b; 2026a; 2026b; Warnes et al., 1998). They have also prepared a biographical conference poster about Rod Austin (Drummond et al., 2026a).
2. Unlike the Meeanee telescope, the historic Wanganui refractor is an internationally significant 'type specimen' as the first relatively large refracting telescope ever built with an all-metal English Equatorial Mounting—earlier telescopes had wooden mountings that were subject to flexure (Orchiston, 2001a). Yet, like the Meeanee telescope, the Wanganui refractor also built an astronomical reputation back in England (Orchiston and Kapoor, 2023; Orchiston and Skilton, 2025), before transferring to New Zealand and subsequently undergoing a timely major overhaul (see Whanganui.govt.nz, 2020).

3. If any dates regarding the formation of the astronomical societies in Table 7 are incurrect, or any societies are missing, can you please inform the lead author.
4. This study is an isolated snapshot of a New Zealand reaction to a periodic astronomical spectacle. Further national studies like this need to be published so we can identify the common denominators of the public, amateur astronomy and professional astronomy responses to Comet 1P/Halley in 1910. Smith (1986) and Kapoor (2023) have already published parallel studies of Canadian and Indian responses to the comet at this time, and we now welcome further comparable perspectives from other former British colonies (such as Australia, Hong Kong and South Africa—all of which were astronomically-active at this time). After this, we welcome similar studies involving other colonial nations, such as France, Germany, Italy, Portugal, Spain and the USA.

## 10 ACKNOWLEDGEMENTS

We would like to sincerely thank the following for help during the researching of this review paper: Elizabeth Charlton (Provincial Archivist at the Marist Archives, Wellington), Ian Cooper (Palmerston North), Gary Kronk and especially his *Cometography* (Volume 3), Jennie McCormick (Auckland), and those from New Zealand astronomical societies who responded to our emails. We found Brian Mackrell's (1985) book *Halley's Comet Over New Zealand* very useful, given the apparent disappearance of all of C.J. Westland's original observational records.

We also acknowledge the extensive use that we made of ADS, and we wish to thank the National Library of New Zealand for providing access to the invaluable Papers Past resource. Without it, this study would have been impossible.

We wish to thank the late Frank Andrews, Marist Archives (Wellington, NZ), Lowell Observatory, Jenny McCormick, Mitchel Library, Sydney), NASA, the late John Seymour, Victoria University of Wellington (NZ) and Wanganui Observatory Archives for kindly providing copies of Figures 7, 8, 13, 22, 24, 25, 33, 35, 38, 41, 45, 46, 50, 57 and 61.

Finally, this is the third and final paper in the first author's PhD by Publication on historical aspects of New Zealand cometary astronomy, though the School of Astrophysics at the University of Southern Queensland in Australia. He is grateful to the University for providing travel funding and other research support.

**John Drummond** became fixated with astronomy at the age of ten when his mother pointed out the Pot in Orion to him. From that moment on he was hooked on the Universe. Joining the Junior Section of the local Gisborne Astronomical Society not long after, John would regularly do group meteor watches, telescope viewing and listen to astronomy talks. He also developed an interest in photography, and it was not long before he combined these two interests and began astrophotography. John's photographs have been used in many overseas books and magazines—and were used on two New Zealand stamps. He was the Director of the Royal Astronomical Society of New Zealand's Astrophotography Section for thirteen years until 2018. He is currently the Director of the Society's Comet Section.

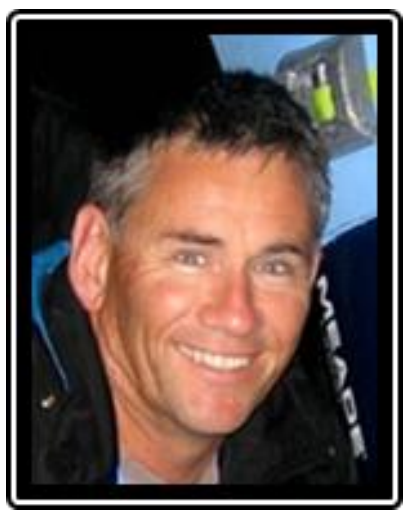

John lives about 10km west of Gisborne, on the east coast of the North Island of New Zealand, and has a range of telescopes up to 0.5 metres in aperture at his Possum Observatory. He regularly images with these telescopes and CCDs, carries out astrometry of comets, asteroids and NEOs, and sends his observations to the IAU Minor Planet Center. He has also confirmed several comets and co-discovered about 20 exoplanets in collaboration with the Ohio State University. He runs Gisborne Astro Tours from his observatories. John has authored or co-authored more than 60 research papers, many of the more recent ones being on the history of New Zealand astronomy.

John is a Past President and Past Secretary of the Royal Astronomical Society of New Zealand, and in 2019 he was made a Fellow of the Society. In 2016 he was awarded an MSc (Astronomy) by Swinburne University in Melbourne (Australia), and currently he is researching aspects of the history of cometary astronomy in New Zealand as a part-time off-campus internet-based PhD student affiliated with the Centre for Astrophysics at the University of Southern Queensland (Australia). His supervisors are the co-authors of this paper.

When he is not doing astronomy, John is a secondary school science teacher. He also enjoys surfing the great waves of Gisborne and pottering around on his small farm tending to his sheep.

**Professor Wayne Orchiston** was born in Auckland (New Zealand) in 1943, and has BA First Class Honours and PhD degrees from the University of Sydney. Currently, he is employed by the University of Science and Technology of China in Hefei as the Co-editor of the *Journal of Astronomical History and Heritage*. He is also an Adjunct Professor of Astronomy in the Centre for Astrophysics at the University of Southern Queensland (USQ) in Toowoomba, Australia. Formerly, Wayne worked at observatories, research institutes and universities in Australia, New Zealand and Thailand.

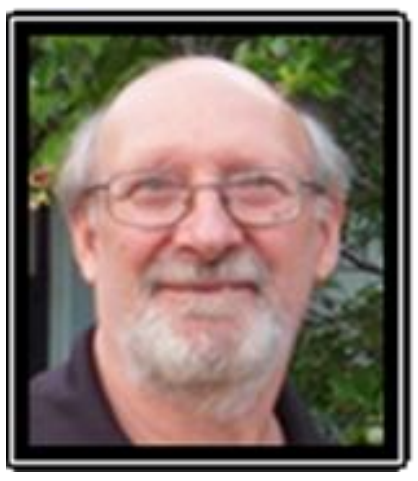

Over the past two decades Wayne has supervised more than 35 Master of Astronomy and PhD history of astronomy research projects through three different Australian universities.

Wayne has wide-ranging research interests and more than 500 publications, mainly about historic transits of Venus; historic solar eclipses; historic telescopes and observatories; the emergence of astrophysics in Asia and Oceania; the history of cometary and meteor astronomy; the astronomy of James Cook's three voyages to the Pacific; amateur astronomy and the amateur–professional interface; the history of meteoritics; Indian, Southeast Asian and Māori ethnoastronomy; and the history of radio astronomy in Australia, France, India, Japan, New Zealand and the USA.

Recent books by Wayne include *Exploring the History of New Zealand Astronomy …* (2016, Springer); *John Tebbutt: Rebuilding and Strengthening the Foundations of Australian Astronomy* (2017, Springer); *The Emergence of Astrophysics in Asia …* (2017, Springer, co-edited by Tsuko Nakamura); *Exploring the History of Southeast Asian Astronomy …* (2021, Springer, co-edited by Mayank Vahia); *Golden Years of Australian Radio Astronomy: An Illustrated History* (2021, Springer, co-authored by Peter Robertson and Woody Sullivan); and *Histoire de la Radioastronomie Française* (2025, EDP Sciences, co-authored by James Lequeux). Wayne has also edited or co-edited a succession of conference proceedings.

Since 1985 Wayne has been a member of the IAU, and he is a former President of Commission C3 (History of Astronomy). In 2003 he founded the IAU's Historical Radio Astronomy Working Group, and is the current Radio Astronomy Subject Editor for the Third Edition of Springer's *Biographical Encyclopedia of Astronomers*. He also founded the IAU Working Group on Historic Transits of Venus, and is the founding Director of the large and dynamic Historical Section of the Royal Astronomical Society of New Zealand. Wayne is also an Editor of Springer's book series on Cultural and Historical Astronomy. In 2014 he founded the History & Heritage Working Group of the Southeast Asian Astronomy Network and ran this successfully until 2024. In the process he organised three different conferences on SE Asian Astronomical History; some of the presented papers ended up in the aforementioned Southeast Asian Springer book. Since 2004 Wayne has also served on the Executive Committee of the ICOA series of Asian–Oceanic conferences, and currently he is a Councillor of the Royal Astronomical Society of New Zealand and a member of the Editorial Board of its journal, *Southern Stars*.

In 1998 Wayne Orchiston and John Perdrix co-founded the *Journal of Astronomical History and Heritage*. After John's death, Wayne was the Managing Editor until 31 July 2022 when he passed ownership of the journal to the University of Science and Technology of China. In 2013 the IAU named minor planet 48471 'Orchiston', and in 2019 former PhD student Stella Cottam, and Wayne, were awarded the Donald E. Osterbrock Prize by the American Astronomical Society for their 2015 Springer book, *Eclipses, Transits and Comets of the Nineteenth Century* … In 2023 Wayne was elected an Honorary Member of the Royal Astronomical Society of New Zealand, and two of his former doctoral students edited the following Festschrift in his honour: Gullberg, S., and Robertson, P. (eds.), 2023. *Essays in Astronomical History and Heritage: A Tribute to Wayne Orchiston on His 80th Birthday* (Springer, 2023). In January 2024 the American Astronomical Society also awarded Wayne their LeRoy E. Doggett Prize for lifetime contributions to history of astronomy, and in August 2026 he received an Honorary DSc from the University of Southern Queensland.

Wayne and Darunee Lingling Orchiston live in a quiet village near Chiang Mai in northern Thailand. When not involved in astronomy Wayne particularly enjoys following Australian and New Zealand athletics and Formula 1 and Indycar racing.

**Dr. Carolyn Brown** was born in Toowoomba, Queensland in Australia and has been interested in astronomy since she was four years old. After watching Stephen Hawking's "A Brief History of Time" in 1991, she proclaimed that she would become an astrophysicist. Which she did. Undertaking degrees in Physics and Astronomy at the University of Southern Queensland, Carolyn took on part-time work in the University's Physics and Astronomy Department while she completed her PhD. In 2009, Carolyn became a full-time academic in Physics and Engineering at the University of Southern Queensland while continuing to follow her passion for astronomy through research and outreach programs.

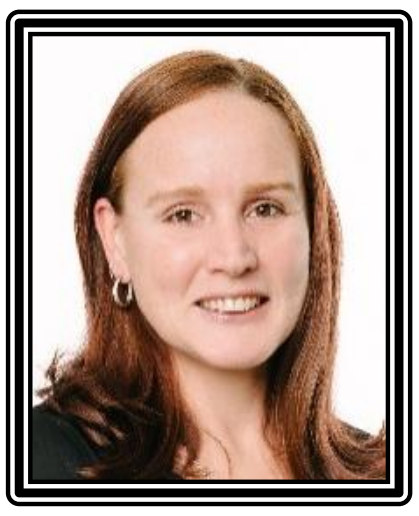

As Carolyn's career progressed, she found her love of teaching grew beyond her love of research, and she began to concentrate on educating the new generation and instilling a passion for astronomy in them, just like others had done for her. Taking on a primary teaching role at the University, Carolyn continued to foster this passion for astronomy in her students, especially her postgraduate students, until one day, convinced by one of her Doctoral students, she made the decision to move into the secondary school sector where she is hoping to make an impact on our youth to help drive them to follow their passions.

When not educating the next generation, Carolyn enjoys restoring classic Australian muscle cars where she can put her theoretical physics and engineering knowledge into a practical (and fast) application. Now all she needs is a bigger shed, with an observatory on the roof.

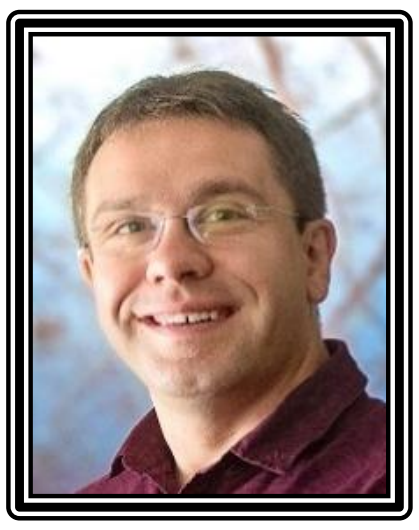

**Professor Jonathan (Jonti) Horner** was born in Wakefield, Yorkshire, in the United Kingdom, in 1978. At the age of five he saw part of an episode of "The Sky At Night" and became hooked on all things astronomical. He joined his local astronomical society—the West Yorkshire Astronomical Society (WYAS)—at the age of eight and began giving regular talks and writing for the society's journal, *Phobos*, by the age of ten. Thanks to the support and advice he received from both the members and the guest speakers at WYAS, Jonti moved to the University of Durham in 1996 to study for an undergraduate Masters' degree in Physics and Astronomy. After a summer project working at Armagh Observatory in 1999, with Professor Mark Bailey, Jonti moved to the University of Oxford in 2000, to begin work on his doctoral studies. His DPhil was conferred in 2004 for a thesis titled "The Behaviour of Small Bodies in the Outer Solar System." He spent time as a Postdoctoral Research Fellow at the University of Bern (Switzerland) and the Open University (UK), and a year as a teaching-only Fellow at Durham University (UK) before moving to Australia in 2010, to take up a Postdoctoral Fellowship at the University of New South Wales in Sydney. Finally, in 2014, Jonti moved to take up a position as Vice-Chancellor's Senior Research Fellow at the University of Southern Queensland in Toowoomba, where he remains to this day.

Jonti has a diverse range of research interests and has published more than 230 papers. His primary research focusses through his career have been the study of the Solar System's small bodies (particularly the Centaurs and planetary Trojans), the search for and characterisation of planets orbiting other stars (Exoplanets), and also the investigation of the various features that could render one planet more or less suitable as a target for the search for life beyond the Solar System. He is particularly proud of the work he did with his former mentor, Professor Barrie Jones, investigating the role that giant planets like Jupiter play in controlling the impact rate for terrestrial worlds—work that definitely shattered the long-held myth that Jupiter serves as Earth's celestial guardian and protector. Instead, Jupiter's role is more nuanced, with the giant planet actively throwing new objects to threaten the Earth with one hand whilst taking them away with the other. More recently, Jonti organised and led a lengthy review titled "Solar System Physics for Exoplanet Research", which has become a *de facto* textbook for undergraduate and Masters' courses around the world. His full publication list can be accessed through a library on the NASA ADS system, here: https://ui.adsabs.harvard.edu/public-libraries/YUefu-IlSQSn4d6JVwKokQ

Jonti is a member of the Astronomical Society of Australia, and an ongoing Committee Member of the Astrobiology Society of the Britain. He is a past member of Australia's National Committee for Space and Radio Science, and serves as the Honorary President of the West Yorkshire Astronomical Society. He is an active and enthusiastic science communicator, regularly appearing in national and international media to discuss stories about planetary and exoplanetary science, and astrobiology. He has written more than one hundred articles for "The Conversation", and is currently serving as a guest presenter on the globally popular "SpaceNuts" podcast. Jonti's ORCID is 0000-0002-1160-7970.

When he is not working as a professional astronomer, Jonti is a keen photographer, both of wildlife and as an astrophotographer. He is also a member of the Toowoomba Philharmonic Society's chorus, and enjoys gaming with friends and family at his home in the Darling Downs.